\documentclass[12pt]{iopart}
\usepackage{graphicx}

\newcommand{\ve}[1]{\bi{#1}}  

\begin{document}

\topical[On the nature of surface roughness with applications]{On the nature
 of surface roughness with application to contact mechanics, sealing,
 rubber friction and adhesion}
\author{B N J Persson\dag\ddag, O Albohr\P, U Tartaglino\dag$\S\sharp$,
 A I Volokitin\dag\ and E Tosatti\ddag\S$\sharp$}

\address{\dag\
 IFF, FZ-J\"ulich, 52425 J\"ulich, Germany}
\address{\ddag\
 International Center for Theoretical Physics (ICTP),
 P.O.Box 586, I-34014 Trieste, Italy}
\address{\P\
 Pirelli Deutschland AG, 64733 H\"ochst/Odenwald, Postfach 1120, Germany}
\address{\S
 International School for Advanced Studies (SISSA),
 Via Beirut 2, I-34014 Trieste, Italy}
\address{$\sharp$
 INFM Democritos National Simulation Center, Trieste, Italy}

\begin{abstract}
Surface roughness has a huge impact on many important phenomena.
The most important property of rough surfaces is the surface roughness
power spectrum $C(q)$.
We present surface roughness power spectra
of many surfaces of practical importance, obtained from the surface
height profile measured using optical methods and the Atomic Force
Microscope.
We show how the power spectrum determines
the contact area between two solids. We also present
applications to sealing, rubber friction and adhesion
for rough surfaces, where the power spectrum enters as an important
input.
\end{abstract}
\vfill

\begin{center}
\small
\noindent
Published on \emph{J.~Phys.:~Condens.~Matter} \textbf{17} (2005) R1-R62 \\
URL: http://stacks.iop.org/JPhysCM/17/R1 \\
DOI: 10.1088/0953-8984/17/1/R01
\end{center}

\maketitle

\clearpage


{\bf Contents}
\vskip 0.3cm

{\bf 1. Introduction}

{\bf 2. Surface roughness power spectrum: definition and general properties}

{\bf 3. Surface roughness power spectrum: experimental results}

     3.1. Surfaces produced by crack propagation

     3.2. Polished crack surfaces

     3.3. Surfaces with long-distance roll-off

     3.4. Road surfaces

     3.5. Other surfaces of practical interest

{\bf 4. Contact mechanics}

     4.1. Elastic contact mechanics

     4.2. Surface stiffness of fractal surfaces

     4.3. Viscoelastic contact mechanics

     4.4. Tack

{\bf 5. Seals}

{\bf 6. Rubber friction}

     6.1. Basic theory of rubber friction

     6.2. Rubber friction and the influence of polishing

     6.3. Rubber friction on wet road surfaces

     6.4. Lubricated rubber O-ring seals

{\bf 7. Adhesion}

     7.1. Adhesion between rough surfaces

     7.2. The adhesion paradox

     7.3. Adhesion in rubber technology

     7.4. Adhesion in biology

     7.5. The role of liquids on adhesion between rough solid surfaces

{\bf 8. Summary and outlook}

{\bf Appendix A: more about surface roughness}

{\bf Appendix B: Hurst exponent and fractal dimension}

{\bf Appendix C: Moments of power spectra}

{\bf Appendix D: Numerical recipes for calculating power spectra}

\clearpage

\section{Introduction}
\label{sec1}

Surface roughness has an enormous influence on many important physical
phenomena such as contact mechanics, sealing, adhesion and friction.
Thus, for example, experiments have shown that already a substrate with
a root-mean-square ({\it rms}) roughness of order $\sim 1 \ {\rm \mu m}$ can
completely remove the adhesion between a rubber ball and a substrate,
while nanoscale roughness will remove the adhesion between
most hard solids, e.g., metals and minerals; in other words, roughness is the main reason why
adhesion is usually not observed in most macroscopic phenomena.
Similarly, rubber friction
on most surfaces of practical interest, e.g., road surfaces, is mainly due
to the roughness-induced pulsating forces which act on the rubber
surface as it slides over the substrate asperities.

\begin{figure}[hbt]
\begin{center}
 \includegraphics[width=0.80\textwidth]{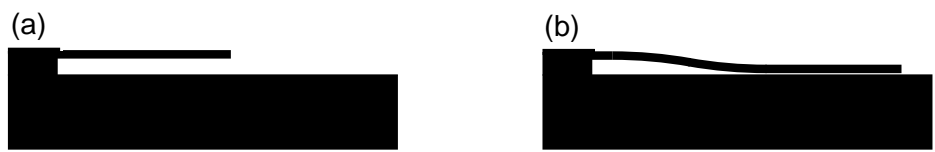} 
\end{center}
\caption{\label{beam}
(a) Micrometer sized cantilever beam. (b) If the beam is too long or too
thin the minimum free energy state corresponds to the beam partly bound
to the substrate.  Surface roughness lowers the binding energy (per unit
area) and hence stabilizes the non-bonded state in (a).}
\end{figure}

Let us illustrate the importance of surface roughness with three modern
applications.  At present there is a strong effort to produce small
mechanical devices, e.g., micromotors.  The largest problem in the
development of such devices is the adhesion and, during sliding, the
friction and wear between the contacting surfaces \cite{YU2003}. As an
example, in Fig.~\ref{beam} we show the simplest possible micro device,
namely a micrometer cantilever beam.  (Suspended micromachined
structures such as plates and beams are commonly used in the manufacturing
of pressure and accelerator sensors.)
If the beam is sufficiently long or thin the free beam state in
(a) will be unstable, and the bound state in (b)
will correspond to the minimum free energy state \cite{Mastrangelo}.
Roughly speaking, the state (b) is stable if the binding energy to the
substrate is higher than the elastic energy stored in the bent
beam. The binding energy to the substrate can be strongly reduced by
introducing (or increasing) the surface roughness on the substrate (see
Sec.~\ref{sec7.1}).  In addition, if the surfaces are covered by appropriate
monolayer films, these surfaces can be made hydrophobic thus eliminating
the possibility of formation of (water) capillary bridges.

\begin{figure}
\begin{center}
 \includegraphics[width=0.5\textwidth]{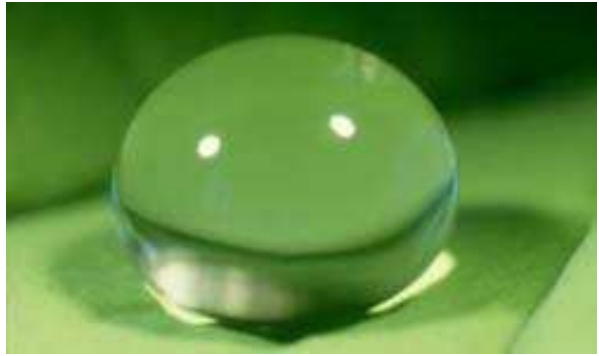} 
\end{center}
\caption{\label{plant1}
A water droplet on a superhydrophobic surface: The droplet touches the leaf
only in a few points and forms a ball. It completely rolls off at the
slightest declination \cite{Inter1}. Reproduced with permission.}
\end{figure}

\begin{figure}
\begin{center}
 \includegraphics[width=0.5\textwidth]{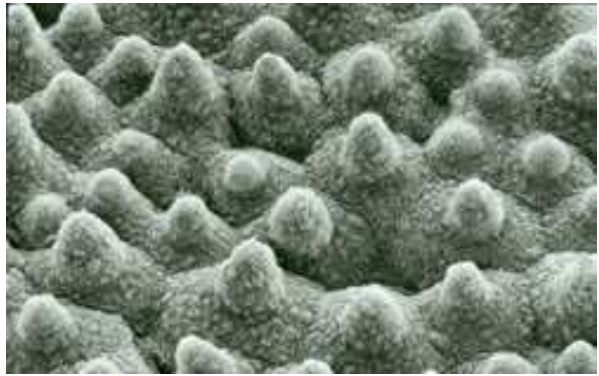} 
\end{center}
\caption{\label{plant2}
A leaf surface with roughness on several length scales
optimized (via natural selection) for hydrophobicity and self-cleaning.
Through the combination of microstructure (cells) and nanostructure
(wax crystals) the water contact angle $\theta_0$ is maximized \cite{Inter1}.
Reproduced with permission.}
\end{figure}

A second application is the formation of hydrophobic coatings on surfaces by
creating the appropriate type of surface roughness \cite{Inter1}.
This amounts to copying Nature where many plant surfaces are
found to be highly hydrophobic (Fig.~\ref{plant1}) as a result of the
formation of special types of surface roughness (Fig.~\ref{plant2}).
The surface roughness allows air to be trapped between the liquid and the substrate,
while the liquid is suspended on the tips of the asperities.
Since the area of real liquid-substrate contact is highly reduced, the contact angle of the drop is
determined almost solely by the surface tension of the liquid, leading to a very large contact angle.
New commercial products based on this ``Lotus effect'' \cite{Quere},
such as self-cleaning paints and glass windows, have been produced.

Finally, we mention the effect of surface roughness on direct wafer
bonding \cite{Gui1999}. Wafer bonding
at room temperature is due to relatively weak interatomic attraction forces, e.g., the van der
Waals interaction or hydrogen bonding, giving (for perfectly flat surfaces) an interfacial
binding energy of order $6\,{\rm meV/\mbox{\AA}^2}$.
The wafer surface roughness is the most critical parameter determining the
strength of the wafer bonding. In particular, when the surface roughness exceeds a critical
value, the wafers will not bind at all, in agreement with the theory presented in Sec.~\ref{sec7.1}.
Primary grade polished silicon wafer surfaces have {\it rms} roughness of order $\sim 0.1 \ {\rm nm}$
when measured over a $10\times 10 \ {\rm \mu m}$ surface area, and such surfaces bind spontaneously.
However, when the surface roughness amplitude is of order $1 \ {\rm nm}$ the surfaces either bind
(slowly) when squeezed together at high enough pressure,
or they do not bind at all depending on the
detailed nature of the surface roughness power spectra.

Surfaces with ``ideal'' roughness, e.g., prepared by fracture or by some growth process, have been
studied intensively for many years \cite{review,[b],[86],Krim}.
However, much less information has been available for more
common surfaces of engineering interest. In this article we discuss the
nature of the power
spectra of some surfaces of practical importance. As illustrations we
discuss contact mechanics, sealing, rubber friction and adhesion.

This paper is organized as follows: In Sec.~\ref{sec2} we define the surface roughness power spectrum $C(q)$,
and discuss some of its properties. In Sec.~\ref{sec3} we present power spectra deduced from
surface topography measurements for different surfaces of technological importance.
Sec.~\ref{sec4} illustrates how the surface roughness power spectrum determines the contact area
between elastic and viscoelastic solids. In Sec.~\ref{sec5} we briefly consider the influence of surface roughness
on sealing.
In Sec.~\ref{sec6} we discuss rubber friction on rough substrates,
and show some results for how the friction depends on the
surface roughness power spectrum. We also discuss the influence of
tire-road polishing and discuss rubber friction
on wet road surfaces. In Sec.~\ref{sec7} we  discuss adhesion between
rough surfaces and present two applications
which illustrate the importance of surface roughness for adhesion
in technology and biology.
Sec.~\ref{sec8} contains our summary and an outlook.

\section{Surface roughness power spectrum: definition and general properties}
\label{sec2}

The influence of roughness on
the adhesional and frictional properties introduced above
is mainly determined by the surface roughness
power spectrum $C(q)$ (or power spectral density) defined by \cite{Nayak}
\begin{equation}
 C(q)= {1\over (2\pi )^2 } \int \rmd^2x \ \langle h(\ve{x})h(\ve{0})\rangle
 \rme^{-\rmi\ve{q}\cdot \ve{x}}
\label{eq1}
\end{equation}
Here $\ve{x}=(x,y)$ and $z=h(\ve{x})$ is the substrate height measured
from the average surface plane, defined so that $\langle h \rangle = 0$.
The $\langle \ldots \rangle $ stands for ensemble averaging, i.e.,
averaging over a collection of different surfaces with identical
statistical properties.
We have assumed that the statistical properties of the substrate
are translationally invariant, so that the correlation
$\langle h(\ve{x}+\ve{x}_0) h(\ve{x}_0)\rangle$ does not depend on the
choice of $\ve{x}_0$, but only on the in-plane distance vector $\ve{x}$.
Alternatively the height correlation can be defined through a spatial
average over $\ve{x}_0$:
\[
  \lim_{L\to\infty} {1 \over L^2} \int_{-L/2}^{+L/2}\rmd x_0
      \int_{-L/2}^{+L/2}\rmd y_0 \  
        h(x_0+x,y_0+x)h(x_0,y_0)
\]
The latter definition is quite popular in the engineers'
community \cite{Nayak}, and it is closer to the experimental procedure to
measure the power spectrum. We notice, however,
that the lateral size $L$ of the surface area is always finite
in any experimental measurement or numerical calculation.
In some special cases, as the ones discussed in Sec.~\ref{sec3.1},
the spatial average is not equivalent to the ensemble average
for any practical choice of $L$.
Finally we point out that the power spectrum can also be defined
through the square modulus of the Fourier transform of $h$; this is
indeed the conventional approach adopted in the context of signal
theory and electronics, and it explains the origin of the name
``power spectral density''. Nonetheless such approach is equivalent to
Eq.~(\ref{eq1}) because of the Wiener-Khintchine theorem (see
\ref{AppendixC} and particularly Eq.~(\ref{eqC6})).

Together with translational invariance, in (\ref{eq1}) we have also
assumed that the statistical properties of
the substrate are isotropic, so that $C(q)$ only depend on the
magnitude $q=|\ve{q}|$ of the wave vector $\ve{q}$. Note that from (\ref{eq1}) follows
\[
 \langle h(\ve{x})h(\ve{0})\rangle =
   \int \rmd^2q \ C(q) \rme^{\rmi\ve{q}\cdot \ve{x}}
\] 
so that the root-mean-square roughness amplitude
$\sigma = \langle h^2\rangle^{1/2}$ is determined by
\begin{equation}
  \langle h^2 \rangle =
  \int \rmd^2q \ C(q) = 2\pi \int_0^\infty \rmd q \  q C(q)
  \label{eq2}
\end{equation}
In reality, there will always be an upper limit and a lower limit to the
$q$-integral in (\ref{eq2}). Thus, the largest possible wave vector will be of
order $2\pi /a$, where $a$ is a short wavelength cutoff corresponding perhaps
to some lattice constant, while the smallest possible wave vector is 
of order $2\pi /L$ where $L$ is the linear size of the surface.
In general, one may define a root-mean-square roughness amplitude which depends 
on the range of roughness $(q_0, q_1)$ involved in the integral in (\ref{eq2}):
\begin{equation}
 \langle h^2 \rangle (q_0,q_1) =
 2\pi \int_{q_0}^{q_1} \rmd q \  q C(q)
 \label{eq3}
\end{equation}

\begin{figure}
\begin{center}
\includegraphics[width=0.80\textwidth]{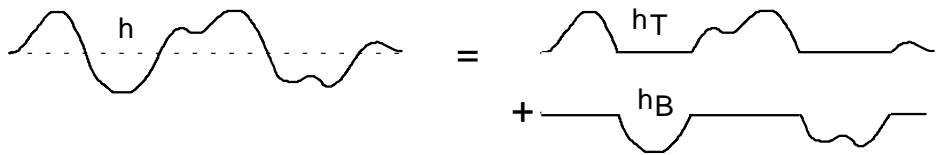} 
\end{center}
\caption{\label{TB}
The surface profile $h(x)$ is decomposed into a top
$h_T(x)$ and a bottom $h_B(x)$ profile.}
\end{figure}

\begin{figure}
\begin{center}
\includegraphics[width=0.70\textwidth]{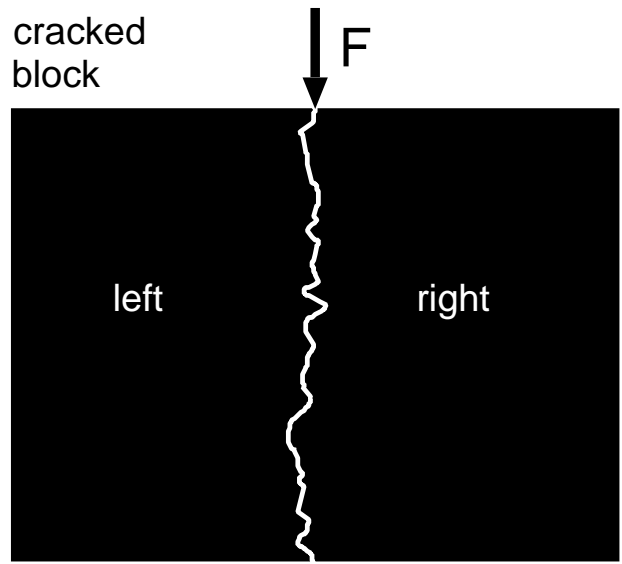} 
\end{center}
\caption{\label{cracked}
Rough surfaces prepared by crack propagation have surface roughness with statistical properties
which must be invariant under the replacement of $h\rightarrow -h$. This follows from the fact that
what is a valley on one of the crack surfaces (say the left) is
an asperity with respect to the other crack surface (right). Thus the top and bottom
power spectra must obey $C_T(q)=C_B(q)$.}
\end{figure}

For a randomly rough surface, when $h(\ve{x})$ is a Gaussian random variable, the statistical
properties of the surface are completely defined by the power spectrum $C(q)$
(see \ref{AppendixA}). In this case
the height probability distribution
\[ P_h = \langle \delta [h-h(\ve{x}) ]\rangle \]
will be a Gaussian
\[ P_h = {1\over ( 2\pi)^{1/2} \sigma } \rme^{-h^2/ 2 \sigma^2} \]
The height distribution of many
natural surfaces, e.g., surfaces prepared by fracture, or surfaces
prepared by bombardment by small particles (e.g., sand blasting or ion sputtering)
are usually nearly Gaussian. On the other hand, rough surfaces, e.g.,
prepared by fracture, and then (slightly)
polished, display a non-symmetric height distribution (i.e., no symmetry as $h \rightarrow -h$) since
the asperity tops have been polished more than the bottom of the valleys.
These surfaces, typically of considerable practical importance--see below,  
have non-Gaussian height distribution. For such surfaces it is interesting to study 
the {\it top}, $C_T$, and the {\it bottom}, $C_B$, power spectra
defined by
\numparts
\begin{equation}
  C_T(q)= {1\over (2\pi )^2 } \int \rmd^2x
   \ \langle h_T(\ve{x})h_T(\ve{0})\rangle
   \rme^{-\rmi\ve{q}\cdot \ve{x}}
 \label{eq4a}
\end{equation}
\begin{equation}
  C_B(q)= {1\over (2\pi )^2 } \int \rmd^2x
   \ \langle h_B(\ve{x})h_B(\ve{0})\rangle
   \rme^{-i\ve{q}\cdot \ve{x}}
\label{eq4b}
\end{equation}
\endnumparts
where $h_T(\ve{x}) = h(\ve{x})$ for $h > 0$ and zero otherwise, while
$h_B(\ve{x}) = h(\ve{x})$ for $h < 0$ and zero otherwise, the latter being
`rectified' profiles, see Fig.~\ref{TB}. It is easy to show that
$C\approx C_T+C_B$. It is also clear by symmetry that for a surface prepared by fracture,
$C_T(q)=C_B(q)$, since what is top on one of the cracked block surfaces is the bottom
on the other (opposite) crack surface, and vice versa, see Fig.~\ref{cracked}.
However, if the cracked surface is slightly polished then, because the
contact pressure with the polishing object
(e.g., sand paper) is highest at the asperity tops, the polishing at the asperity 
tops will be stronger  than at the valley bottoms, and thus $C_B > C_T$.
If $n_T$ and $n_B$ are the fraction of the nominal surface area (i.e., the surface area projected on the $xy$-plane)
where $h>0$ and $h<0$, respectively, with $n_T+n_B=1$, then we may  also define
$C_T^*(q)= C_T/n_T$ and $C_B^*=C_B/n_B$. In general, $n_T\approx n_B \approx 0.5$ and for surfaces
prepared by fracture $n_T=n_B=0.5$. Roughly speaking, $C_T^*$ would be the power
spectrum resulting if the actual bottom profile (for $h<0$) was replaced
by a mirrored top profile (for $h>0$).
A similar statement holds for $C_B^*$.

\begin{figure}
\begin{center}
\includegraphics[width=0.70\textwidth]{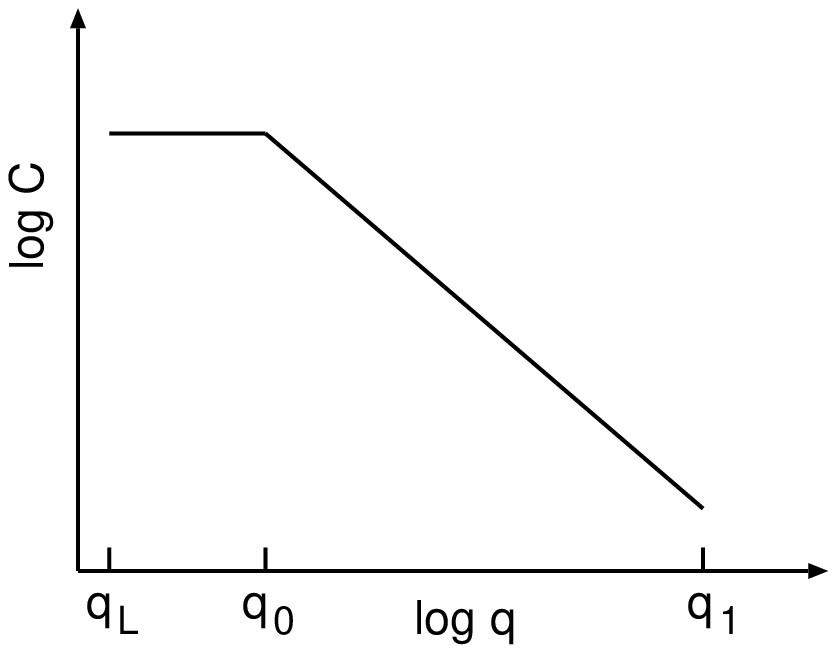} 
\end{center}
\caption{\label{Cq1}
Surface roughness power spectrum of a surface which is a self-affine fractal for
$q_0<q<q_1$. The long-distance roll-off wave vector $q_0$ and the short distance cut-off
wave vector $q_1$ depend on the system under consideration. The slope of
the $\log C-\log q$  plot for $q > q_0$ determines the fractal
exponent of the surface (see text). 
The lateral size $L$ of the available surface region
determines the smallest possible wave vector $q_L=2\pi /L$.}
\end{figure}

Many surfaces tend to be nearly self-affine fractal. A self-affine fractal
surface has the property that if part of the surface is magnified, with a magnification
which in general is appropriately different in the perpendicular direction to the surface as compared
to the lateral directions, then the surface ``looks the same'', i.e., the statistical
properties of the surface are invariant under the scale transformation (see \ref{AppendixB}).
For a self-affine
surface the power spectrum has the power-law behaviour
\[ C(q) \sim q^{-2(H+1)}, \]
where the Hurst exponent $H$ is related to the fractal dimension $D_{\rm f}$ of the 
surface via \cite{review}
$H=3-D_{\rm f}$. 
Of course, for real surfaces this relation only holds in some finite
wave vector region $q_0 < q < q_1$, and in a typical case $C(q)$ has the form shown
in Fig.~\ref{Cq1}. Note that in many cases there is a roll-off wavelength $q_0$ below which
$C(q)$ is approximately constant. We will discuss this point further below.

Finally, note that while the root-mean-square roughness usually
is dominated by the longest wavelength surface
roughness components, higher order moments of the power spectra such
as the average slope or the average surface
curvature are dominated by the shorter wavelength components. For example,
assuming a self affine fractal surface, Eq.~(\ref{eq3}) gives
\[
  \langle h^2\rangle (q_0,q_1) \sim
   \int_{q_0}^{q_1} \rmd q \ q^{-2H-1} \sim
   q_0^{-2H}-q_1^{-2H} \approx
   q_0^{-2H}
\]
if $q_1/q_0 \gg 1$. However, the average slope and the average
curvature have additional factors of $q^2$ and $q^4$,
respectively, in the integrand of the $q$-integral,
and these quantities are therefore dominated by the
large $q$ (i.e., short wavelength) surface roughness components (see \ref{AppendixC}).

\section{Surface roughness power spectrum: experimental results}
\label{sec3}

In this section we present power spectra for different surfaces of
practical importance. The power spectra have been
calculated using Eqs. (\ref{eq1}), (\ref{eq4a}) and (\ref{eq4b}) (see
\ref{AppendixD}), where the height profile $h(\ve{x})$ has been measured
using either optical methods\footnote{
  Many methods for measurement of surface topography have been developed
  see, e.g., http://www.michmet.com, http://www.solarius-inc.com or
  http://www.schmitt-ind.com for experimental equipment and some
  illustrative results. 
  }
or by Atomic Force Microscopy \cite{Krim}. 

\begin{figure}
\begin{center}
\includegraphics[width=0.70\textwidth]{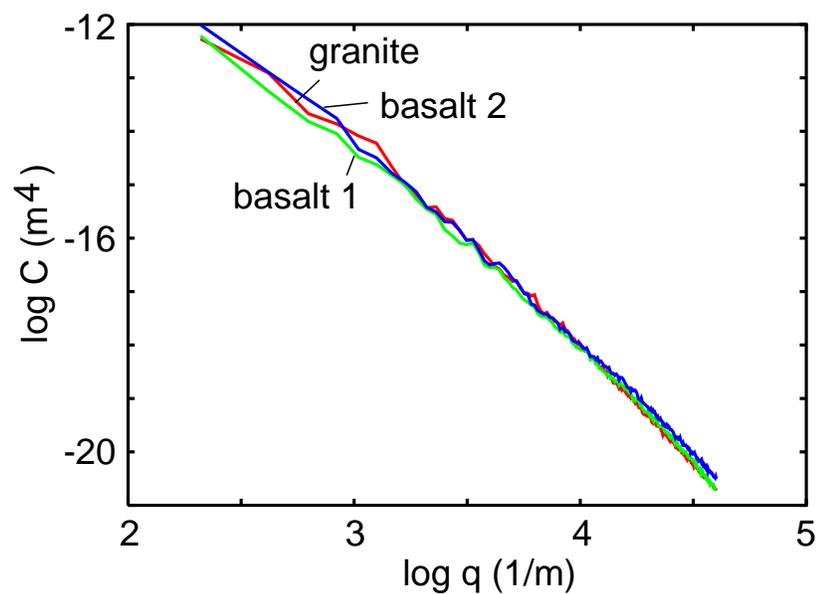} 
\end{center}
\caption{\label{GranitBasalt1}
The surface roughness power spectra for two freshly cleaved basalt surfaces
and a fresh granite surface.}
\end{figure}

\subsection{Surfaces produced by crack propagation}
\label{sec3.1}

Fig.~\ref{GranitBasalt1} shows the power spectra $C(q)$ for three
freshly cleaved stone surfaces, namely a granite
and two basalt stone surfaces. Here, and in what follows, we show the
power spectra on a log-log scale, where log stands for the base 10 logarithm.
Note that the granite and basalt
surfaces, in spite of the rather different mineral microstructure
(see below), give identical power spectra within the accuracy of the
measurement. It has been stated (see, e.g., Ref.~\cite{Bouch}) that
surfaces produced by crack propagation have self affine
fractal structure with the universal fractal dimension
$D_{\rm f} \approx 2.2$. However, our
measured $\log C-\log q$ relations are not perfectly straight lines,
i.e., the surfaces in the length-scale range studied
cannot be accurately described as self-affine fractals. Moreover
the average slope of the curves
in Fig.~\ref{GranitBasalt1} correspond to the fractal dimension
$D_{\rm f} \approx 2$ rather than $2.2$.

The similarity of the power spectra for the basalt and granite
surfaces in Fig.~\ref{GranitBasalt1} is striking. Granite and basalt both result
from magma and have a similar composition, consisting mainly of minerals
from the silicate group.  However, granite originates from magma which was
trapped deep in the crust, taking a very long time to cool down and
crystallize into solid rock. As a result granite is a coarse-textured
rock where individual mineral grains are easily visible. Basalt, on
the other hand, results from fast cooling of magma from, e.g., volcanic
eruptions, is fine grained, and nearly impossible to resolve into
the individual mineral grains without magnification. In spite of these
differences, the surface roughness power spectra of freshly cleaved
surfaces are nearly identical. This may indicate some kind of universal
power spectrum for surfaces resulting from cleaving of mineral stones of
different types, a point that could deserve further investigation. 

\begin{figure}
\begin{center}
\includegraphics[width=0.70\textwidth]{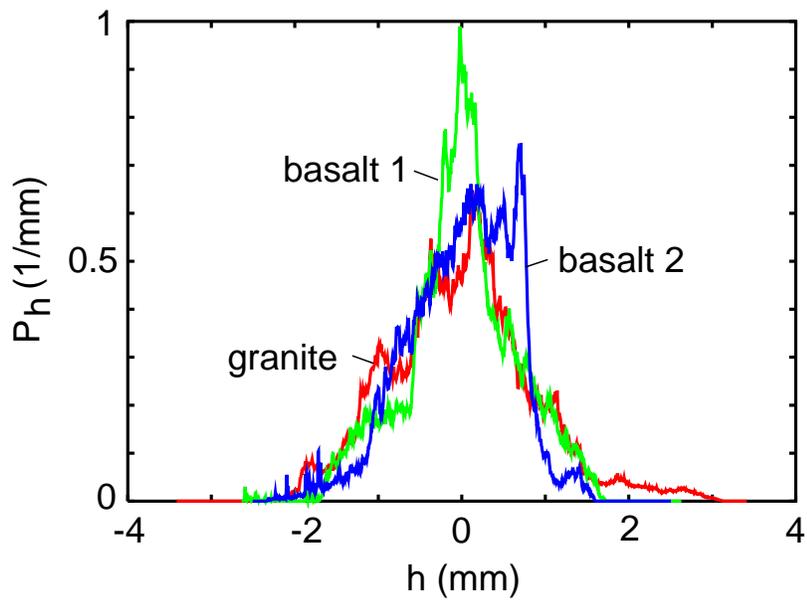} 
\end{center}
\caption{\label{Ph}
The height distribution $P_h$ for two freshly cleaved (cobble stone)
basalt surfaces and a fresh granite surface. Note the random
non-Gaussian nature of the height profiles.}
\end{figure}

Note that there is no roll-off region for these fracture-produced 
surfaces, whose behaviour appears fractal-like up to the
longest length scale studied, determined by the lateral size $L$ of the
surfaces (or of the regions experimentally studied, 
of the order of $1 \ {\rm cm}$ in our case), i.e., with
reference to Fig.~\ref{Cq1}, $q_0 = q_L \equiv 2\pi/L$. One consequence of this is
that the {\it rms}-roughness amplitude is determined mainly by the
$\lambda \sim L$ wavelength fluctuations of the surface height, and will
therefore depend on the size $L$ of the surface. Furthermore, the height
distribution $P_h$ obtained for any given realization of the rough
surface will not be Gaussian, but will exhibit random fluctuations as
compared to other realizations (see Fig.~\ref{Ph}, which illustrate this
point for the three stone surfaces discussed above).  However, the
ensemble averaged height distribution (not shown) should be Gaussian or
nearly Gaussian. Thus, when there is no roll-off region in the measured
power spectra, averaging over the surface area is not identical to
ensemble averaging. However, when there is a roll-off wave vector
$q_0 = 2\pi /\lambda_0$, and if the surface is studied over a region
with the lateral size $L\gg \lambda_0$, ensemble averaging and averaging
over the surface area $L\times L$  will give identical results for
$P_h$, and the {\it rms}-roughness amplitude will be independent of $L$
for $L \gg \lambda_0$.

\begin{figure}
\begin{center}
\includegraphics[width=0.70\textwidth]{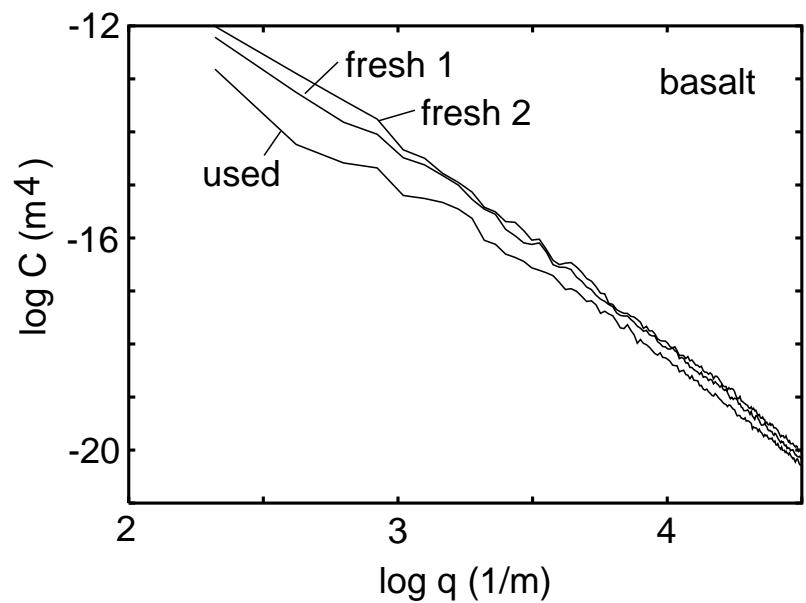} 
\end{center}
\caption{\label{Cq77}
The surface roughness power spectra $C(q)$ for two freshly cleaved
cobble stone (basalt) surfaces, and for a wear-polished (used)  surface.}
\end{figure}

\subsection{Polished crack surfaces}
\label{sec3.2}

\begin{figure}
\begin{center}
\includegraphics[width=0.70\textwidth]{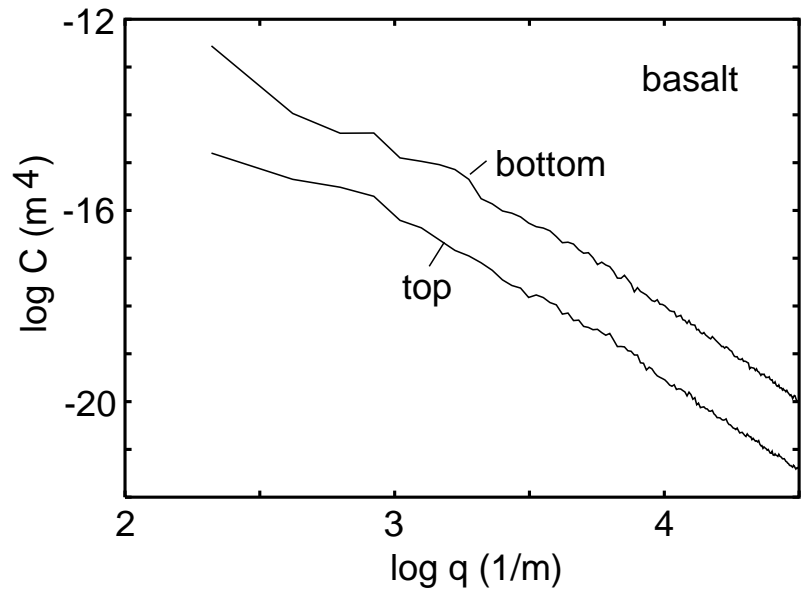} 
\end{center}
\caption{\label{CqTB}
The top $C_{\rm T}^*$ and the bottom $C_{\rm B}^*$
surface roughness power spectra $C(q)$ for a used cobble stone (basalt)
surface.}
\end{figure}

In the past, cobble stones, made of granite or basalt,
were frequently used for road surface pavements.
However, these surfaces do not exhibit good frictional properties
against rubber. In particular, with increasing time of use, the cobble
stone surfaces become polished by slipping tires (see
Sec.~\ref{sec6.2}), and that polishing results in a reduced rubber-road
friction, even during dry driving conditions\footnote{
  Cobble stones made from porphyry (a volcanic rock), thanks to the
  difference in hardness of the minerals it contains (see
  Sec.~\protect\ref{sec6.2}), have better polishing properties than
  cobble stones made of granite and basalt. See, e.g.,
  http://www.bourgetbros.com.
  }.
Fig.~\ref{Cq77} illustrate this polishing
effect. It shows the power spectrum
of a strongly used (basalt) cobble stone, and
of two freshly cleaved surfaces (from Fig.~\ref{GranitBasalt1}), from the same cobble stone. At long wavelength the
power spectrum of the strongly used surface is nearly one decade smaller than that of the freshly prepared
surfaces. The effect of polishing is further emphasized by
calculating the top and bottom power spectra,
$C_T^*$ and $C_B^*$, as shown in Fig.~\ref{CqTB}. The top power spectrum of the worn surface
is a factor $\sim 30$ times smaller than the bottom spectrum for {\em all wave vectors}. 
As anticipated, the asymmetry arises from the higher polishing of asperities
relative to valleys.
It is of course crucial to take this polishing effect
into consideration when designing road pavements, and we will
discuss this point further in Sec.~\ref{sec6.2}.

\begin{figure}
\begin{center}
\includegraphics[width=0.70\textwidth]{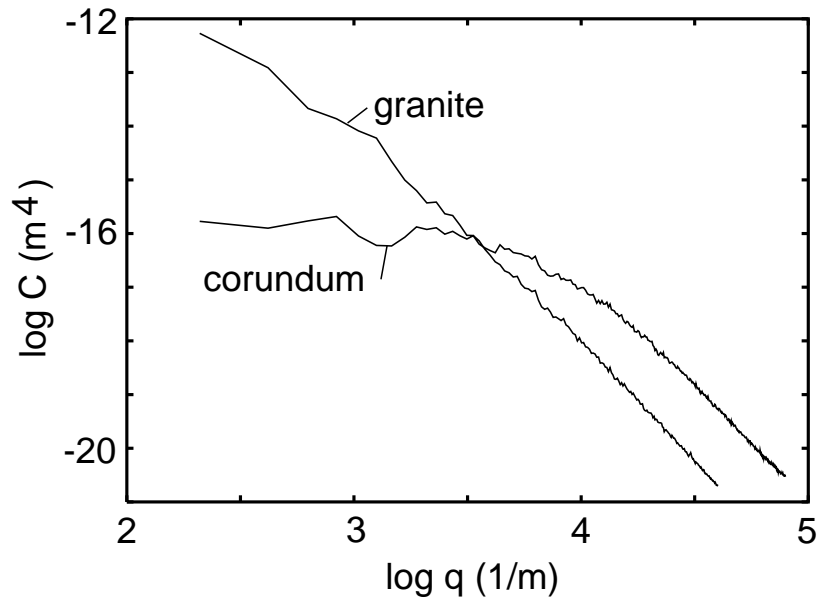} 
\end{center}
\caption{\label{GranitBasalt2}
The surface roughness power spectra for a fresh
granite surface and a fresh sintered corundum surface.}
\end{figure}

\subsection{Surfaces with long-distance roll-off}
\label{sec3.3}

\begin{figure}
\begin{center}
\includegraphics[width=0.70\textwidth]{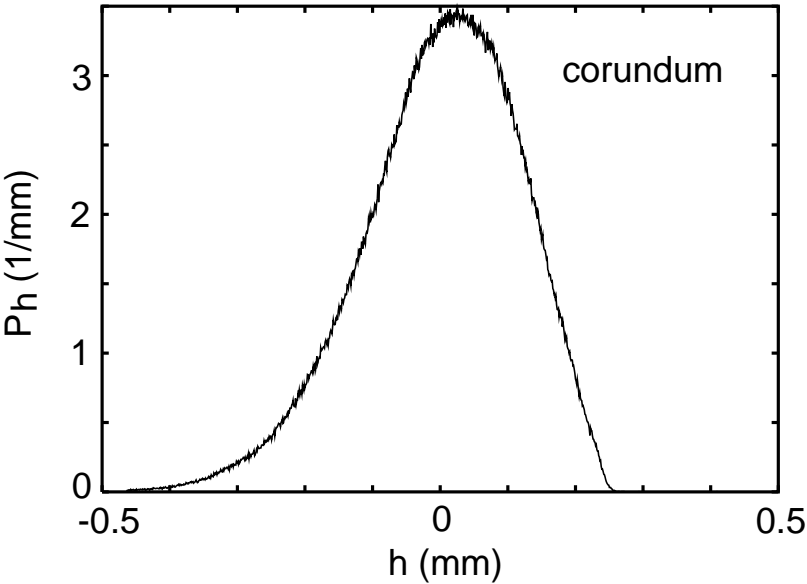} 
\end{center}
\caption{\label{GranitBasalt}
The height distribution $P_h$ as a function of the height $h$ for a
sintered corundum surface.}
\end{figure}

\begin{figure}
\begin{center}
\includegraphics[width=0.70\textwidth]{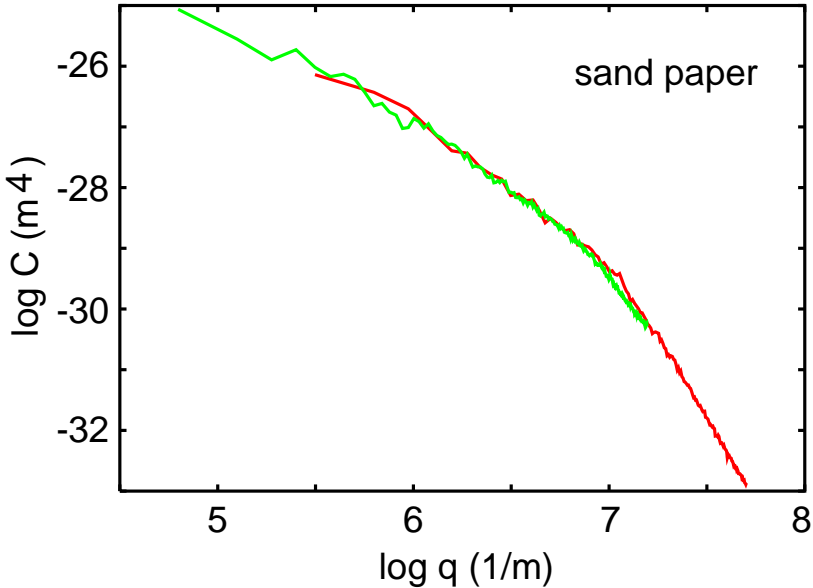} 
\end{center}
\caption{\label{CqOpel2}
The surface roughness power spectra $C(q)$ for a sandpaper surface.
The two curves are based on the height profiles measured with an
AFM at two different spatial resolution over $20\times 20$ (red)
and $100\times 100$ ${\rm \mu m}$ (green) square areas.}
\end{figure}

\begin{figure}
\begin{center}
\includegraphics[width=0.70\textwidth]{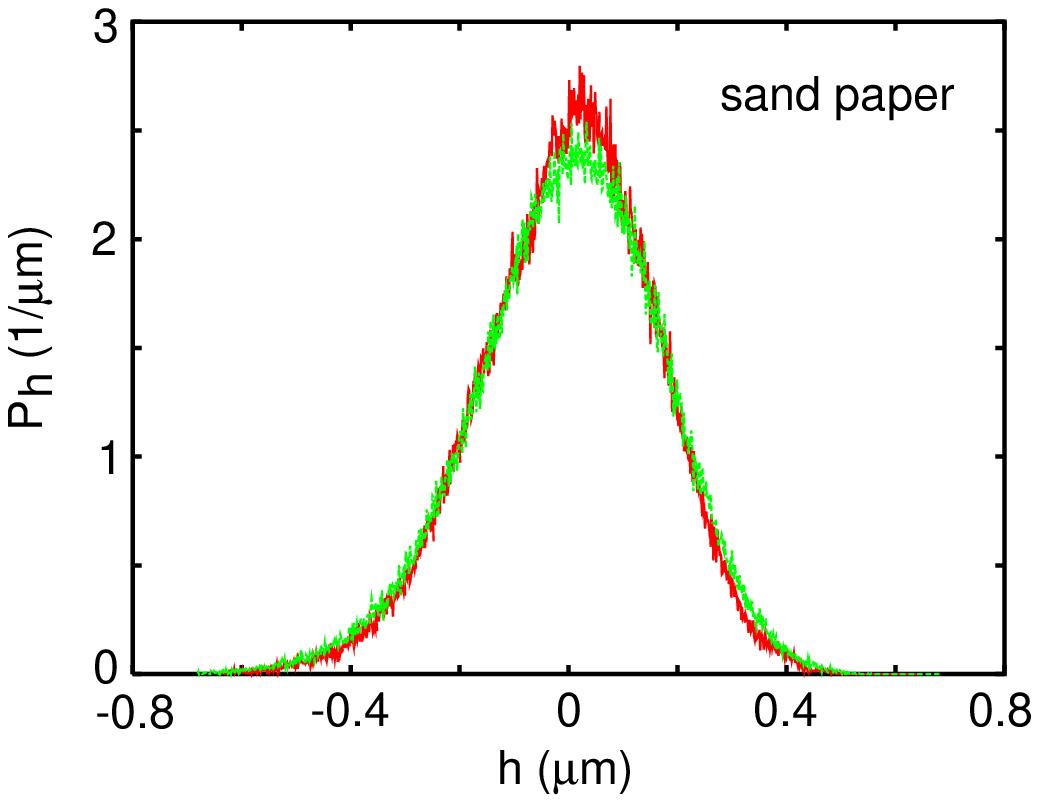} 
\end{center}
\caption{\label{CqOpel1}
The surface roughness height probability distribution $P_h$ for a sandpaper
surface.
The two curves are based on the height profiles measured with an
AFM at two different spatial resolution over $20\times 20$ (red)
and $100\times 100$ ${\rm \mu m}$ (green) square areas.}
\end{figure}

As pointed out above, surfaces prepared by fracture have no natural
long-distance cut-off and the {\it rms} roughness amplitude increases
continuously and without limit as the probed surface area increases. This
is similar to Brownian motion where the mean square displacement
increases without limit (as $\sim t^{1/2}$) as the time $t$ increases.
However, most surfaces of engineering interest have a long distance
cut-off or roll-off wavelength $\lambda_0$ corresponding to a wave
vector $q_0 = 2\pi /\lambda_0$, as shown in Fig.~\ref{Cq1}. For example,
if a flat surface is sand blasted for some time the resulting rough
surface will have a long distance roll-off length, which increases with
the time of sand blasting. Similarly, if atoms or particles are
deposited on an initially flat surface the resulting rough surface will
have a roll-off wavelength which increases with the deposition time, as
has been studied in detail in recent growth models.
Another way to produce a surface with a long-distance roll-off
wavelength is to prepare the solid as a conglomerate of small particles. A nominally
flat surface of such a solid has still roughness on length scales
shorter than the diameter of the particles, which therefore may act as a
long distance roll-off wavelength. We illustrate this here with a solid
produced by sintering
together corundum particles at high temperature
and pressure (Fig.~\ref{GranitBasalt2}), and for a sandpaper surface
(Fig.~\ref{CqOpel2}).  For both surfaces the height distribution $P_h$
is smooth
(see Figs.~\ref{GranitBasalt} and
\ref{CqOpel1}), since averaging over a surface area with lateral size
$L\gg\lambda_0$ is equivalent to ensemble averaging.

The sandpaper surface in Fig.~\ref{CqOpel2} was studied using the AFM
at two different resolutions over square areas $20\times 20 \ {\rm \mu m}$
and $100\times 100 \ {\rm \mu m}$ as indicated by the two different lines in
Fig.~\ref{CqOpel2}. The height distribution $P_h$ (and hence also the
{\it rms}-roughness amplitude) calculated from these two different
measurements over different surface areas, see Fig.~\ref{CqOpel1}, are
nearly identical, as expected when $L$ is larger than the
roll-off length $\lambda_0$.

\subsection{Road surfaces}
\label{sec3.4}

\begin{figure}
\begin{center}
\includegraphics[width=0.70\textwidth]{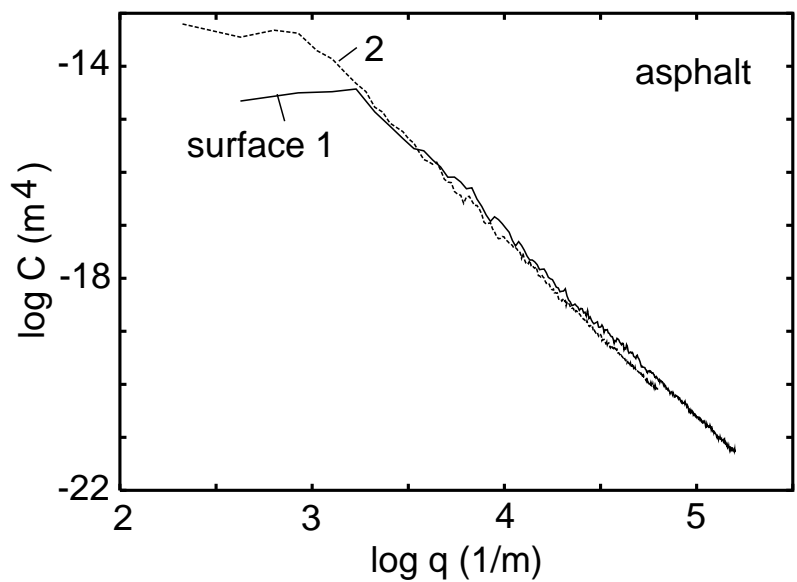} 
\end{center}
\caption{\label{CqOpel}
The surface roughness power spectra $C(q)$ for two asphalt road surfaces.}
\end{figure}

\begin{figure}
\begin{center}
\includegraphics[width=0.70\textwidth]{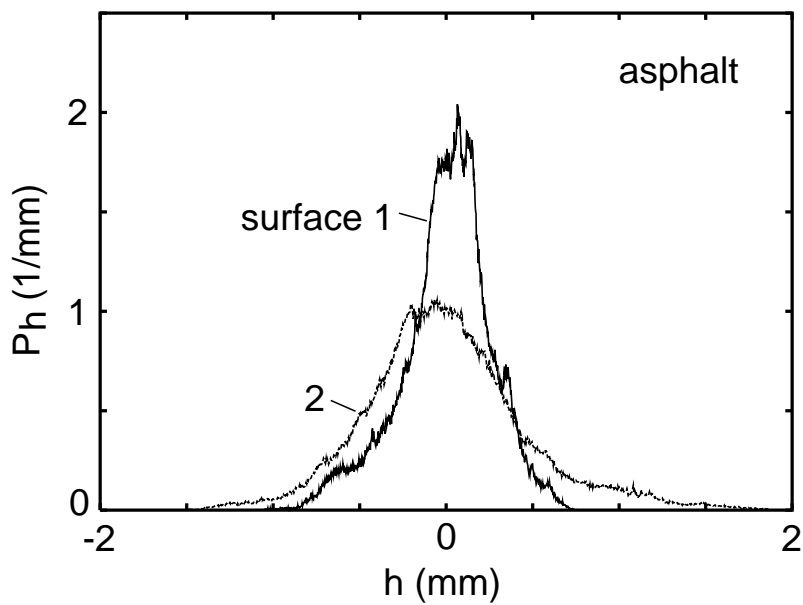} 
\end{center}
\caption{\label{asphaltPh}
The height distribution $P_h$ for two different asphalt road surfaces.
}
\end{figure}

Asphalt and concrete road pavements have nearly perfect self-affine
fractal power spectra, with a very well-defined roll-off wave vector $q_0
= 2\pi /\lambda_0$ of order $1000 \ {\rm m}^{-1}$, corresponding to
$\lambda_0 \approx 1 \ {\rm cm}$, which reflects the largest stone
particles used in the asphalt. This is illustrated in Fig.~\ref{CqOpel}
for two different asphalt pavements. From the slope of the curves for
$q>q_0$ one can deduce the fractal dimension $D_{\rm f} \approx 2.2$,
which is typical for asphalt and concrete road surfaces. The height
distributions of the two asphalt surfaces are shown in
Fig.~\ref{asphaltPh}.
Note that the {\it rms} roughness amplitude of surface {\bf 2} is nearly
twice as high as for surface {\bf 1}.  Nevertheless, the tire-rubber
friction is slightly higher on the road surface {\bf 1} because it has
slightly larger power spectra for most $q$-values in Fig.~\ref{CqOpel}.
Thus there is in general {\it no} direct correlation between the {\it
rms}-roughness amplitude and the rubber friction on road surfaces, as
will be further discussed in Sec.~\ref{sec6}.

Many attempts have been made to relate rubber friction on road surfaces to
the so called ``sand filling number''. The sand filling number is the
amount of very fine-grained sand needed to fill out all the
road surface cavities in a given surface area. However, no correlation
between the sand filling number and rubber friction on dry road surfaces
has been found \cite{Pirelli}.  In the light of modern rubber friction
theories, this result is not unexpected since the rubber friction depends on
the power spectrum for {\em all wave vectors}, while only the long-wavelength
components contribute appreciably to the sand filling number.
Thus, for example, of the two asphalt surfaces in Fig.~\ref{CqOpel}, the
sand filling factor of road {\bf 2} is nearly twice as large as for road
{\bf 1}, but the rubber friction is slightly higher on road surface {\bf 1}.

\subsection{Other surfaces of practical interest}
\label{sec3.5}

\begin{figure}
\begin{center}
\includegraphics[width=0.70\textwidth]{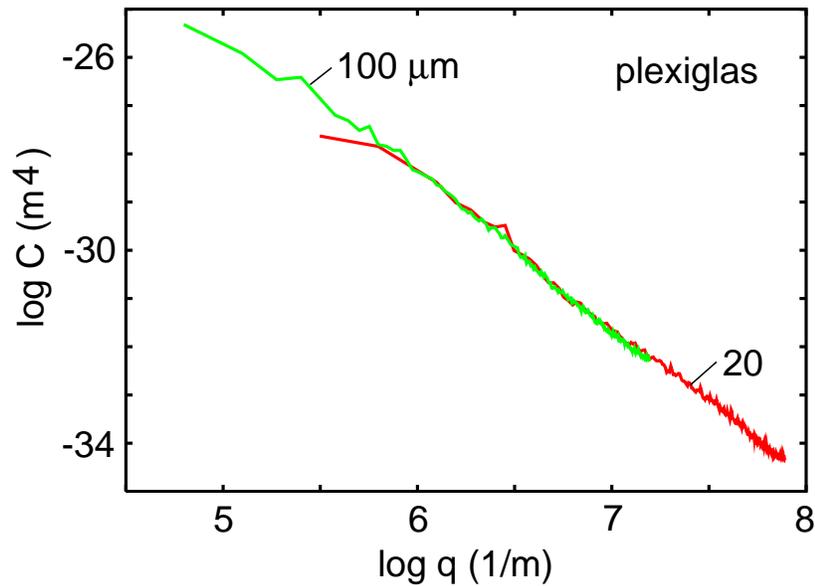} 
\end{center}
\caption{\label{CqOpel10}
The surface roughness power spectra $C(q)$ for a plexiglas surface.
The two curves are based on the height profiles measured with an
AFM at two different spatial resolution over $20\times 20$ and $100\times 100$
${\rm \mu m}$ square areas.}
\end{figure}

\begin{figure}
\begin{center}
\includegraphics[width=0.70\textwidth]{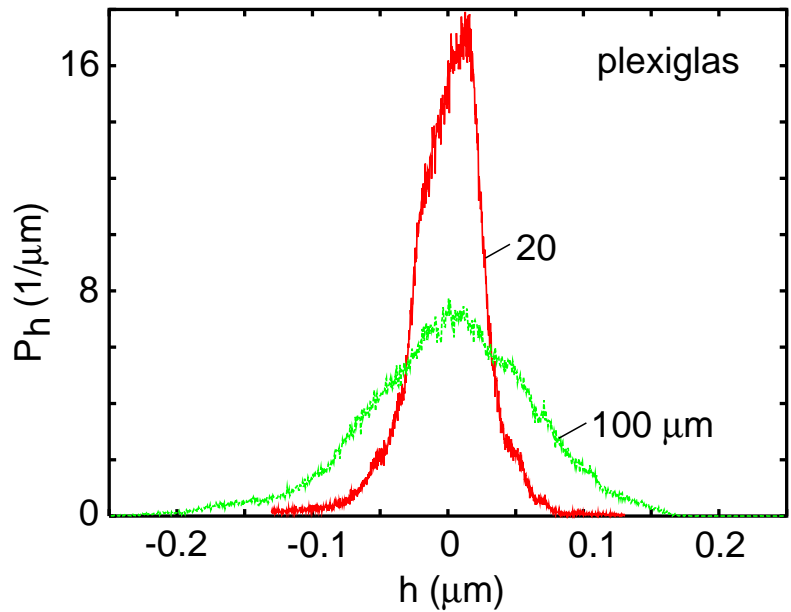} 
\end{center}
\caption{\label{CqOpel8}
The surface roughness height probability distribution $P_h$ for a
plexiglas surface.
The two curves are based on the height profiles measured with an
AFM at two different spatial resolution over $20\times 20$
and $100\times 100$ ${\rm \mu m}$ square areas.}
\end{figure}

Finally, let us consider two other surfaces of practical importance.
Fig.~\ref{CqOpel10} shows the power spectrum of a plexiglas surface
measured (using an AFM) at two different resolutions over two different
surface areas $20\times 20 \ {\rm \mu m}$ and $100\times 100 \ {\rm \mu
m}$ wide. This surface does not exhibit a roll-off wave vector in the
studied wave vector range and the height distributions deduced from the
two different surface areas differ strongly, see Fig.~\ref{CqOpel8},
with the {\it rms} roughness amplitude being more than twice as large
for the measurement over the larger surface area.  The $\log C$-$\log q$
relation in Fig.~\ref{CqOpel10} is a nearly perfect straight line, and
the slope correspond to the fractal dimension $D_{\rm f} \approx 2.4$.

\begin{figure}
\begin{center}
\includegraphics[width=0.70\textwidth]{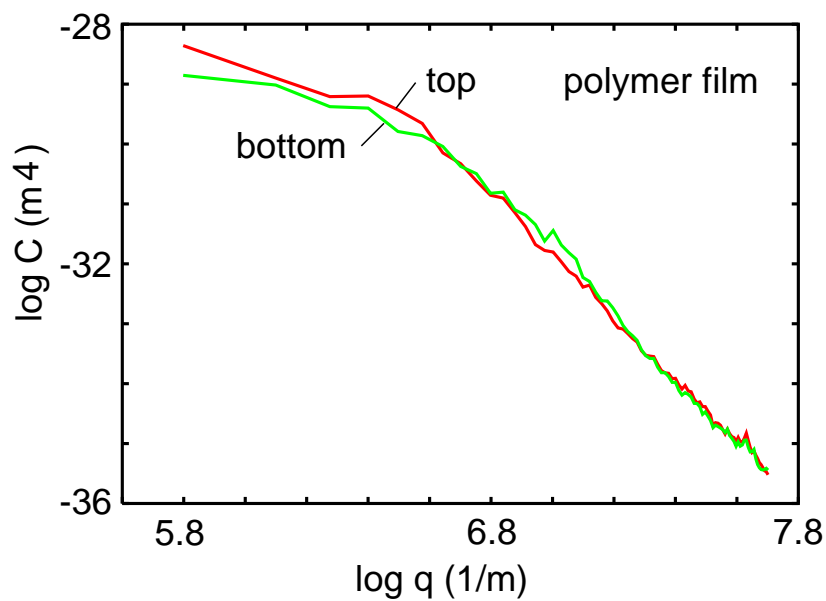} 
\end{center}
\caption{\label{CqOpel11}
The top $C_T^*$ and the bottom $C_B^*$
surface roughness power spectra for a polymer
film spin coated on a very smooth substrate and dried.
 From Ref.~\protect\cite{PlusIsrael}.}
\end{figure}

Fig.~\ref{CqOpel11} shows the top and bottom power spectra $C_T^*$ and
$C_B^*$ for a polymer film which was spin coated and dried
on a flat silicon surface. In this case there is a roll-off
wavelength $\lambda_0 \approx 1 \ {\rm \mu m}$ which is probably
related to the average polymer film thickness. Note that
within the accuracy of the experiment, $C_T^* = C_B^*$,
proving qualitatively that the
short-wavelength roughness in the large valleys is very similar to the
short-wavelength roughness on the large asperities.
The $\log C -\log q$ relation in Fig.~\ref{CqOpel11} for $\log q > 6.6$
is a nearly perfect straight line, and the slope corresponds to a
roughness exponent $H=1.6$. This is larger than unity, which 
implies that the
surface cannot be described as fractal at any scale.

\section{Contact mechanics}
\label{sec4}

Practically all macroscopic bodies have surfaces with
roughness on many different length scales.
When two bodies
with nominally flat surfaces are brought in contact,
real (atomic) contact will only occur in small
randomly distributed areas, and the area of real
contact is usually an extremely small fraction of the nominal
contact area.
We can visualize the contact regions as small areas where asperities
from one solid are squeezed against asperities of the other solid;
depending on the conditions the asperities may deform elastically or
plastically.

How large is the area of {\it real} contact between a solid block and
a substrate? This fundamental question has extremely important
practical implications. For example, it determines the contact
resistivity and the heat transfer between the solids. It is also of
direct importance for wear and sliding friction \cite{[1]}, e.g., the
rubber friction between a tire and a road surface, and has a major
influence on the adhesive force between two solid blocks in direct
contact.

\begin{figure}[htb]
\begin{center}
\includegraphics[width=0.70\textwidth]{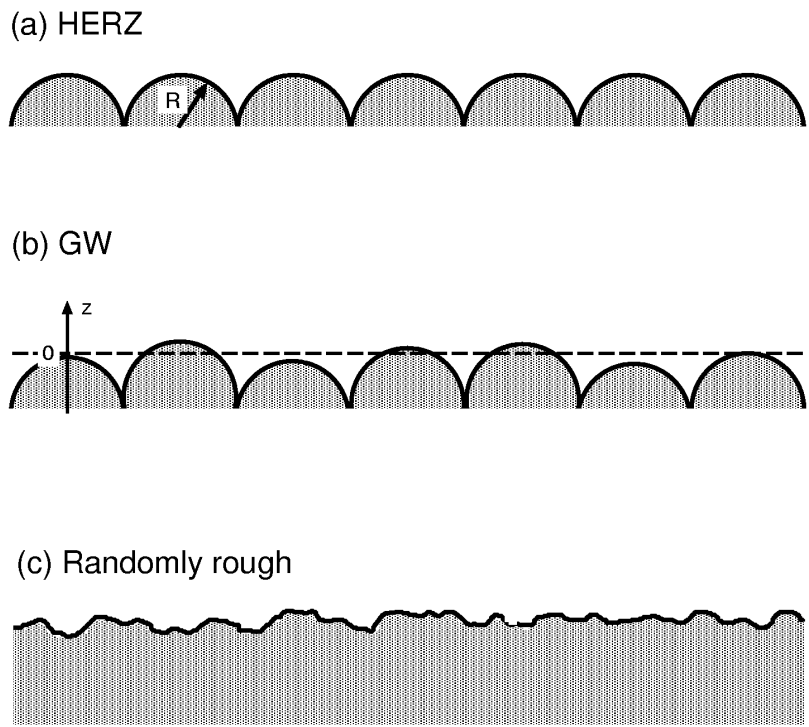} 
\end{center}
\caption{ \label{KKK}
Three models of ``rough'' surfaces. In case (a) all the ``asperities'' are
equally high and have identical radius of curvature. In this case, according
to the Hertz contact theory,
the area
of real contact $\Delta A$ between a solid with a flat surface
and the surface depends
non-linearly on the squeezing force (or load) $F_{\rm N}$ according to
$\Delta A \sim F_{\rm N}^{2/3}$. If the asperities have instead random height distribution 
as in (b) then, for small $F_{\rm N}$,
$\Delta A$ is {\it nearly} proportional to the squeezing force.
If the surface roughness is random with ``asperities'' of different heights
and curvature radii as in (c), the area of real contact for small $F_{\rm N}$
is {\it exactly}
proportional to the squeezing force.}
\end{figure}

Contact mechanics has a long history. The first study was presented by
Hertz \cite{Hertz}. He gave the solution for the frictionless normal contact of two elastic
bodies of quadratic profile. He found that the area of real contact
$\Delta A$ varies nonlinearly
with the load or squeezing force: $\Delta A \propto F_N^{2/3}$.
In 1957 Archard \cite{[6]} applied the Hertz
solution to the contact between rough surfaces and showed that for
a simple fractal-like model, where small spherical bumps (or asperities)
where distributed on top of larger spherical bumps
and so on, the area of real contact varies {\it nearly linearly} with $F_N$.
A similar conclusion was reached by Greenwood and Williamson \cite{[4],[5],[7]}
who again assumed asperities
with spherical summits (of identical radius) with a Gaussian distribution
of heights, as sketched in Fig.~\ref{KKK}(b).
A more general contact mechanics theory has been developed by Bush et al.\ \cite{Bush}
They approximated the summits by paraboloids and applied the
classical Hertzian solution for their deformation. The height distribution was
described by a random process, and they found that at low squeezing force $F_{\rm N}$
the area of real contact increases linearly with $F_{\rm N}$.

\begin{figure}[htb]
\begin{center}
   \includegraphics[width=0.6\textwidth]{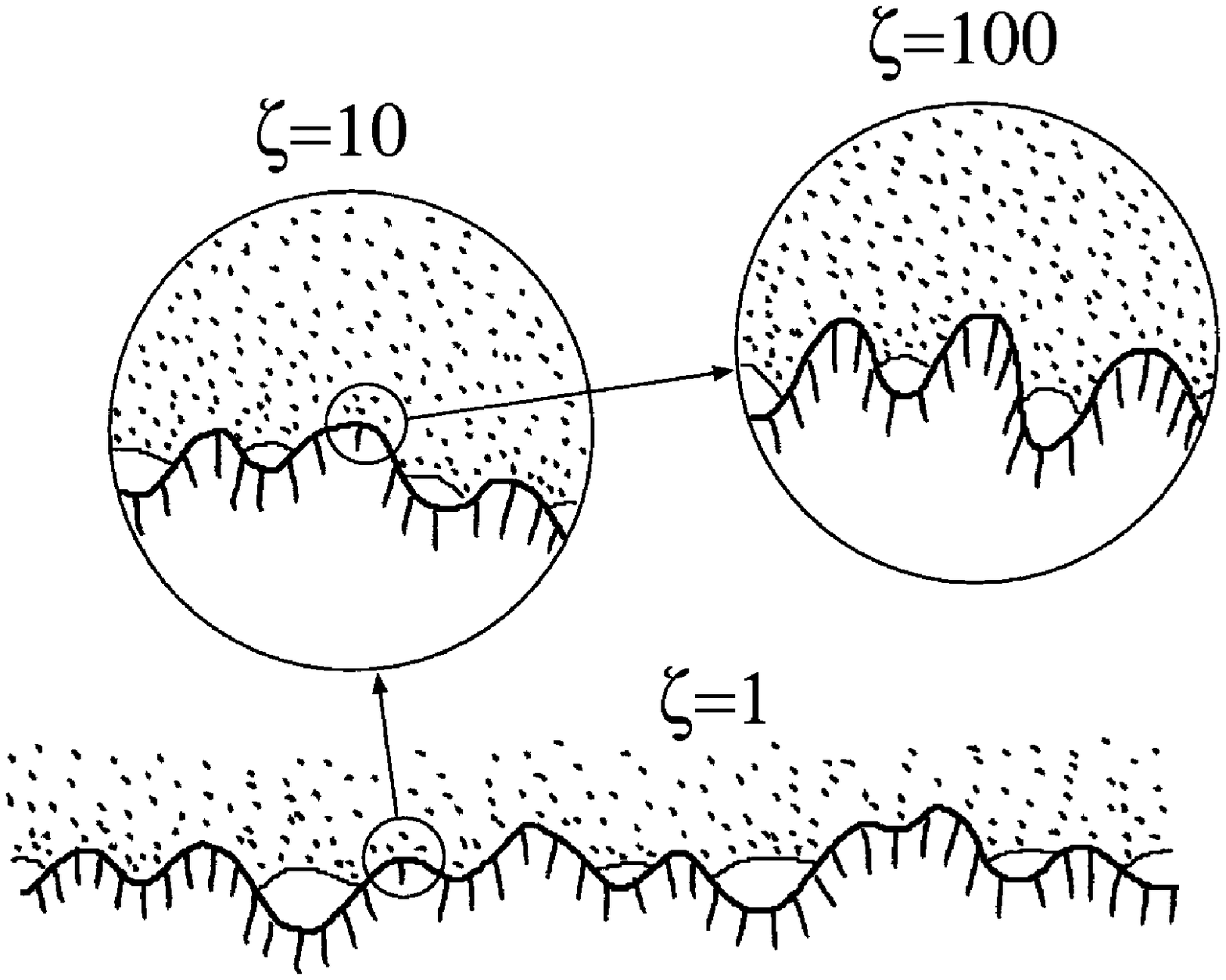} 
\end{center}
\caption{ \label{1x}
A rubber block (dotted area) in adhesive contact with a hard rough
substrate (dashed area). The substrate has roughness on many different
length scales and the rubber makes partial contact with the substrate on
all length scales.  When a contact area is studied at low magnification
($\zeta=1$) it appears as if complete contact occurs in the macro
asperity contact regions, but when the magnification is increased it is
observed that in reality only partial contact occurs.}
\end{figure}

Fig.~\ref{1x} shows the
contact between two solids at increasing magnification $\zeta$. At low magnification
($\zeta = 1$)
it looks as if complete contact occurs between the solids at many {\it macro asperity}
contact regions. When the magnification is increased and smaller length scale roughness is detected,
it can be observed that only partial contact occurs at the asperities.
In fact, if there were no short distance cut-off the true contact area
would eventually vanish. In reality, 
a short distance cut-off always exists, e.g., the interatomic distance. 
In many cases the local pressure in the contact regions at the asperities 
may become so high
that the material yields plastically before reaching the atomic dimension.
In these cases
the size of the real contact area will be determined mainly by the yield stress
of the solid.

\subsection{Elastic contact mechanics}
\label{sec4.1}

 From contact mechanics (see, e.g., Ref.~\cite{[7]}) it is known that
in the frictionless contact of elastic solids with rough surfaces
the contact stresses depend only upon the
shape of the gap between them before loading. Thus,
without loss of generality,
the actual system may then be replaced
by a flat elastic surface
[elastic modulus $E$ and Poisson ratio $\nu$,
related to the original quantities via
$(1-\nu^2)/E = (1-\nu_1^2)/E_1+(1-\nu_2^2)/E_2$]
in contact with
a rigid body having a surface roughness profile
which result in the same undeformed gap between the surfaces.

One of us (Persson) has recently developed a theory of contact
mechanics \cite{[2]}, valid for randomly rough (e.g., self affine
fractal) surfaces.  In the context of rubber friction, which motivated
this theory, mainly elastic deformation occurs.  However, the theory can
also be applied when both elastic and plastic deformations occur in the
contact areas. This case is of course relevant to almost all materials
other than rubber.

\begin{figure}[htb]
\begin{center}
   \includegraphics[width=0.9\textwidth]{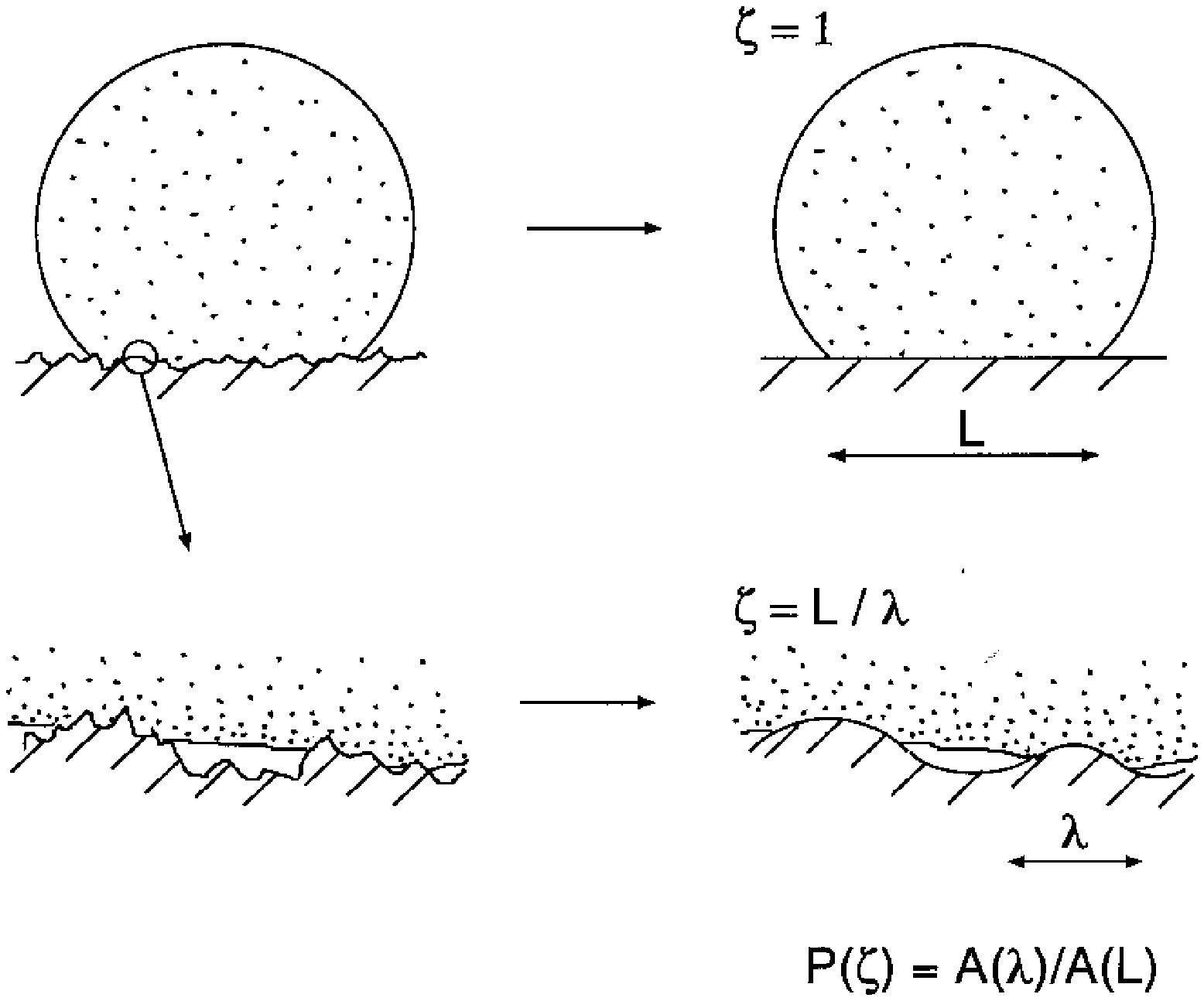} 
\end{center}
\caption{\label{P2}
An elastic  ball squeezed against a hard, rough, substrate. Left: the system
at two different magnifications. Right: The area of contact $A(\lambda)$ on the
length scale $\lambda$ is defined as the area of real contact when the surface
roughness on length scales shorter than $\lambda$ has been removed.}
\end{figure}

The basic idea behind the new contact theory
is that it is very important not to exclude {\it a priori}
any roughness length scale from the analysis.
Thus, if $A(\lambda)$ is the (apparent) area of contact
on the length scale\footnote{
  We define the apparent contact area $A(\lambda)$ on the length scale
  $\lambda$ to be the area of real contact if the surface would be smooth
  on all length scales shorter than $\lambda$. That is, considering the
  Fourier transform of the surface profile, all the components whose wave
  vector is larger that $2\pi/\lambda$ have to be set to 0, and the
  contact area with this new surface is by definition $A(\lambda)$.
  }
$\lambda$ (see Fig.~\ref{P2}), then we study the
function $P(\zeta )=A(\lambda)/A(L)$
which is the
relative fraction of the surface area
where contact occurs on the length scale
$\lambda=L/\zeta$ (where $\zeta \geq 1$), with $P(1) =1$.
Here $A(L)=A_0$ denotes the macroscopic
contact area [$L$ is the diameter of the
macroscopic contact area so that $A_0\approx L^2$].

Consider the system
at the length scale $\lambda =L/\zeta$, where $L$ is the
diameter of the nominal contact area.
We define $q_L = 2\pi /L$ and write $q=q_L\zeta$.
Let $P(\sigma,\zeta)$ denote the stress distribution
in the contact areas under the magnification $\zeta$.
The function $P(\sigma,\zeta )$ satisfies the differential equation (see
Ref.~\cite{[2]}):
\begin{equation}
  {\partial P \over \partial \zeta}  =
  f(\zeta ) {\partial^2 P \over \partial \sigma^2}
  \label{eq5}
\end{equation}
where $f(\zeta)=G'(\zeta)\sigma_0^2$, $\sigma_0$ being the average
pressure in the nominal contact area and
\begin{equation}
 G(\zeta) =
 {\pi \over 4} \left( {E^* \over \sigma_0} \right)^2
   \int_{q_L}^{\zeta q_L} \rmd q \ q^3 C(q),
  \label{eq6}
\end{equation}
with $E^*=E/(1-\nu^2)$.

Eq.~(\ref{eq5}) is a diffusion-type equation, where time is replaced by
magnification $\zeta$, and the spatial coordinate with the stress
$\sigma$ (and where the ``diffusion constant'' depends on $\zeta$).
Hence, when we study $P(\sigma, \zeta)$ on shorter and shorter length
scales (corresponding to increasing $\zeta$), the $P(\sigma,\zeta)$
function will become broader and broader
in $\sigma$-space. We can take into account that detachment will occur when
the local stress reaches $\sigma = 0$ (we assume no adhesion) via the boundary
condition \cite{PC}:
\begin{equation}
  P(0,\zeta)=0.
 \label{eq7}
\end{equation}
In order to solve the equation (\ref{eq5}) we also need an ``initial''
condition.  This is determined by the pressure distribution at the
lowest magnification $\zeta=1$.
If we assume
a constant pressure $\sigma_0$ in the nominal contact area, then
$P(\sigma, 1) = \delta(\sigma-\sigma_0)$.

We assume that only elastic deformation occurs (i.e., the yield stress
$\sigma_Y\rightarrow \infty$).
In this case
\[
  P(\zeta ) =
  \int_0^\infty \rmd\sigma P(\sigma,\zeta)
\]
When adhesion is taken into account, tensile stresses can occur at the interface
between the two solids, and the boundary
condition (\ref{eq7}) is no longer valid \cite{Persson}, see Sec.~\ref{sec7.1}.
It is straightforward to solve (\ref{eq5}) with the boundary conditions
$P(0,\zeta)=0$ and $P(\infty,\zeta)=0$ to get
\begin{equation}
 P(\zeta )= {2\over \pi} \int_0^\infty \rmd x\, {\sin x \over x}
 \rme^{-x^2 G(\zeta)}= {\rm erf} \left ({1\over 2 \sqrt{G}}\right ).
 \label{eq8}
\end{equation}
Note that for small load $\sigma_0$, $G \gg 1$ and in this case (\ref{eq8})
reduces to $P(\zeta) \approx P_1(\zeta)$ where
\begin{equation}
 P_1(\zeta) = \left[ \pi G(\zeta ) \right]^{-1/2}.
 \label{eq9}
\end{equation}
Since $G\sim 1/\sigma_0^2$ it follows that the area of real contact is {\it proportional} to the load
for small load. Using (\ref{eq8}) and (\ref{eq9}) we can write in a general case
\begin{equation}
  P(\zeta) = {\rm erf} \left({\sqrt{\pi} \over 2} P_1(\zeta )\right)
  \label{eq10}
\end{equation}

\begin{figure}[htb]
\begin{center}
\includegraphics[width=0.80\textwidth]{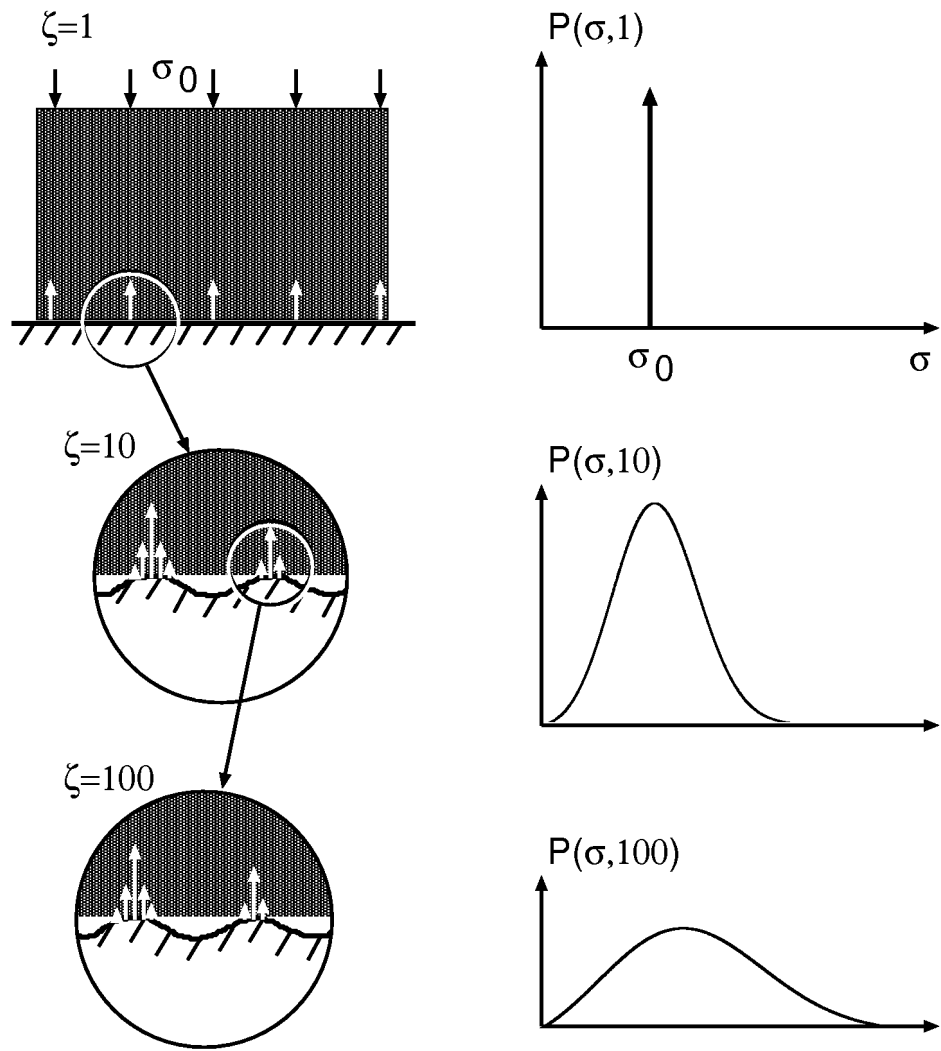} 
\end{center}
\caption{\label{LAST}
The stress distribution $P(\sigma, \zeta)$ in the contact region
between a (rigid) block and an elastic substrate at increasing magnification
$\zeta$.  At the lowest (engineering) magnification $\zeta=1$ the substrate
surface looks smooth and the block makes (apparent) contact with the substrate
in the whole nominal contact area. As the magnification increases,
we observe that the area of (apparent) contact decreases, while
the stress distribution becomes broader and broader.}
\end{figure}

The physical meaning of the diffusion-like Eq.~(\ref{eq5}) is as follows:
When the system is studied at the lowest magnification $\zeta = 1$ no surface roughness can be
observed and the block makes (apparent) contact with the substrate everywhere in the nominal contact area.
In this case, if we neglect friction at the interface, the stress at the interface will everywhere
equal the applied stress $\sigma_0$, see Fig.~\ref{LAST}(a),
so that the distribution will initially be 
delta function-like,  $P(\sigma, 1) = \delta (\sigma - \sigma_0)$.
Increasing the magnification, we include surface roughness with wavelength down to $\lambda =
L/\zeta$, and here one may observe some non-contact
regions as shown in Fig.~\ref{LAST}(b). Since the
stress must go continuously to zero at the edges of the boundary between the contact and non-contact
regions, it follows that the stress distribution $P(\sigma, \zeta)$ will have a tail extending the whole
way down to the zero stress as indicated in
Fig.~\ref{LAST}(b) (right). There will also be a tail
toward larger stresses $\sigma > \sigma_0$ because the average stress must be
equal to $\sigma_0$. This distribution broadens as in a diffusion problem.
With increasing magnification,
the stress distribution will broaden further and without limit as indicated in
Fig.~\ref{LAST}(right).

The theory presented above predicts that the area of contact increases linearly
with the load for small load. In the standard theory of Greenwood and
Williamson \cite{[5]} this result holds only approximately and a comparison of
the prediction of their theory with the present theory is therefore difficult.
Bush et al.\ \cite{Bush} have developed a more general and accurate contact
theory.  They assumed that the rough surface consists of a mean plane with
hills and valleys randomly distributed on it. The summits of these hills are
approximated by paraboloids, whose distribution of heights and principal
curvatures are obtained from the random process theory.  This is to be compared
with the GW assumption that the caps of the asperities are spherical, each
having the same mean radius of curvature.  As a result of the more random
nature of the surface, Bush et al found that at small load the area of contact
depends linearly on the load accordingly to
\begin{equation}
 A = \kappa {F_N \over E^*} \left( \int \rmd^2q \ q^2 C(q) \right)^{-1/2}
 \label{eq11}
\end{equation}
where $F_N$ is the normal load, $E^*=E/(1-\nu^2)$, and
$\kappa= (2\pi )^{1/2}$.
This result is very similar to the prediction of the present theory where, for small load,
from (\ref{eq6}) and (\ref{eq9}), $A$ is again given by Eq.~(\ref{eq11}) but now with $\kappa = (8/\pi )^{1/2}$.
Thus our contact area is a factor of $2/\pi$ smaller
than the one predicted by the
theory of Bush et al. Both the theory of Greenwood and Williamson and that of Bush et al.,
assume that the asperity contact regions are
independent. However, as discussed in Ref.~\cite{PC},
for real surfaces (which always have
surface roughness on many different length scales) this will never be
the case even at a very low nominal contact pressure. As we argued \cite{PC}, this may be the
origin of the $2/\pi$-difference between our theory (which assumes roughness on many
different length scales) and the result of Bush et al.

The predictions of the contact theories of Bush et al.\ \cite{Bush} and Persson \cite{[2]}
were compared to
numerical calculations (see Ref.~\cite{PC}\cite{Mark2004}).
Borri-Brunetto et al.\ \cite{[a5]} studied the contact between self affine fractal surfaces
using an essentially exact numerical method. They found that the contact area is proportional to the squeezing
force for small squeezing forces. Furthermore, it was found that the slope $\alpha (\zeta)$
of the line $A=\alpha (\zeta) F$
decreased with increasing magnification $\zeta$. This is also predicted by the analytical theory
[Eq.~(\ref{eq11})]. In fact, good agreement was found between theory and 
computer simulations for the change in the slope with magnification and its
dependence on the fractal dimension $D_{\rm f}$.

\begin{figure}[htb]
\begin{center}
\includegraphics[width=0.80\textwidth]{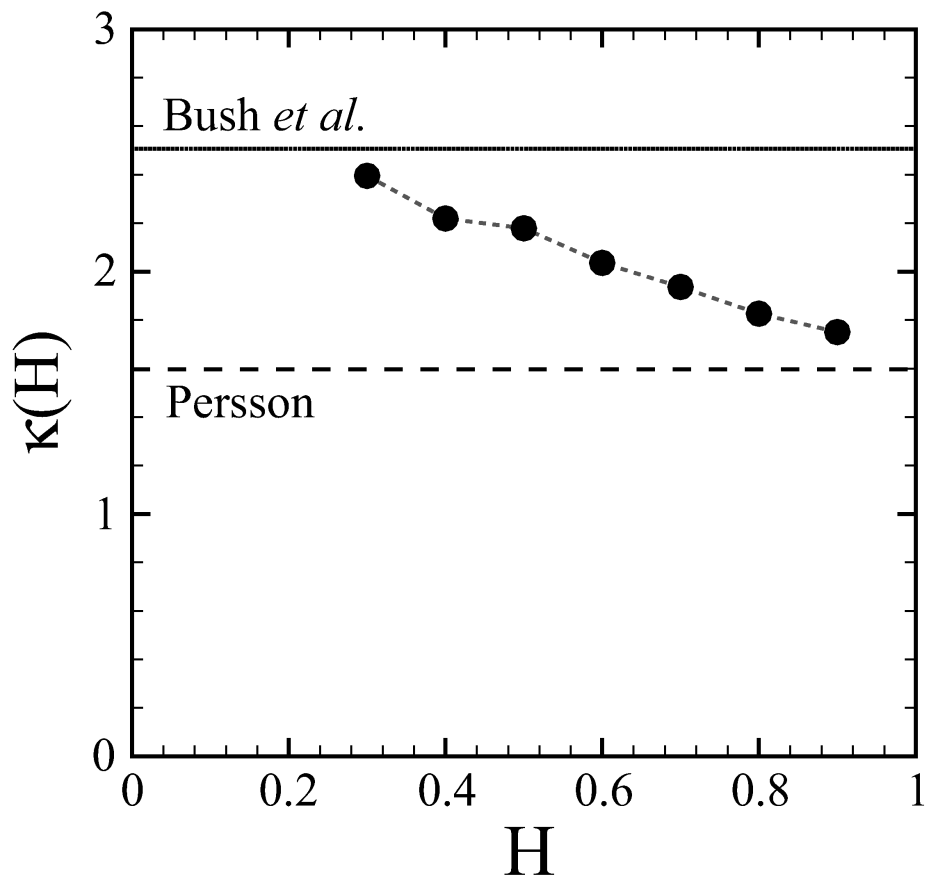} 
\end{center}
\caption{\label{Mark}
Dots: the factor $\kappa$ as a function of Hurst's exponent $H$ for
self affine fractal surfaces. The two horizontal lines
are the predictions of the theories of Bush et al. (solid line)
and Persson (dashed line). From Ref.~\protect\cite{Mark2004}.}
\end{figure}

Hyun et al. performed a finite-element analysis of contact between
elastic self-affine surfaces. The simulations were done for a rough elastic
surface contacting a perfectly rigid flat surface.  The elastic solid was
discretized into blocks and the surface nodes form a square grid.  The contact
algorithm identified all nodes on the top surface that attempt to penetrate the
flat bottom surface. The total contact area $A$ was obtained by multiplying the
number of penetrating nodes by the area of each square associated with each
node.  As long as the squeezing force was so small that the contact area
remained below $10\%$ 
of the nominal contact area, $A/A_0 < 0.1$, the area of real contact was found 
to be proportional to the squeezing force
in accordance with Eq.~(\ref{eq11}).
Fig.~\ref{Mark} shows Hyun et al.'s results for
the factor $\kappa$ in (\ref{eq11}) as a function of Hurst's exponent $H$ for
self affine fractal surfaces. The two horizontal lines
are predictions of the theories of Bush et al. (solid line)
and Persson (dashed line). The agreement with the analytical predictions is quite
good considering the ambiguities in the discretization of the surface. The algorithm only
consider nodal heights and assumes that contact of a node implies contact over the entire corresponding
square. This procedure would be accurate if the spacing between nodes where much
smaller than the typical size of asperity contacts.
However, the majority of the contact area consists of
clusters containing only one or a few nodes.
Since the number of large clusters grows as $H\rightarrow 1$, this may
explain why the numerical results approach Persson's prediction in this limit.

Hyun et al. also studied the distribution of connected contact regions and the contact morphology.
In addition, the interfacial stress distribution was considered, and it was found that the stress
distribution remained non-zero as the stress $\sigma \rightarrow 0$.
This violates the boundary condition
(\ref{eq7}) that $P(\sigma, \zeta)=0$ for $\sigma = 0$. However,
it has been shown analytically \cite{PC} that for
``smooth'' surface roughness this latter condition
must be satisfied, and we believe that the
violation of this boundary condition in the numerical
simulations may reflect the way the solid was
discretized and the way the contact area was defined in Hyun et al.'s numerical procedure.

Elastic contact theory and numerical simulations show that in the region where the
contact area is proportional to the squeezing force, the stress distribution at the interface is
independent of the squeezing force. In addition, for an infinite system the distribution
of sizes of the contact regions does not depend on the squeezing force (for small squeezing forces).
Thus, when the squeezing
force increases, new contact regions are formed in such a way that the distribution of
contact regions and the pressure distribution remains unchanged. This is the physical origin
of Coulomb's friction law which states that the friction force is proportional to the
normal (or squeezing) force \cite{[1]}, and which usually holds accurately as long as the block-substrate
adhesional interaction can be neglected \cite{WEAR2003}.

\subsection{Surface stiffness of fractal surfaces}
\label{sec4.2}

\begin{figure}
\begin{center}
\includegraphics[width=0.90\textwidth]{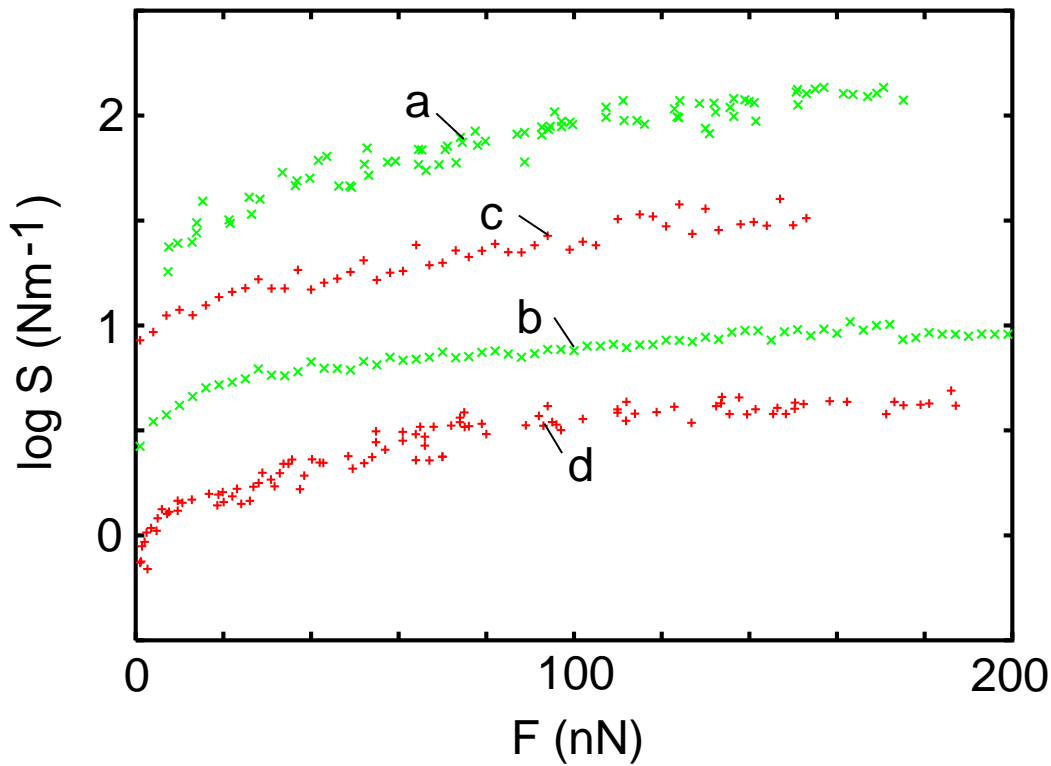} 
\end{center}
\caption{\label{S}
Surface stiffness $S=\rmd F/\rmd u$ as a function of the squeezing force $F$ for a flat hard micrometer
sized surface in contact with self affine fractal substrates made from two different
materials (${\bf +}$ and ${\bf \times }$ symbols, respectively). The fractal dimension $D_{\rm f}$
and the {\it rms}-roughness amplitude $\sigma$ of the surfaces are: (a)
$D_{\rm f}=2.10$, $\sigma= 20 \ {\rm nm}$; (b) $2.30$, $80 \ {\rm nm}$;
(c) $2.15$, $36 \ {\rm nm}$ and (d) $2.26$, $160 \ {\rm nm}$.
 From Ref.~\protect\cite{Renato}}
\end{figure}

The contact behaviour of realistic surfaces with random multiscale roughness
remains largely unknown. Recently experimental results of the surface stiffness
for self-affine fractal surfaces were presented \cite{Renato}.
A hard micrometric flat probe surface was squeezed against different substrates,
and the vertical
displacement $u$ was measured as a function of the squeezing force $F$. It was found
that the surface stiffness $S=dF/du$ depends remarkably on the fractal
dimension of the substrate surface, {\it decreasing by more than an order of magnitude
as the substrate fractal dimension is increased by only $10 \%$}.
As expected,
the surface stiffness was nearly
proportional to the inverse of the
{\it rms}-roughness amplitude $\sigma$.
Fig.~\ref{S} shows the surface stiffness as a function of the squeezing force $F$
for fractal substrate surfaces produced from two different materials (${\bf +}$ and ${\bf \times}$ symbols,
respectively). Note the strong decrease in the surface stiffness $S$ with increasing
fractal dimension $D_{\rm f}$ of the substrate.

\subsection{Viscoelastic contact mechanics}
\label{sec4.3}

\begin{figure}
\begin{center}
\includegraphics[width=0.50\textwidth]{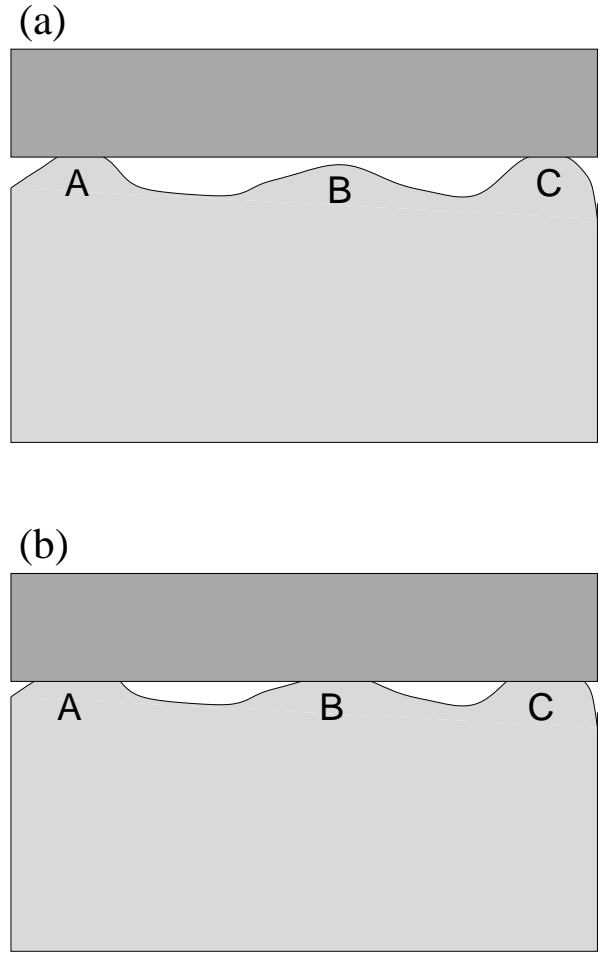} 
\end{center}
\caption{\label{Two.Block.t}
With increasing time, the contact area between the viscoelastic substrate and a flat hard solid surface 
increases through growth of existing asperity contact areas (A and C), as well as by the formation
of new asperity contact areas (B).}
\end{figure}

The contact between a viscoelastic solid and hard, randomly rough, substrates is a topic of
great practical importance, e.g., for pressure sensitive adhesives, rubber friction and
rubber seals. When a viscoelastic solid is squeezed with a constant force
against a rough substrate,
the area of
real contact will increase monotonically with the contact time, see Fig.~\ref{Two.Block.t}.
Since rubber-like materials have a wide
distribution of relaxation times, the area of real contact will usually increase over a very long
time period (which, e.g., could be a year or more).
Since the pull-off force depend on the area of real contact, contact theories for viscoelastic solids
are important for estimating
how the pull-off force (or tack) depends on the applied squeezing pressure and the squeezing
time, see Sec.~\ref{sec4.4}.

Rubber-like materials have elastic modulus $E(\omega )$
which depend strongly on frequency $\omega$ and temperature $T$. Thus, at
very low frequencies or high temperatures
they behave as very soft \emph{rubbery} materials, with typical
elastic modulus in the range $0.01-1 \ {\rm MPa}$. At high frequencies or low temperatures
they instead behave
as hard glassy materials with the elastic modulus of order $1 \ {\rm GPa}$ or more.
Thus, as a function of increasing frequency (or decreasing temperature)
the elastic modulus may increase by a factor of 1000 or more.
The transition from the rubbery region to the glassy region is very wide, usually
extending over more than three frequency-decades. In a contact experiment, the inverse of
the contact time is a characteristic frequency; thus for long contact times
rubber behaves as a
\emph{soft} solid and the contact area is \emph{large},
while for short contact times it is relatively \emph{hard}
and the contact area is \emph{small}.

\begin{figure}
\begin{center}
\includegraphics[width=0.8\textwidth]{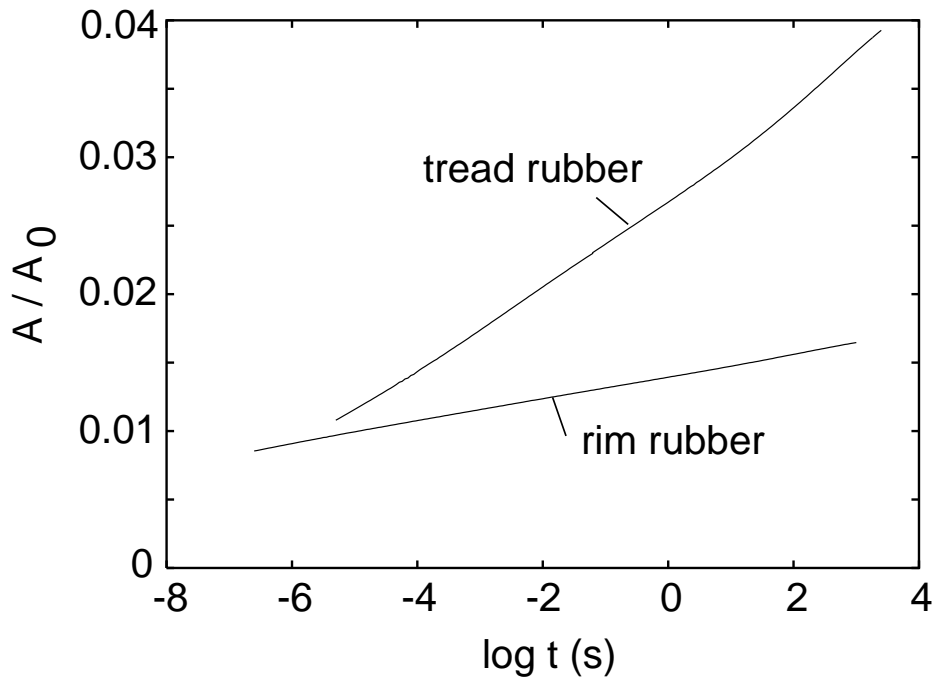} 
\end{center}
\caption{\label{Area.time}
The relative contact area [obtained from Eqs. (\ref{eq12}) and (\ref{eq15})] at the
magnification $\zeta = 10^5$,
as a function of the logarithm of the contact time $t$,
for the
nominal pressures $\sigma_0 = 0.1 \ {\rm MPa}$, for a tread tire rubber compound,
and for a compound used in the tire-rim area,
at the temperature $T= 60^{\circ} {\rm C}$. The substrate is self affine fractal
with the fractal dimension $D_{\rm f}=2.2$, and with
the rms roughness $1 \ {\rm \mu m}$, and with the roll-off wave vector
(see Fig.~\ref{Cq1}) $q_0 = 6\times10^4 \ {\rm m}^{-1}$ (this is a typical $q_0$ for a polished
steel surface). The magnification $\zeta=1$ refer to the length scale
$\lambda_0 = 2\pi /q_0 \approx 0.1 \ {\rm mm}$, so that the magnification $\zeta=10^5$ correspond
to the length scale (resolution) $\lambda \approx 1 \ {\rm nm}$.}
\end{figure}

When a viscoelastic solid is squeezed with a constant force
against a hard rough substrate, the contact area
will increase continuously with time \cite{JCP2004,Cornell}. The relative contact area
$P(\zeta,t)=A(\zeta,t)/A_0$ is approximately given by Eq.~(\ref{eq10}):
\begin{equation}
 P(\zeta , t) = {\rm erf} \left( { \sqrt{\pi} \over 2} P_1(\zeta,t) \right)
 \label{eq12}
\end{equation}
where \cite{JCP2004}
\begin{equation}
  P_1(\zeta,t) =
  {2\over \pi Q(\zeta )} \int_{-\infty}^\infty \rmd\omega \
  {\tilde \sigma_0 (\omega) \over E^*(\omega ) } \rme^{-\rmi\omega t},
 \label{eq13}
\end{equation}
denotes the relative contact area to linear order in $\sigma_0$.
Here $\tilde \sigma_0(\omega )$ is the Fourier transform
of the squeezing pressure $\sigma_0(t)$, and
\[ Q(\zeta)=\left( \int_0^{q_L\zeta} \rmd q \ q^3C(q) \right)^{1/2}. \]
Now, let us assume that $\sigma_0 (t) = \sigma_0$ for $0 < t < t_1$ and zero otherwise.
In this case
\begin{equation}i
 \tilde \sigma_0 (\omega) =
 {\sigma_0/2\pi \over 0^+ -\rmi\omega} \left( 1-\rme^{\rmi\omega t_1}\right)
 \label{eq14}
\end{equation}
Substituting this into (\ref{eq13}) gives
\begin{eqnarray}
 P_1(\zeta,t) & = &
  {2 \sigma_0 \over \pi Q(\zeta )} {1\over 2 \pi}
  \int_{-\infty}^\infty \rmd\omega
       \  {1- \rme^{\rmi\omega t_1}\over -\rmi\omega}
          {\rme^{-\rmi\omega t} \over  E^*(\omega ) } \nonumber \\
 & = &
  {2 \sigma_0 \over \pi^2 Q(\zeta )} {\rm Re}
  \int_{0}^\infty \rmd\omega \  {1- \rme^{\rmi\omega t_1} \over -\rmi\omega}
  {\rme^{-\rmi\omega t} \over E^*(\omega )}
 \label{eq15}
\end{eqnarray}
This result does, of course, not depend on $t_1$ as long as $t<t_1$ (causality).
Thus we are free to choose for $t_1$ any value larger
than the time $t$ under consideration. The equations presented
above are only valid as long as the contact area
increases with time, which is the case, e.g., if a constant squeezing force is
applied at time $t=0$.
As an illustration,
in Fig.~\ref{Area.time} we show the relative contact area
calculated from Eq.~(\ref{eq15}) using the measured viscoelastic modulus for two types of rubber.
The rubber block is squeezed with the nominal pressure $\sigma_0=0.1 \ {\rm MPa}$ against a
``steel'' surface. We show results for two typical rubbers used in the construction of
tires, namely a tread rubber compound, and a rubber compound
from the tire rim region involved in the tire air
sealing (see Sec.~\ref{sec5}).
The substrate was a self affine fractal
with the fractal dimension $D_{\rm f}=2.2$, and with
the {\rm rms}-roughness amplitude $1 \ {\rm \mu m}$.
The tread compound exhibits the largest contact area because it is elastically softer 
than the tire rim compound. Note also that the change in the contact area is only a
factor $\sim 2-4$ when the contact time changes by $\sim 9$  decades.

\subsection{Tack}
\label{sec4.4}

Pressure sensitive adhesives (PSA) are used in many important applications, e.g.,
for Scotch tapes, post-it pads and self-adhesive labels and envelopes \cite{Creton1,[16c]}.
The adhesive consists of a very thin layer
(usually of order $\sim 20-100 \,{\rm \mu m}$)
of a very soft, weakly cross-linked rubber compound. The low frequency elastic
modulus is typically only $0.01 \,{\rm MPa}$ which is $\sim 100$ times lower than
the rubber used for tires.
As a result of the low elastic modulus, nearly complete
contact will occur in the apparent contact area even for relatively low squeezing pressures
and large surface roughness. This is in contrast to tire rubber which under similar condition
would give a contact area of only a few \% of the nominal contact area,
as discussed in Sec.~\ref{sec6.1}.

We carried out calculations [using Eq.~(\ref{eq15})] 
of the time-dependent contact area for a
standard rubber tack compound [acrylic PSA
namely poly(2ethylhexyl acrylate) with 2\% acrylic acid
(PEHA-AA)].

\begin{figure}
\begin{center}
\includegraphics[width=0.6\textwidth]{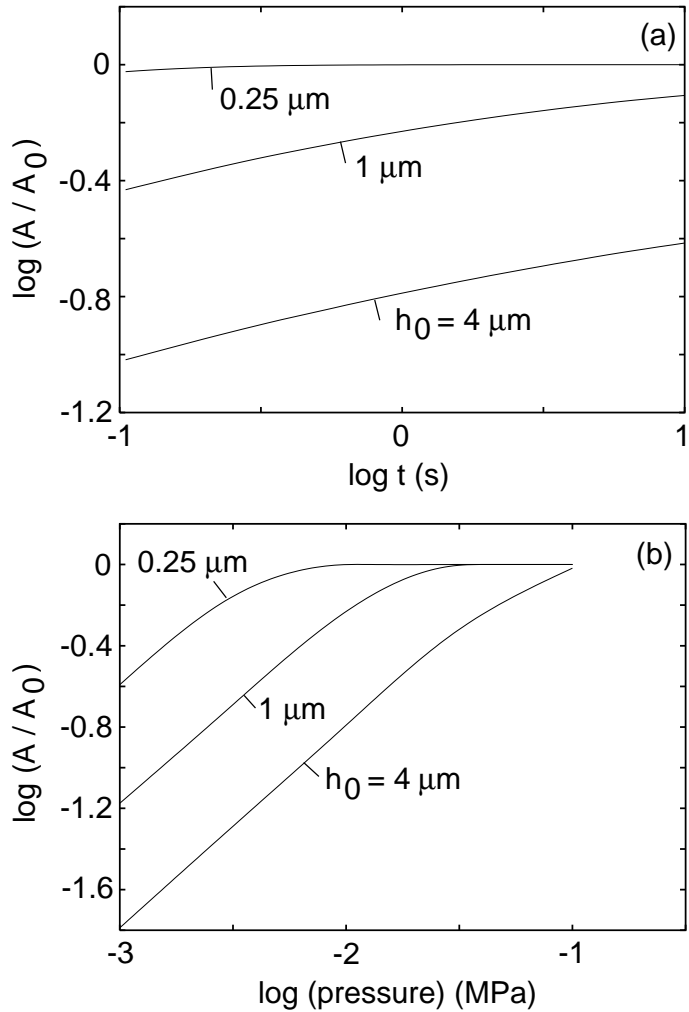} 
\end{center}
\caption{\label{Area.Time.Pressure}
(a) Calculated logarithm of the relative contact area
at the magnification $\zeta=10^5$,
as a function of the logarithm of the
contact time. Contact pressure $\sigma_0 = 0.01 \ {\rm MPa}$.
(b)
Calculated logarithm of the relative contact area
after 1s of contact
at the magnification $\zeta=10^5$,
as a function of the logarithm of the
contact pressure. 
All calculations for PEHA-AA at
temperatures $T = 20^{\circ} {\rm C}$ on 
a self affine fractal substrate with
fractal dimension $D_{\rm f} = 2.2$, roll-off wave vector
$q_0=6\times 10^4 \ {\rm m}^{-1}$ and
{\it rms} roughness amplitudes $h_0=0.25$, 1 and $4 \ {\rm \mu m}$.}
\end{figure}

Fig.~\ref{Area.Time.Pressure} shows the results at
$T=20^{\circ} {\rm C}$ when the PSA is squeezed in contact with
``steel'' surfaces with
{\it rms} roughness amplitudes $h_0=0.25$, 1 and $4 \ {\rm \mu m}$.
Fig.~\ref{Area.Time.Pressure}(a) shows
the logarithm of the relative contact area,
as a function of the logarithm of the
contact time,
for a contact pressure $\sigma_0 = 0.01 \ {\rm MPa}$.
Fig.~\ref{Area.Time.Pressure}(b)
shows
the logarithm of the relative contact area
after 1s of contact,
as a function of the logarithm of the
contact pressure.

\begin{figure}
\begin{center}
\includegraphics[width=0.6\textwidth]{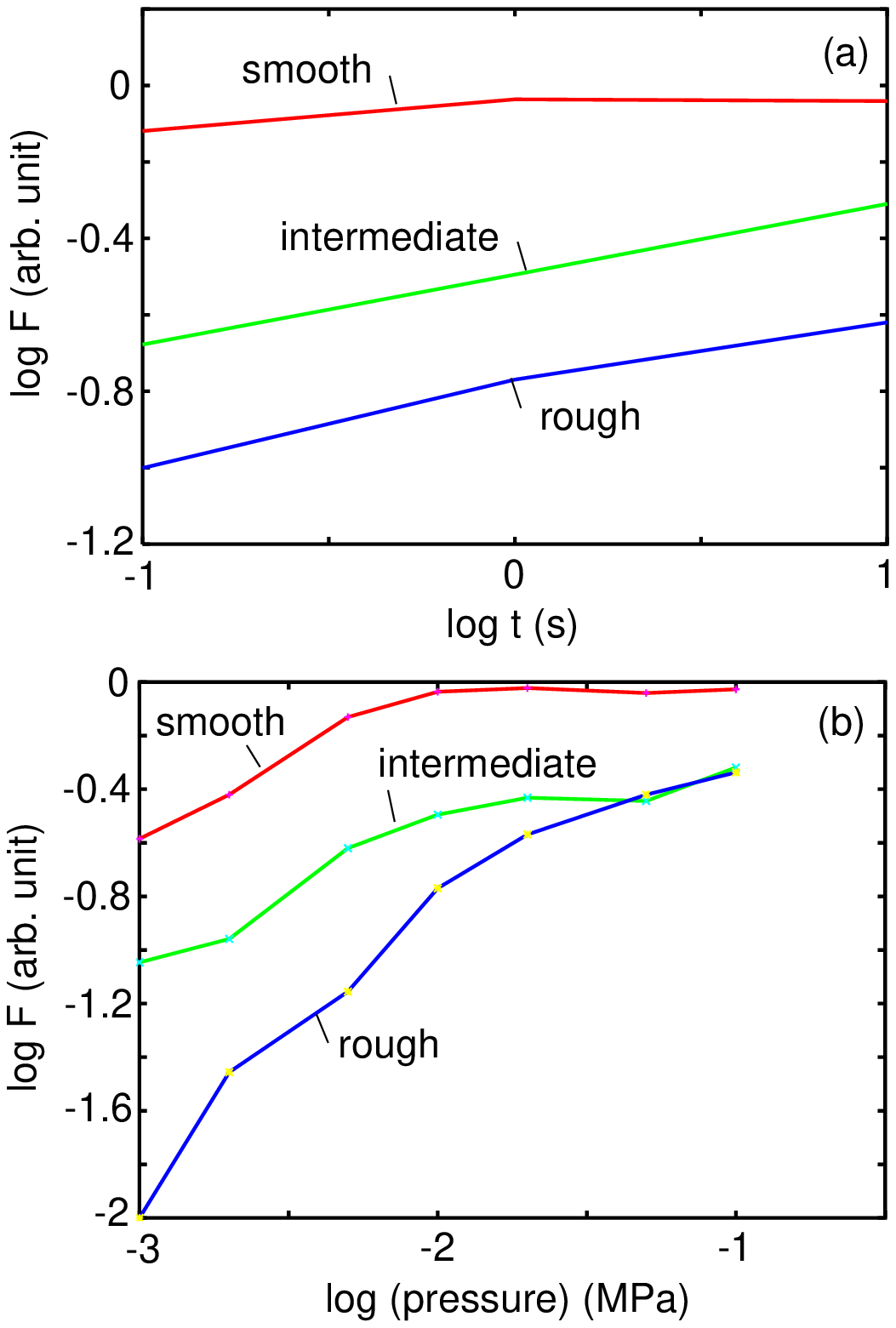} 
\end{center}
\caption{\label{F.t.pre}
(a) Logarithm of the experimental pull-off force $F$
(in arbitrary units)
as a function of the logarithm of the
contact time. Contact pressure $\sigma_0 = 0.01 \ {\rm MPa}$.
 From Ref.~\protect\cite{Hamm}
(b) Logarithm of the experimental pull-off force
[same units as in (a)] after 1s of contact,
as a function of the logarithm of the
contact pressure, for three different surface roughnesses.
 Ref.~\protect\cite{Hamm}}
\end{figure}

It is interesting to compare the results in Fig.~\ref{Area.Time.Pressure}
with the experimental data reported in Ref.~\cite{Hamm}
and shown in Fig.~\ref{F.t.pre}.
The data show
the dependence of the maximum pull-off
force $F$ on the contact time and contact pressure
for smooth and rough PSA on the same smooth steel surface.
The tack film is a standard polymer compound similar to PEHA.
In Ref.~\cite{Hamm} no numerical values
of the {\it rms} roughness was presented for any of the studied surfaces.
If one assumes, as a first approximation, that the pull-off force
is proportional to the area of real contact, and moreover that
the smooth and rough PSA films
in the measurements correspond, say, to the surfaces
in Fig.~\ref{Area.Time.Pressure} with
the {\it rms} roughness $0.25$, $1$
and $4 \ {\rm \mu m}$, then the agreement between theory
and the experiment is remarkably good. 

\section{Seals}
\label{sec5}

Surface roughness is an important factor which influences the rate of 
leakage through seals.
The exact mechanism of roughness induced
leakage is not well understood \cite{[8c]}. In this section we
present a new way of looking at this problem \cite{JCP2004}.

Viscoelastic materials such as rubber are often used for sealing. Here we consider the
tire-rim sealing. We are interested in the air (or gas) flow from inside the tire
to the outside via the roughness-induced non-contact area (pore channels) 
in the rubber-steel rim area.
The rim is made from steel. We assume that the
steel surface is a self-affine fractal with the
fractal dimension $D_{\rm f} = 2.2$.
The surface root-mean-square roughness is assumed to be $1\,{\rm \mu m}$,
a rather typical value for polished
steel surfaces.

\begin{figure}
\begin{center}
\includegraphics[width=0.6\textwidth]{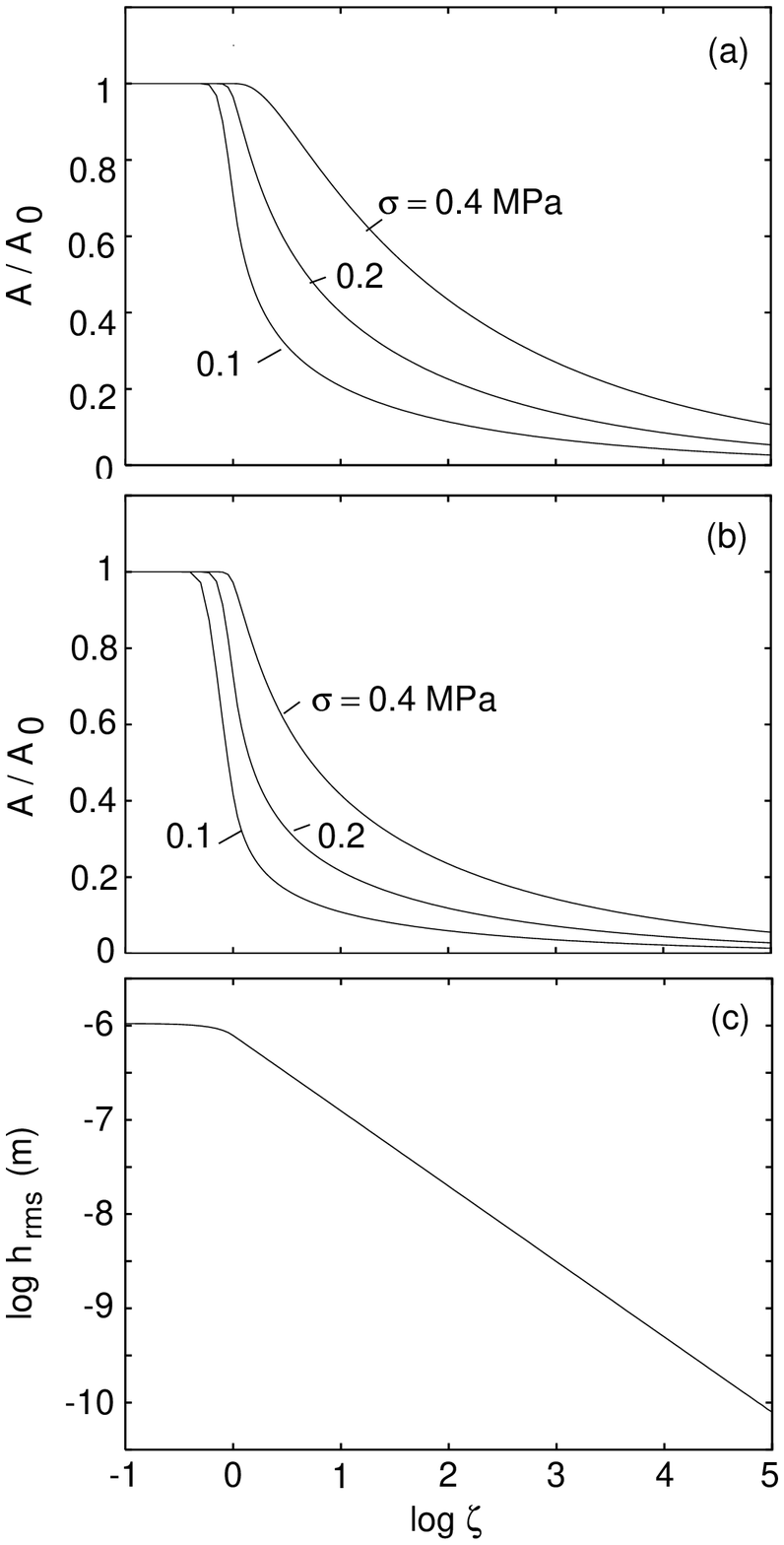} 
\end{center}
\caption{\label{Magn.Area}
Calculated relative contact area after one second of contact
for tread tire rubber (a),
for rim tire rubber (b), and the root-mean-square roughness
as a function of magnification (c), where $\zeta = 1$ corresponds to
a wavelength $\lambda_0 = 2 \pi /q_0 \approx 100 \,{\rm \mu m}$.
The highest magnification
$\zeta = 10^5$ corresponds to $\lambda \approx 1 \,{\rm nm}$.
Results are shown for a
nominal pressures $\sigma_0 = 0.1$, 0.2 and $0.4 \,{\rm MPa}$,
at the temperature $T= 60^{\circ} {\rm C}$. Substrate surface with 
{\it rms} roughness $h_0=1 \,{\rm \mu m}$ and roll-off wave vector
$q_0 = 6\times10^4 \,{\rm m}^{-1}$.}
\end{figure}

Fig.~\ref{Magn.Area} shows the calculated [using Eq.~(\ref{eq15})] 
relative contact area after one second of contact for tread tire rubber (a),
and for rim tire rubber (b), and the root-mean-square roughness
(c), as a function of the magnification (the time dependence of the contact area was
presented in Fig.~\ref{Area.time}).
Here we defined
\[
 h_{\rm rms} (\zeta ) =
  \left( 2\pi \int_{\zeta q_0}^{q_1} \rmd q \ q C(q) \right)^{1/2}
\]
The results are for a
nominal pressures $\sigma_0 = 0.1$, $0.2$ and $0.4 \ {\rm MPa}$,
at the temperature $T= 60^{\circ} {\rm C}$.

\begin{figure}
\begin{center}
\includegraphics[width=0.8\textwidth]{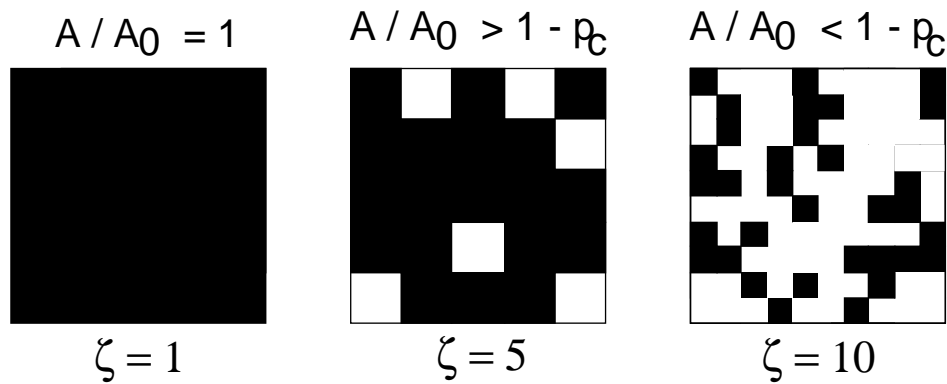} 
\end{center}
\caption{\label{Percolation}
When the interface between the solids is studied at low magnification, there
appear to be complete contact (black area) between the solids. When the magnification is increased
it is observed that only partial contact occur. At high enough magnification the non-contact
(white) surface area will percolate and one (or several) air flow channels will be visible at the
interface.}
\end{figure}

We consider now a rubber-steel
interface at increasing magnification. At the lowest magnification $\zeta = 1$
complete contact occurs at the interface, see Fig.~\ref{Percolation} (left).
When we increase the magnification we observe non-contact areas or islands.
Magnification is now increased until
the non-contact areas percolate, i.e., until
a channel of non-contact surface area, extending
from the high-pressure internal region of the tire to the outside (atmospheric pressure region),
is first observed. As the magnification is increased further, new non-contact region
will be observed, but the separation between the surfaces in these new non-contact areas will be smaller
than along the percolation channel. Since the gas flow $\dot N$ 
(number of molecules per unit time)
through a rectangular pore of height $h$
depends as $\dot N \sim h^3$ we will assume that most
gas will leak through the percolation channel.

Assuming that the contact regions at any magnification are
approximately randomly distributed in the apparent contact region,
we expect from percolation theory that the non-contact region will percolate when $A/A_0
\approx 1-p_c$, where $p_c$ is the site percolation number \cite{Perk}. For a hexagonal
lattice, which is the most plausible lattice structure in the present case, one has
$p_c \approx 0.7$, while for a square lattice
(as in Fig.~\ref{Percolation})
percolation occur at $p_c\approx 0.6$. Thus, the exact value
of the percolation threshold does not depend sensitively on the symmetry of the unit cell.

We assume that the main
gas leakage comes from gas flow through the percolation channel.
The narrowest passage in this channel can be considered as a rectangular pore of height $h$,
and of width and
length $\lambda$, where $\lambda$ is determined by the magnification $\zeta_c$ at the point where
$A/A_0 = 1-p_c \approx 0.3$. The height $h$ of the pore is (approximately)
determined by the {\it rms} roughness
at the magnification $\zeta_c$. In the present case, if the tire gas pressure is in the range
$0.2-0.3 \ {\rm MPa}$, from Fig.~\ref{Magn.Area}(b)
we get $\zeta_c \approx 10$ and from Fig.~\ref{Magn.Area}(c),
$h\approx 0.1 \ {\rm \mu m}$.

We divide the tire-rim contact area into $m$ square areas $B\times B$,
where $B$ is the width of the tire-rim contact area (we expect $B$ to be of order a few cm).
The number of squares is $m=2\pi R/B$, where $R$ is the radius of the tire at the rim.
We expect $m\approx 100$.

Let us study the gas flow through a rectangular pore of
height $h$ and width and length (in the flow direction) $\lambda$.
We assume stationary and laminar flow, and that $h \ll \lambda$. In this case,
if $N_1(t)$ denote the number of gas phase molecules in the tire (which is proportional to the pressure
$P_1$ in the tire),
the basic equations of hydrodynamics give
for the typical case $P_1 \gg P_0$:
\[ \dot N_1 \approx -{ m P_1^2 h^3 \over 24 \mu k_BT} \]
Here we have implicitly assumed that the full pressure drop
$P_1-P_0$ occurs over the pore.
Thus, the time it takes for the pressure in the tire to drop with $\approx 4 \%$ is
\[ \Delta t = {\mu \over P_1} {V_1\over h^3} \]
where $V_1=V_0/m$ (where $V_0$ is the air volume in the tire)
is the volume of air in an angular section of the tire of width $B$.
With $B\approx 3 \ {\rm cm}$ we get $V_1 \approx 3\times 10^{-4} \ {\rm m}^3$
and using the viscosity of air $\mu \approx 17\times 10^{-6} \ {\rm Ns/m^2}$ gives
$\Delta t \approx 1 \ {\rm year}$. This is an upper limit of the leakage time, since
when the interfacial contact area is studied at higher magnification new
pore channels through which the air can leak will be detected.

It is interesting to note that the adhesional interaction between the rubber surface and the
steel rim is likely to have negligible influence on the leakage rate. Adhesion will affect the
(apparent) contact area only at very high magnification \cite{JCP2004}, but most of the gas leakage
occurs via the much larger air flow channels which can be observed at
low magnification.

\begin{figure}
\begin{center}
\includegraphics[width=0.8\textwidth]{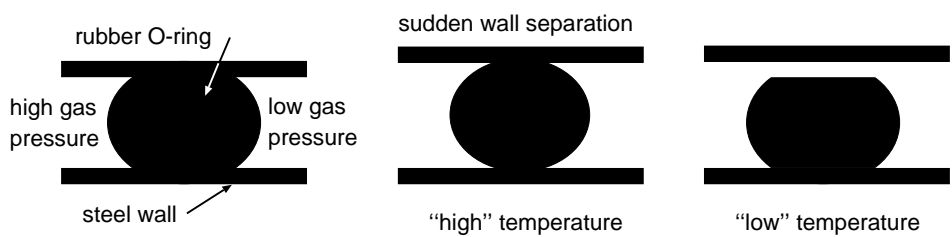} 
\end{center}
\caption{\label{Oring}
Rubber O-ring seal. Left: The rubber ring is compressed between two solid walls.
Middle: If the rubber temperature is high enough,
when the separation between the walls is suddenly increased the rubber will quickly expand and 
still provide tight sealing. Right: At low temperature the rubber relaxation times are very long and the
rubber will not be able to expand quickly enough. Here the sealing fails.}
\end{figure}

When a rubber block is squeezed with a constant force against a rough substrate, the
area of real contact will increase continuously, and the (average) space between the surfaces
will decrease continuously with increasing time. This is due to stress relaxation in the rubber
and was illustrated
in Fig. \ref{Area.time}. In some applications,
such as tire-air sealing, this relaxation effect is clearly
beneficial as it will reduce the size of the interfacial air-flow channels.
However, in other applications, such as
O-ring seals, stress relaxation may result in catastrophic events. One prominent example
was the Challenger catastrophe \cite{Feynman}.
In this case rubber O-rings where used to seal the hot gas inside
the solid fuel rockets and prevent it from leaking through the horizontal joints used to hold
the different vertical rocket wall-segments together.
O-rings
normally operate with about $\sim 15\%$ compression to ensure a tight seal, see Fig. \ref{Oring} (left).
However, the high pressure inside the rocket resulted in expansion of the steel
tube, and to a (small) separation of the wall-surfaces at the joints.
At high enough rubber temperature this is not a problem
since the compressed rubber O-ring would then quickly expand and seal tightly even if
the space between the walls would increase, see Fig. \ref{Oring} (middle).
However, the temperature at the time of the Challenger launch was exceptionally
low (around zero Celsius degrees), which resulted in very
long rubber relaxation times, so that the compressed form of the
O-seals was ``frozen in'', see Fig. \ref{Oring} (right). This effect was
demonstrated by Feynman in a famous
experiment where part of an O-ring was kept in a deformed state in a glass with ice water.
Thus, as the space between the surfaces increased, one or several
O-rings failed to seal-off the hot gas, resulting in leak of hot
fuel gas through the rocket joints, and finally to the catastrophic failure
of the rocket.

Stress relaxation is an important aspects of O-ring seals even when no
separation of the solids walls occurs at the
seal \cite{article}. When a seal is under constant compression (fixed solid walls), the initial stress decays with
increasing time, roughly proportional to the logarithm of the time of contact. Thus the peak compressive
stress may eventually drop below the system (gas or liquid) pressure, and the seal leaks.
Clearly, stress relaxation effects must be taken into account when determining the type of O-ring to be used
in a particular application.

\section{Rubber friction}
\label{sec6}

First principle calculations of frictional forces for realistic systems
are generally impossible. The reason is that friction
usually is an interfacial property, often determined by the last few
uncontrolled monolayers of atoms or molecules at the interface. An extreme
illustration of this is offered by diamond. The friction between two clean
diamond surfaces in ultra high vacuum is huge because of the strong
interaction between the surface dangling bonds. However, when the
dangling bonds are saturated by a  hydrogen monolayer
(as they generally are in real life conditions), friction becomes
extremely low \cite{Flipse}. Since most surfaces of practical use are
covered by several monolayers of contaminant molecules of unknown
composition, a quantitative prediction of sliding friction coefficients
is out of the question. An exception to this may be rubber friction
on rough surfaces, which is the subject we address here.

Rubber friction on smooth substrates, e.g., on smooth glass
surfaces, has two contributions, namely an adhesive (surface) and a hysteretic
(bulk) contribution \cite{Moore,PerssonSS}. The adhesive contribution
results from the attractive binding forces between the rubber
surface and the substrate. Surface forces are often dominated by
weak attractive van der Waals interactions.
For very smooth substrates, because of the low elastic moduli of rubber-like materials,
even when the applied squeezing force is very gentle this weak attraction
may result in a nearly complete contact at the
interface \cite{Persson,Sophie},
leading to the large sliding friction force usually observed \cite{Grosch}.
For rough surfaces, on the other hand, the adhesive contribution to rubber
friction will be much smaller because of the small contact area. The actual
contact area between a tire and the road surface, for example, is
typically only $\sim 1 \%$ of the
nominal footprint contact area \cite{[2],PV,Heinrich}.
Under these conditions the bulk (hysteretic) friction mechanism
is believed to prevail \cite{[2],Heinrich}.
For example, the exquisite sensitivity of tire-road friction to temperature
just reflects the strong temperature dependence of the viscoelastic bulk
properties of rubber.

Here we discuss how rubber friction depends
on the surface roughness power spectra,
and we consider the influence of wear-induced polishing and of water on the road track
on rubber friction.

\subsection{Basic theory of rubber friction}
\label{sec6.1}

\begin{figure}
\begin{center}
\includegraphics[width=0.80\textwidth]{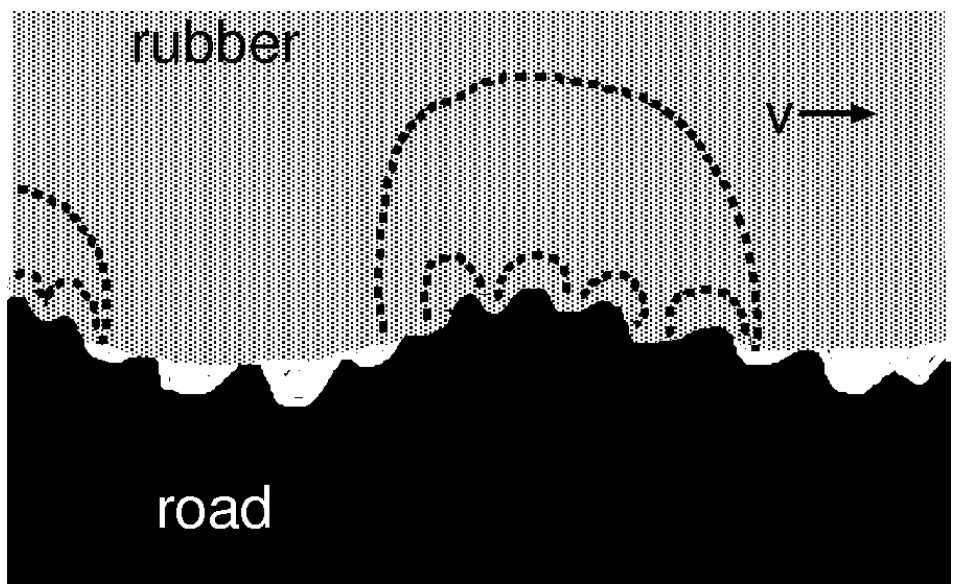} 
\end{center}
\caption{\label{deformation}
Hysteretic friction of a rubber block on a rough road surface. 
The road asperities exert pulsating forces on the sliding rubber block,
leading to energy dissipation in the rubber via the rubber internal friction.
Most of the energy dissipation occurs in the volume elements bounded by the dashed lines.
The rubber viscoelastic deformations in the
large volume elements are induced by the large road asperities, while the smaller dissipative regions
result from the smaller asperities distributed
on top of the large asperities. In general,
in calculating the rubber friction,
the viscoelastic energy dissipation induced
by all the
asperity sizes must be included, and the local temperature increase (flash temperature) in the
rubber resulting from the energy dissipation should also be taken into account in the analysis.
}
\end{figure}

The main contribution to rubber friction when a rubber block is sliding on a rough substrate, such as a tire
on a road surface, is due to the viscoelastic energy dissipation in the surface region of the rubber as a
result of the pulsating forces acting on the rubber surface from the
substrate asperities, see Fig.~\ref{deformation}. Recently
one of us has developed a theory
which accurately describes this energy dissipation process, and which predict the
velocity dependence (and, in a more general case, the time-history dependence) of the rubber friction
coefficient \cite{[2],PV}.
The results depend only on the (complex) viscoelastic modulus $E(\omega )$ of the rubber, and on
the substrate surface roughness power spectra $C(q)$.
Neglecting the flash temperature effect (the term \emph{flash temperature}
refers to a local and sharp temperature rise occurring in the tire-road
asperity contact regions during slip), 
the kinetic friction coefficient at velocity $v$ is determined
by \cite{[2]}
\[
  \mu_{\rm k} = {1 \over 2}\int_{q_0^*}^{q_1^*} \rmd q \ q^3 C(q) P(q)
  \int_0^{2\pi} \rmd\phi \ \cos \phi
    \ {\rm Im} \frac{E(qv \ \cos\phi )}{(1-\nu^2)\sigma}, \]
where
\[ P(q)=
 {\rm erf}\left({1 \over 2\sqrt{G}} \right ), \]
with
\[
  G(q)=\frac{1}{8}\int_{q_0^*}^q \rmd q \ q^3C(q)\int_0^{2\pi} \rmd\phi
   \ \left|\frac{E(qv \ {\rm cos}\phi )}{(1-\nu^2)\sigma}\right|^2,
\]
where $\sigma$ is the mean perpendicular pressure (load divided by the
nominal contact area), and $\nu $ the Poisson ratio,
which equals 0.5 for rubber-like materials.

The theory takes into account the substrate roughness in the range $q_0^* < q < q_1^*$,
where $q_0^* $ is the smallest relevant wave vector
of order $2\pi /L$, where (in the case of a tire) $L$ is the
lateral size of a tread block, and where $q_1^*$ may have
different origins (see below). Since $q_0^*$ for a tire tread block is smaller than the
roll-off wave vector $q_0$ of the
power spectra of most road surfaces (see Fig.~\ref{CqOpel}), rubber friction
is very insensitive to the exact value of $q_0^*$.

The large wave vector cut off $q_1^*$ may be related to road contamination, or may be an intrinsic property
of the tire rubber. For example, if the road surface is covered by small contamination particles
(diameter $D$) then $q_1^* \approx 2\pi /D$. In this case,
the physical picture is that when the tire rubber surface is covered by
hard particles of linear size $D$, the rubber will not be able to penetrate 
into surface roughness ``cavities'' with diameter
(or wavelength) smaller than $D$, and such short-range roughness will therefore not contribute to the rubber friction.
For perfectly clean road surfaces we believe instead that the cut-off $q_1^*$ is related to the tire rubber properties.
Thus, the high local (flash) temperatures during braking, and the high local stresses which occur
in the tire rubber-road asperity contact regions, may
result in a thin (typically of order a few micrometer)
surface layer of rubber with modified properties (a ``dead'' layer), which would contribute very little
to the observed rubber friction. Since the stresses and temperatures which develop 
in the asperity contact regions depend somewhat
on the type of road [via the surface roughness power spectra $C(q)$],
the thickness of this ``dead'' layer may vary from
one road surface to another, and some run-in time period will be necessary 
for a new ``dead'' layer to form when a car
switches from one road surface to another. Such ``run-in'' effects are well known experimentally.

In the theory that one of us has developed, the thickness of the dead layer
is determined by studying (via computer simulations) the temperatures and stresses which
develop on the surfaces of the tire tread blocks during ABS
breaking.

\begin{figure}
\begin{center}
\includegraphics[width=0.50\textwidth]{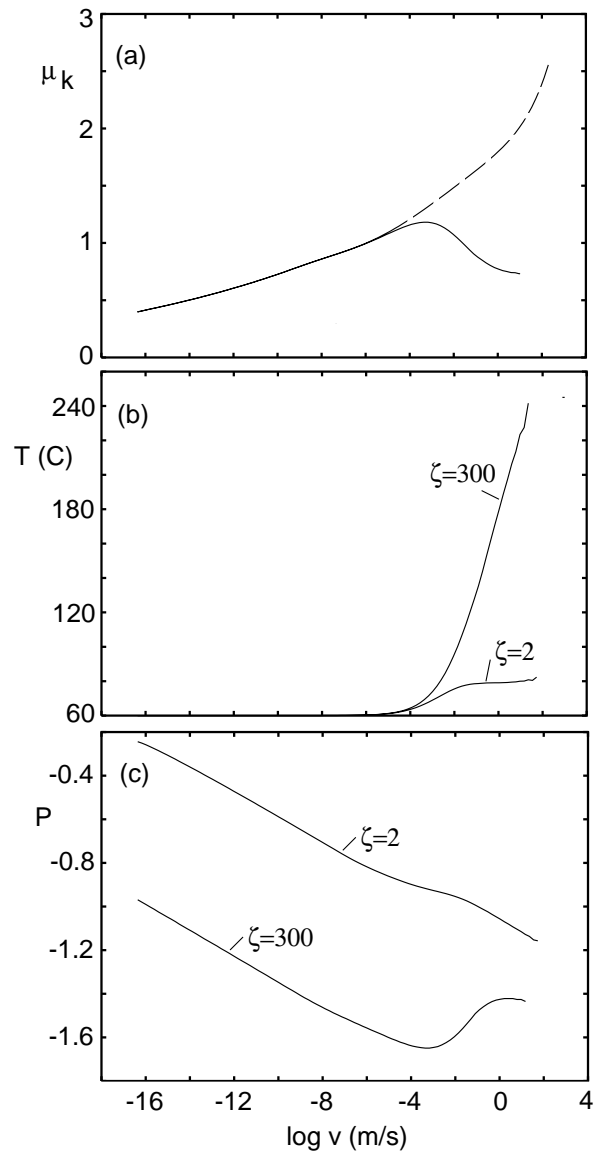} 
\end{center}
\caption{\label{muk2}
Calculated friction coefficients for a rubber block sliding 
on a self affine fractal surfaces with $H=0.8$, roll-off wavevector $q_0 = 1000 \,{\rm m^{-1}}$, 
largest wave vector $q_1^*=q_0 \times 320.0$ (corresponding to the magnification $\zeta$ =320)
rms roughness $\sigma=1 \,{\rm mm}$, and the background temperature 
assumed was $T_0=60^\circ {\rm C}$.
(a) The kinetic friction coefficient as a function of the logarithm of the
sliding velocity. Solid and the dashed lines are with and without flash
temperature effects, respectively.
(b) Flash temperature calculated at two different magnifications. The highest
magnification $\zeta=300$ correspond to the temperature very close to the surface while the
low magnification curve correspond to the temperature deeper into the rubber.
(c) Relative contact area $P(\zeta )= A(\zeta )/A_0$ calculated for two different magnifications.
}
\end{figure}

Since this is not the proper place for a full presentation of all
details of the theoretical calculations of tire-road friction, we 
shall simply present some numerical results to illustrate how
the rubber friction depends on the surface roughness
power spectra.
Fig.~\ref{muk2}(a) shows the kinetic friction coefficient as a function of the sliding velocity
calculated for a rubber block sliding
on a self affine fractal surface with the Hurst exponents $H=0.8$,
roll-off wave vector $q_0 = 1000 \,{\rm m^{-1}}$, and large wave vector cut-off
$q_1^*=320.0 \times q_0$, typical for road surfaces.
We show results both with and without the flash temperature effect.

For low sliding velocities,
the kinetic sliding friction depends very weakly on the sliding velocity,
and only at extremely
low velocities is friction strongly reduced. Thus
the present calculation (as well as experiments) shows that even sliding
velocities as small as $10^{-10} \,{\rm m/s}$,
may give rise to a relative large kinetic friction.
The physical reason for this is the very wide distribution of relaxation times in
most rubber materials, extending to very long times.

For typical tire rubber compounds, calculations (neglecting temperature effects)
as well as  measurements at low sliding velocities and different temperatures
(subsequently shifted to a common temperature utilizing the frequency-temperature 
Williams-Landel-Ferry (WLF) transform \cite{WLF}),
indicate that the maximum friction coefficient $\mu_{\rm k}(v)$ will typically
occur at very high velocities, of order 1000 m/s
[see dashed line in Fig.~\ref{muk2}(a)].
However, direct experiments without the WLF transform show
a maximum in $\mu_{\rm k}$ at much lower velocities.
This is a result of the flash temperature, see solid line in Fig.~\ref{muk2}(a).
Thus when the flash temperature
effect is taken into account the maximum of $\mu_{\rm k}(v)$
shifts at much lower sliding velocities
in the typical range $0.1-1 \ {\rm cm/s}$, in agreement with experiment.

\begin{figure}
\begin{center}
\includegraphics[width=0.35\textwidth]{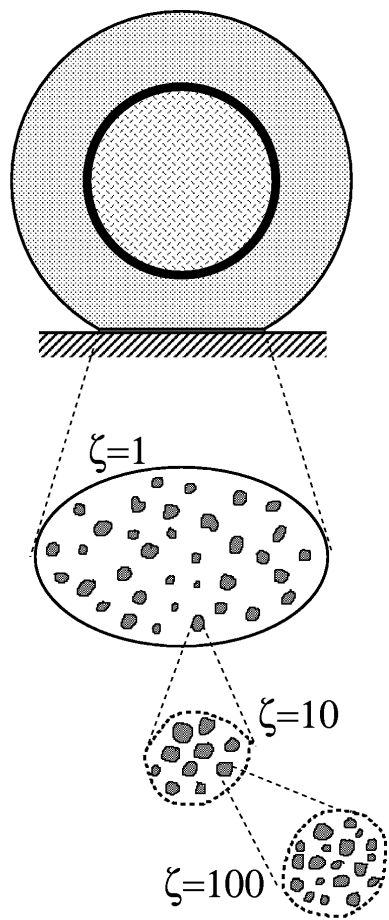} 
\end{center}
\caption{\label{tire}
A tire in contact with a road surface does not make perfect contact with
the road in the whole footprint area. At the lowest magnification the
asperity contact regions have diameters
of order $\sim 0.3-1 \ {\rm cm}$. When the magnification is increased
these macrocontact regions break up into smaller contact regions separated by noncontact areas.
At a magnification of order $\sim 100-1000$ the tire rubber-road contact area
is just a few $\%$ of the nominal footprint contact area.}
\end{figure}

Fig.~\ref{muk2}(b) and (c) show the flash temperature and the relative contact area
at two different magnifications. For velocities $v< 10^{-4} \,{\rm m/s}$ the contact area decreases
with increasing sliding velocity. This is a result of the increasing frequencies of the
deformations of the tire surface with increasing sliding velocity.
However, for $v > 10^{-3} \,{\rm m/s}$ the contact area {\it increases}.
This is a result of the increased
temperature of the rubber in the contact areas shifting the viscoelastic dissipation
maximum to higher frequencies and hence making the rubber elastically softer
at any given frequency. It is also of interest to consider the relative contact area.
For sliding velocities of order $1 \,{\rm cm/s}$ the tire-road contact area
at a magnification of order $100-1000$ is
just a few $\%$ of the nominal footprint contact area.
This is illustrated in Fig.~\ref{tire} which shows the tire-road
contact region at different magnification.

\begin{figure}
\begin{center}
\includegraphics[width=0.6\textwidth]{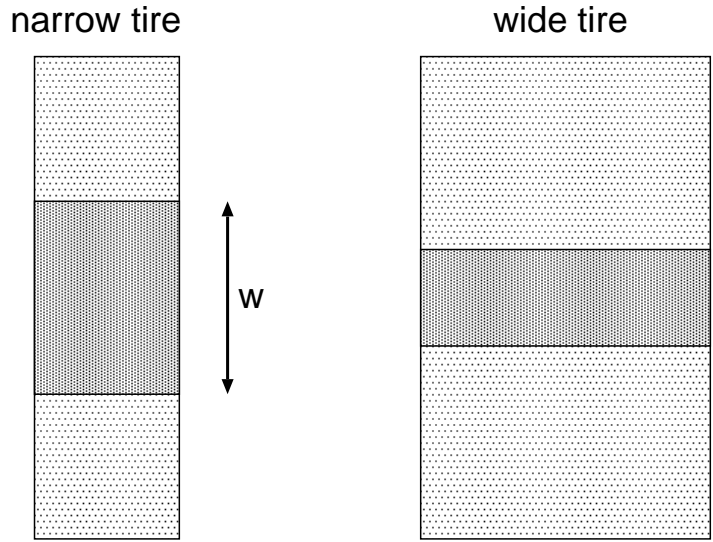} 
\end{center}
\caption{\label{WideNarrow}
The nominal tire-road contact area is approximately the same for a wide and a narrow tire.
Hence the length $w$ of the foot print area in the rolling direction will be smaller for the wide tire.}
\end{figure}

The flash temperature effect described above is of extreme practical importance.
If the excessive buildup of flash temperatures could be avoided or reduced, the effective tire-road
friction coefficient would clearly increase. One way to reduce the flash
temperature is to use wide tires instead of narrow tires\footnote{
  The discovery that wide tires gives higher effective friction than
  narrow tires was made by Jim Hall and coworkers in the middle 60's for
  racing tires. The story is beautifully described in \cite{discovery},
  where Hall stated:
  ``In a matter of less than a year we went from those narrow little
  Dunlop Green Sports to a tire that was almost double the width.  And our
  single-speed torque converter just couldn't handle all that grip. We
  immediately went to a two-speed transmission so we could get more torque
  multiplication off the slow-speed corners.'' ``So that's what happened
  with me on tires. In the middle `60s, probably `64 or `65, Firestone
  took us from less than 6 inches of tread to width up to 12 inches.''
  Paul Haney gives in his book \cite{discovery} an explanation for why
  wide tires give higher effective friction than narrow tires. However,
  this `explanation', and many others given on the Internet, are incorrect.
  }.
This can be understood as follows.
In a first approximation the nominal tire-road contact area is the same for 
narrow and wide tires. In fact, this holds exactly when a tire can be described
as a thin elastic (torus shaped) membrane (which is a good approximation
for airplane tires but less so for car tires) and the tire air pressure and the load
are the same for both tires\footnote{
  For the membrane tire the nominal pressure in the footprint contact area is
  given by the air pressure $P$ in the tire so that the nominal contact area
  $A_0=F_{\rm N}/P$ is the same for both tires if the load $F_{\rm N}$ is the
  same.
  }.
Thus, wide tires will have a footprint area which is shorter
in the longitudinal rolling direction
than a narrow tire, see Fig. \ref{WideNarrow}. In order to fully build up the flash temperature,
a tread block must slide a distance of order the average diameter $D$
of the macro asperity contact regions.
The macro asperity contact regions are the contact between the rubber and the largest
road asperities involving the longest wavelength roughness components characterized by the roll-off
wavelength $\lambda_0 = 2 \pi /q_0$ in the surface roughness power
spectra (see Fig.~\ref{CqOpel}). In a typical case $\lambda_0 \approx 1 \,{\rm cm}$, and the
average diameter of the macro
asperity contact regions $D\approx 0.5 \,{\rm cm}$.
If a tread block in the footprint contact region slips by less
than the distance $D$, the local temperature increase will be below that expected during stationary
sliding at the same slip velocity. If $v$ denotes the average slip velocity of a tread block
then the slip distance $d \approx v t$ where the time the tread block spend in the footprint
area $t=w/v_{\rm R}$, $v_{\rm R}$ being the tire rolling velocity, 
and $w$ the length of the footprint area. Thus, $d \approx (v/v_{\rm R}) w$
and the slip velocity necessary for the slip to be of order $D$ is $v \approx (D/w) v_{\rm R}$. 
Hence, for a wide tire (small $w$) the tread block slip velocity can reach higher 
values than for a narrow tire without building up the flash temperature effect fully. 
Since the kinetic friction coefficient
in the absence of the flash temperature effect increases monotonically with increasing slip velocities
(up to very high slip velocities, perhaps $1000 \,{\rm m/s}$ or more, see dashed line in
Fig.~\ref{muk2}) it follows that wide tires
will give higher effective friction than narrow tires.
The discussion above is only of semi-quantitative validity since in reality
the tread blocks will not slip with a constant velocity in the footprint area,
but with a nonuniform velocity.

Another way to reduce the flash temperature could be to use a filler material which exhibits
a first order phase transition somewhere close to (but above) the
background tire-rubber temperature,
which typically
may be around $60^\circ {\rm C}$. If the phase transition (e.g., a structural, magnetic
or ferroelectric transition) could absorb a sufficiently large latent heat, it 
would strongly reduce the flash temperature.
The filler materials used today
(mainly carbon black and silica) does not exhibit any such phase transition.

The theoretical results presented above are in accordance with experimental
observations \cite{Westermann,Popp,Austria,Westermann2}.
Thus, flash temperature effects have been observed during dry ABS braking test, where
the temperature distribution on the tire surface can be measured with high-speed
IR-cameras \cite{Westermann,Westermann1}.
In one experiment the temperature increase in the hot spots on the tread block surfaces
on the length scale of the measurement was $\sim 25\ {\rm C}$ above the
background temperature. The decrease of the kinetic friction force with increasing sliding
velocity for $v > 0.1 - 1 \,{\rm cm/s}$ (as a result of the flash temperature),
has been reported by several research groups \cite{Westermann,Popp,Austria,Westermann2}
and is in good agreement with the theory.
A decreasing kinetic friction coefficient may result in stick-slip motion; thus
one mechanism of rubber stick-slip is due to the flash temperature.

The theory also predicts that when a rubber block is sliding on a rough substrate the friction
force is nearly proportional to the normal load, i.e., the friction coefficient is
independent of the load. This is again in agreement with experiment,
see, e.g., Ref.~\cite{Popp}. We note that for tires the effective
friction coefficient usually decreases with increasing load.
This does not actually reflect a load dependence of the fundamental rubber-substrate friction coefficient,
but results rather as an effect of the dependence of the size and the pressure 
distribution in the footprint area on the load,
which affects the motion of the tread blocks \cite{PPersson}.

The theory sketched above has been used successfully to predict the ABS
breaking performance of
tires \cite{PirelliT,Westermann,Westermann1,Westermann2}. In one study 6 tires
were produced with variation
only in the tread compound \cite{Westermann,Westermann1}.
The tread rubber compounds were varied so that big variations in ABS rating were
expected. The observed braking distance was found to correlate very well with the theoretical
predicted effective friction coefficient, with a high correlation coefficient
$R^2=0.97$. This result shows that it is possible to estimate the performance of a tire
on physically sound frictional models which focus not only on the viscoelastic properties of
the rubber compound, but take explicitly into account additional
important information like the surface
topography at all length scales as well as ambient and flash temperature conditions.

\subsection{Rubber friction and the influence of polishing}
\label{sec6.2}

\begin{figure}
\begin{center}
\includegraphics[width=0.80\textwidth]{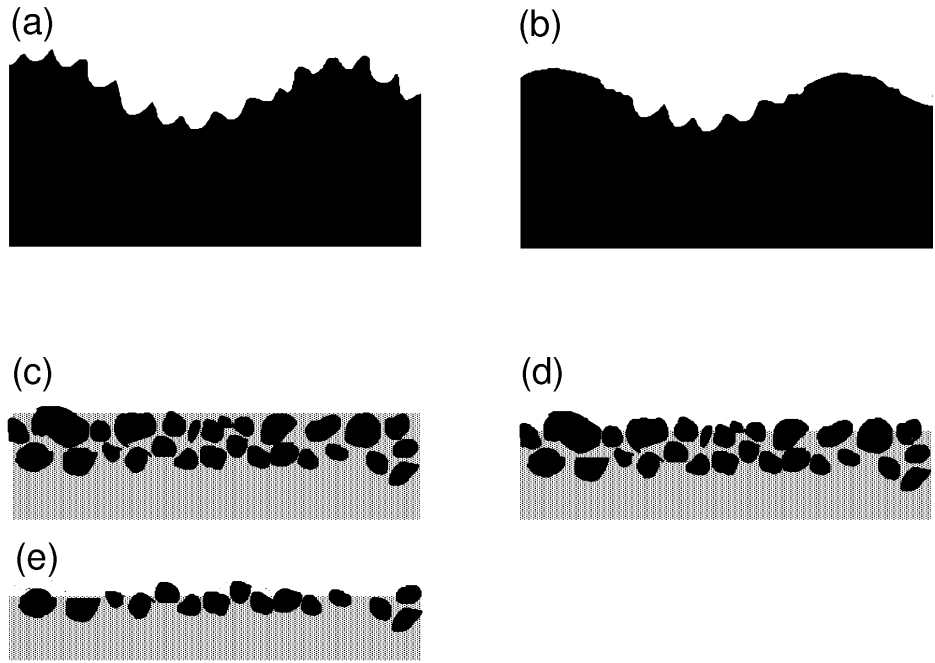} 
\end{center}
\caption{\label{CobbAsp}
(a) and (b): Rolling and braking on cobble stone road surfaces causes the surface asperities
to get polished.
(c)-(e) On asphalt or concrete road surfaces smaller hard stone particles
never get highly polished as they are continuously exposed and/or removed 
by wear of the softer surrounding matrix.}
\end{figure}

Let us now discuss the role of road polishing by sliding tires\footnote{
  See, e.g., {\it http://www.betterroads.com/articles/oct03b.htm}
  for experimental information from road engineers about rubber friction and
  road surfaces, in particular about road polishing.
  }.
It is known that a relative new asphalt roadway has a ``sharper'' surface pavement
with higher coefficient of friction compared to an older well-traveled asphalt roadway.
This is a consequence of the wear action exerted by vehicle tires, which results 
in polishing the exposing aggregate surface, and the state of polish 
is one of the main factors affecting the resistance
to skidding. (Note: the fact that the polishing of the surface asperities
{\it reduces} the friction is in itself an indication that the adhesional contribution to rubber friction
on road surfaces is unimportant, since the latter should increase when the surface becomes smoother.)
Resistance to this polishing action is determined principally by the inherent quality of the
aggregate itself. Rocks composed of minerals of widely different hardness, and rocks that wear by the
pulling out of mineral grains from a relative soft matrix, have high
resistance to polishing, see Fig.~\ref{CobbAsp}(c)-(e).
Conversely, rocks consisting of minerals having nearly the same hardness 
wear uniformly and tend to have lower resistance to polishing, 
see Fig.~\ref{CobbAsp}(a),(b). Thus, sandstones
have a good resistance to polishing, whereas the
limestone and flint groups present the lowest resistance\footnote{
  The influence of polishing of road surfaces is studied in a standardized
  lab test where fragments (of diameter $\sim 1 \ {\rm cm}$)
  of the stone (or other material) to be tested are glued onto a flat
  surface and exposed to accelerated (as compared to polishing on the road) polishing
  using standardized conditions. The rubber friction coefficients obtained after
  a fixed time period of polishing is then measured. The result will depend on the
  rubber used in the friction test (and the laboratory conditions, e.g., the temperature),
  and only relative friction values (for different polished surfaces)
  are analyzed. In one such set of
  measurements (see {\it http://www.wainwright.co.uk/technical.htm})
  it was found that gritstone and sandstone have the best resistance against
  polishing (giving friction coefficients after polishing $\mu = 0.68$ and $0.67$,
  respectively) followed by basalt and granite ($\mu = 0.56$ and
  $0.55$ respectively), while limestone gave the worst result ($\mu = 0.37$).
  }.
Among other groups, basalt, granite and quartzite yield intermediate results.
Microroughness is gradually polished away by the action of heavy traffic so,
over time, the skidding resistance of a road will fall to an ``equilibrium'' level that depends
upon the type of aggregates used in the surfacing.
For concrete and asphalt pavements that have too low friction, diamond grinding will bring
its skid resistance up. Grinding increases the pavement microstructure by dislodging polished
sand particles in the mortar matrix.

Experience by road engineers shows that rubber friction on dry clean roads depends on micro roughness
with wavelength mainly in the range $1 \,{\rm \mu m} < \lambda < 1000 \,{\rm \mu m}$, in accordance
with our calculations.
Longer wavelength roughness $0.1 \,{\rm cm} < \lambda < 5 \,{\rm cm}$
is important on wet road surfaces especially to prevent hydroplaning for velocities above $60 \,{\rm km/h}$.
High dry friction on asphalt or concrete road surfaces are obtained when hard minerals (such as feldspat and
quartz) and aggregates with good microstructure, such as sandstone and slag (usually, air-cooled blast
furnace slag from iron production),
are embedded in a softer matrix, e.g., tar (asphalt) or mortar (concrete). Minerals with rough grains,
or mixture of minerals with different texture, will resist polishing and maintain high frictional
properties.
Special surface coatings with very high skid resistance have been developed.
One such coating consist of epoxy adhesive binding a synthetic aggregate to the
road surfaces\footnote{
  See {\it http://www.new-technologies.org/ECT/Civil/italgrip.htm}
  for high friction coatings for road surfaces.
  }.
The aggregate, manufactured from steel slag, has extremely high polish resistance,
and is claimed to
result in tire-road friction about $40\%$ higher than on normal road surfaces.
In temperate climates such as Sweden, the surface microroughness varies cyclic,
with skidding resistance at its lowest during the summer and autumn,
recovering to some extent during the winter.
This is due in part to weather induced erosion
of the road surface (e.g., freezing water in the road cavities and frost heave
may break-off road surface fragments),
and partly due to the strong wear-induced surface roughening which result from the ``heavy''
equipments used to clean the road surfaces from snow and ice.

\begin{figure}
\begin{center}
\includegraphics[width=0.90\textwidth]{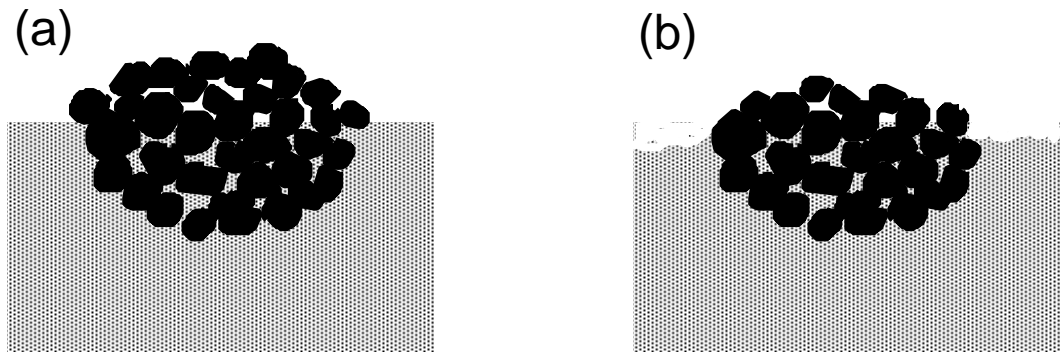} 
\end{center}
\caption{\label{polished.cluster}
Asphalt or concrete roads with aggregates of small
(micrometer) particles, e.g., fragments of sandstone
or slag, have a high resistance against polishing if the tire-road interaction mainly results in
a slow removal of the microparticles in the aggregates, preserving an aggregate particle surface
that is (on the microscale) constantly rough. 
(a) and (b) show an aggregate before and after wear (schematic).}
\end{figure}

Summarizing,
it is clear that the best road pavements, with high resistance against polishing, consist of aggregates
of very small (micrometer) and hard particles, e.g., fragments of sandstones, where the wear occurs by removal
of the individual micro-particles or clusters of micro-particles,
rather than by polishing the particles, see Fig.~\ref{polished.cluster}. However, the
binding of the particles in the aggregates must be so strong that the wear of the road pavement is slow,
i.e., the wear rate by removal of particles should be only slightly larger 
than the wear rate by polishing of the
particles. The optimal case is when the wear process results in
particle aggregates of various sizes being removed from the road surface so that
the road surface remains rough at all the length scales above the size of the
smallest particles.

\subsection{Rubber friction on wet road surfaces}
\label{sec6.3}

\begin{figure}[htb]
\begin{center}
  \includegraphics[width=0.60\textwidth]{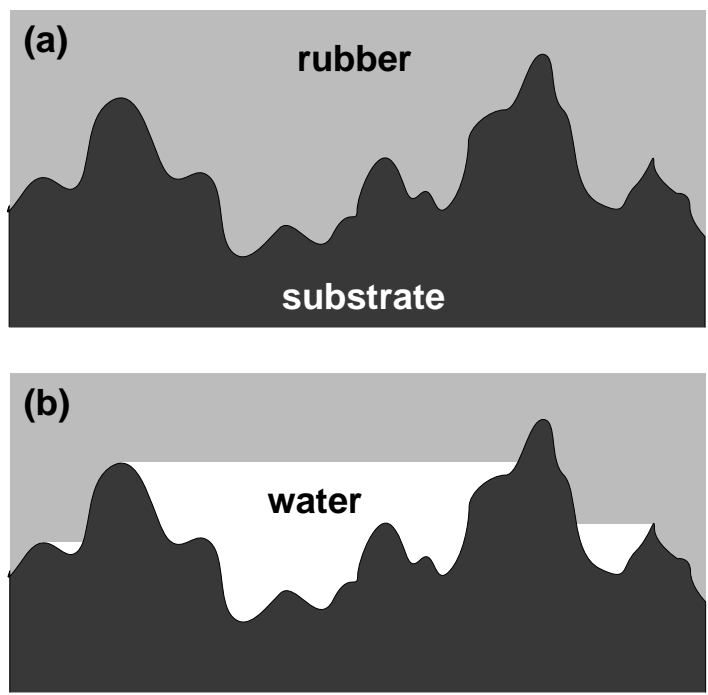} 
\end{center}
\caption{ \label{sealing}
A rubber block sliding on a rough hard substrate. (a) On a dry substrate
the rubber penetrates  a large valley and explores the short wavelength
roughness within. The pulsating rubber deformations induced by the short-wavelength
roughness contributes to the friction force.
(b) On a wet substrate the valley turns into a water pool. Sealing of the
pool now prevents the rubber from entering the valley. By removing
the valley contribution to the frictional force, this {\em sealing effect}
of rubber reduces the overall sliding friction.}
\end{figure}

For rubber friction on {\em wet} rough substrates
at low sliding velocities it is known that the friction
typically drops by as much as $20-30\%$ relative to the corresponding
dry case \cite{Gert,MeyerWalter}.
Owing to the small contact area, this cannot be the result of a
water-induced change of adhesion. On the other hand, as will be discussed below,
the friction decrease cannot be
blamed on a purely hydrodynamical effect either.
That leaves finally
the possibility that water might change precisely the bulk, hysteretic friction. We proposed
recently \cite{tobepubl} that this is indeed the case. Water pools that form in the wet
rough substrate are {\em sealed off} by the rubber, as sketched in Fig.~\ref{sealing},
and that will effectively smoothen the substrate surface. Smoothening reduces
the viscoelastic deformation from the surface asperities, and thus reduces
rubber friction.

As discussed before, rubber friction from the viscoelastic deformation by the substrate asperities
is determined by the complex frequency-dependent bulk viscoelastic modulus
$E(\omega )$ of rubber and by the substrate surface roughness power spectrum
$C(q)$.

\begin{figure}[htb]
\begin{center}
  \includegraphics[width=0.40\textwidth]{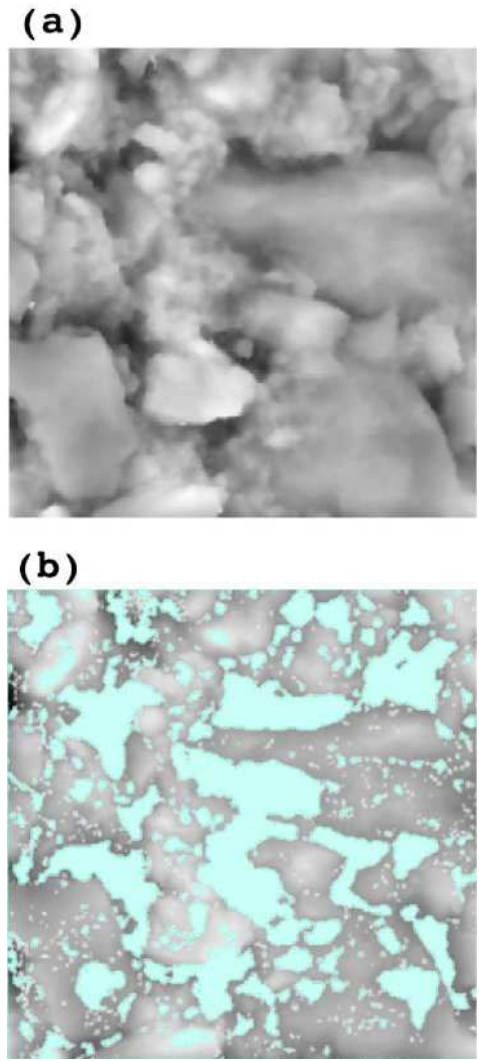} 
\end{center}
\caption{ \label{topo}
(a) Optically observed height profile of a dry asphalt road
($1.5 \,{\rm cm} \times 1.5 \,{\rm cm}$ area), darker areas corresponding to
deeper regions. (b) Calculated wet profile for the same area, with water pools
(light blue).}
\end{figure}

\begin{figure}[htb]
\begin{center}
  \includegraphics[width=0.70\textwidth]{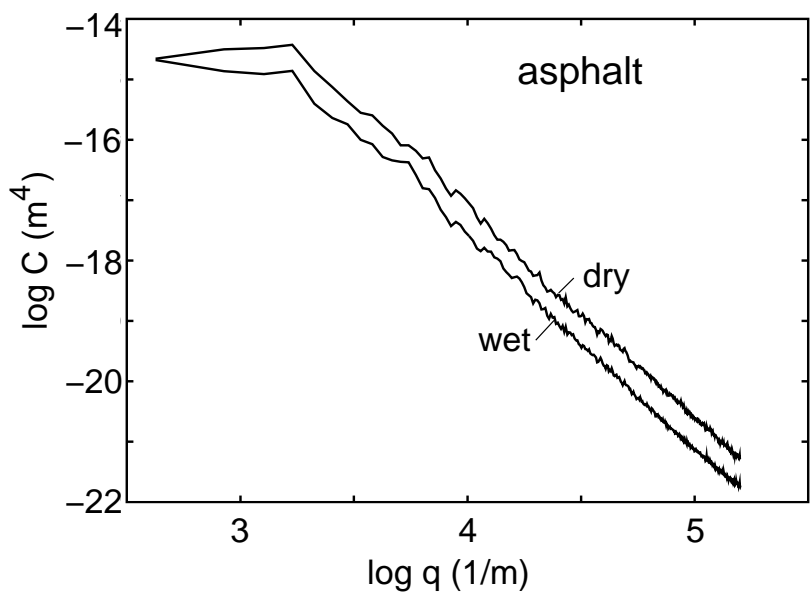} 
\end{center}
\caption{ \label{Cq}
Surface roughness power spectra $C(q)$ (above) extracted
from the measured height profile for a dry asphalt road surface, and (below)
calculated assuming sealing of all pools in the same
surface when wet, as in Fig.~\ref{topo}.
Note the logarithmic scales.}
\end{figure}

The upper curve in Fig.~\ref{Cq} shows the
power spectrum extracted via Eq.~(\ref{eq1}) from the measured height profile $h(\ve{x})$.
The log-log scale shows that for $q > 1600 \,{\rm m}^{-1}$, $C(q)$ drops
as a power law, as expected for a self-affine fractal surface. The fractal
dimension of this surface is determined by the slope
of the curve in Fig.~\ref{Cq}
and is about $D_{\rm f} =2.2$.
The root-mean-square roughness can be obtained directly from the height
profile,
$h_{\rm rms} \approx 0.3 \,{\rm mm}$.

Consider now a tire rolling and sliding on a wet road surface.
At low velocities (say
$v < 60 \,{\rm km/h}$) there will be negligible hydrodynamic water
buildup between the tire and the road surface \cite{tobepubl}.
In essence, if
$v < (\sigma / \rho )^{1/2}$, where $\sigma$ is the perpendicular
stress in the tire-road contact area and $\rho$ the water mass density,
there is sufficient time for the water to be squeezed out
of the contact regions between the tire and the road surface,
{\it except} for water trapped in road cavities. The water pools
will be sealed off by the road-rubber contact at the upper boundaries of the
cavities (see Fig.~\ref{sealing}). Thus, we can focus on the smoothening 
effect on the road profile caused by the sealing effect.

Starting from a dry substrate profile $h(\ve{x})$ we can
numerically build a new wet surface height profile $h'(\ve{x})$
as shown in Fig.~\ref{topo}(b). The algorithm assumes every valley
to be filled with water up to the maximum level where the water will
remain confined, i.e., up to the lowest point of the edge surrounding
the pool.
Any extra water added to the
profile of Fig.~\ref{topo}(b) will flow straight out of the square area.
Once the size of the area considered is at least as large as
$\lambda_0$ this construction becomes unique, with no free parameter.

 From the water-smoothened height profile $h'(\ve{x})$ we obtain a
modified power spectrum $C'(q)$ shown by the lower curve in Fig.~\ref{Cq}.
While the fractal power-law decay and the roll-off wave vector are
essentially the same as for the dry surface, the reduction in the power spectrum
reflect the effective water-induced smoothening
of the rough substrate.

The sealed-off water in the pools (see Fig.~\ref{sealing})
removes the contact with the
interior of the valley, which smoothen the effective substrate roughness
profile. Our basic assumption is therefore that when rubber
slides on the wet rough surface, the friction force will be determined
by the modified power spectrum $C'(q) < C(q)$.

Let us now examine, based on this model, numerical results about tire
friction on dry and wet substrates, calculated using the hysteretic friction theory presented
in Ref.~\cite{[2]}.
The hysteretic friction coefficient at velocity $v$ is determined
by knowledge of the rubber viscoelastic modulus $E(\omega )$ and of the surface
roughness spectrum $C(q)$.

We present results for the friction
of a standard tread compound, sliding on
the asphalt road just characterized. We used the measured rubber complex
viscoelastic modulus (not shown)
along with the power spectra presented in Fig.~\ref{Cq} for the dry and for
the wet road surfaces.

\begin{figure}[htb]
\begin{center}
  \includegraphics[width=0.70\textwidth]{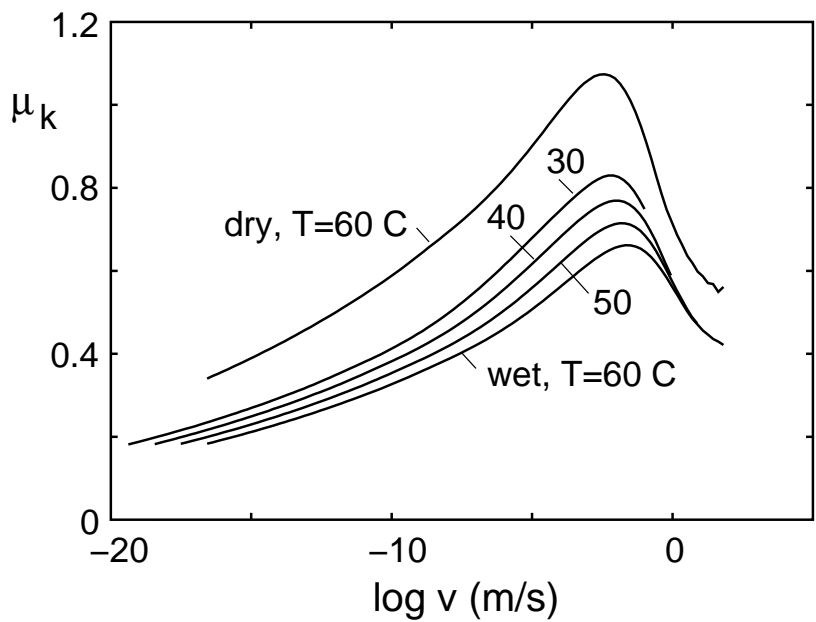} 
\end{center}
\caption{ \label{mukinetic}
Kinetic friction coefficient as a function of the logarithm of
the sliding velocity, calculated for a standard tread compound
and an asphalt substrate with the roughness spectra of Fig.~\ref{Cq}.}
\end{figure}

Fig.~\ref{mukinetic} shows the rubber-asphalt kinetic friction coefficient
calculated for the dry surface (here including also the flash temperature)
at $T=60^\circ {\rm C}$ as a typical
tire temperature while rolling on a dry road, and for the wet surface
at four different temperatures, namely $T=30$, $40$, $50$, and
$60^\circ {\rm C}$ (on a wet road the tire temperature
is typically about $30^{\circ} C$, and generally lower than on the
dry surface).
When dry and wet
frictions are compared, the calculation shows first of all a
water-induces friction decrease of $\sim  30\%$ at $T=60^\circ {\rm C}$.
The water-induced decrease becomes somewhat less ($\sim  20\%$)
if the wet substrate temperature is (realistically) reduced to $T=30^\circ {\rm C}$.
We have also calculated $\mu$-slip curves, and they show a similar reduction in the
friction for wet road surfaces.
These figures are in excellent agreement with the known reduction
of low-speed rubber friction on road surfaces \cite{Gert,MeyerWalter}.
We note in addition that the decreasing friction with increasing
temperature shown in Fig.~\ref{mukinetic} is very commonly observed for rubber.
It results from the shift in the viscoelastic spectrum to higher frequencies
with increasing temperature, making the rubber more elastic and less viscous,
in turn reducing the rubber friction.

The above picture does in our view catch an important novel effect of water
on rubber friction. Yet, it is certainly open to refinements in various
ways. First, dry friction of tires is not pure sliding but also
involves some stick-slip \cite{lowvelocities}. This effect is included in
the calculation of $\mu$-slip curves,
but the observed reduction in the effective friction
is similar to in Fig.~\ref{mukinetic}.
Second, after enough time all sealing regions leak. This will be
particularly true in the present case because the upper boundary of a
water filled pool, which is in contact with the rubber, still has
roughness on many length scales. So one cannot expect the rubber to
make equally perfect contact everywhere, and there will be narrow channels
through which the water slowly leaks out of the pools. As a result, for
sufficiently low sliding velocities the negative water influence on rubber
friction may revert to negligible. Experiments have indeed shown that
for extremely low velocities $v < 0.7 \,{\rm m/s}$ the difference in
$\mu_{\rm k}$ between dry and wet surfaces is very small \cite{lowvelocities}.
We should also stress that the effects addressed here clearly apply only to moderately
wet substrates and for rolling or sliding velocities $v < 60 \,{\rm km/h}$.
For flooded surfaces and $v > 60 \,{\rm km/h}$ aquaplaning may occur,
which originates instead from the inertia of the water. Finally, for
rubber friction on relative smooth wet surfaces,
where the adhesional interaction is important, the so called dewetting transition
may be important \cite{dewetting1,dewetting2,dewetting3}.

\subsection{Lubricated rubber O-ring seals}
\label{sec6.4}

Surface roughness has also a big influence on 
lubricated rubber O-ring seals \cite{article}. Tests have shown that
the longer a lubricated seal sits idle, the higher is static,
or start-up friction coefficient $\mu_{\rm s}$.
Eventually, for smooth surfaces, the friction coefficient reaches a
maximum almost as high as for unlubricated seal. This increase in $\mu_{\rm s}$ with time is
caused by the squeeze-out of most of the lubricant from the contact area \cite{PM2004}.
However, the static friction can be reduced by optimizing the surface roughness and lubricant viscosity.
Thus, in one experiment \cite{article} with
surfaces with the {\it rms} roughness amplitude $\sigma \approx 0.4 \,{\rm \mu m}$
tiny pockets with lubricant was found trapped at the interface,
making it available at startup. Too smooth finish leaves no pockets of trapped lubricant,
while too rough surfaces can cause high wear.

\section{Adhesion}
\label{sec7}

In this section we discuss adhesion between rough surfaces. We point out that
even when the force to separate two solids vanishes, there may still be
a finite contact area (at zero load) between two solids as a result of the adhesional
interaction between the solids.
We also present two applications where surface roughness influences
adhesion in a fundamental way. We first consider an industrial application
related to one step in the fabrication of rubber sheets. For this system a recently developed
viscoelastic contact mechanic theory, combined with surface topography measurements
using the Atomic Force Microscope, was able to explain the
origin of a long-standing technological problem.

The second application is a study of adhesion relevant to biological systems, e.g.,
flies, crickets and lizards, where the adhesive microstructure consists of arrays
of thin fibres and plates. The effective elastic modulus of the fibre-plate arrays can
be very small, which is of fundamental importance for
adhesion on smooth and rough substrates. This application illustrate how nature,
through the process of natural selection, has been able to produce elastically very soft,
but still wear-resistant
layers, which can make good contact and exhibit strong adhesion even to surfaces with
roughness on all length scales down to the atomic dimension.

\subsection{Adhesion between rough surfaces}
\label{sec7.1}

A theory of adhesion between an
elastic solid and a hard randomly rough substrate must take into account
that partial contact may occur
between the solids on all length scales.
For the case where the
substrate surface is a self affine fractal, theory
shows that when the fractal dimension is close to 2
complete contact typically occurs in the macro asperity
contact areas (the contact regions observed when the system
is studied at a magnification corresponding to the
roll-off wavelength $\lambda_0=2\pi/q_0$
of the surface power spectra, see Fig.~\ref{Cq1}),
while when the fractal dimension is larger than 2.5,
the area of (apparent) contact decreases continuously
when the magnification is increased.
An important result is that
even when the surface roughness is so high that no adhesion can be detected
in a pull-off experiment, the area of real contact (when adhesion is included)
may still be several times larger than when the adhesion is neglected.
Since it is the area of real contact which determines the sliding friction force,
{\em the adhesion interaction may strongly affect the friction force even when no 
adhesive force 
can be detected in a pull-off experiment}.

The influence of surface roughness on the
adhesion between rubber (or any other elastic solid)
and a hard substrates has been studied in a classic
paper by Fuller and Tabor \cite{[3]} (see also \cite{[Ken],[333],[8a],[8b],Zilber,Sam,Israel}).
They found that already a relative
small surface roughness can completely remove the adhesion.
In order to understand the experimental data they developed a very simple model
based on the assumption of surface roughness on a single length scale.
In this model the rough surface is modeled by asperities, all possessing
the same radius of curvature, and with heights following a Gaussian distribution.
The overall contact force was obtained by applying the contact theory of
Johnson, Kendall and Roberts \cite{KLJ} to each individual asperity.
The theory predicted that the pull-off force, expressed as a fraction of the
maximum
value, depends upon a single parameter, which may be regarded as
representing the statistically averaged competition between the
compressive forces exerted by the higher asperities trying to
prize the surfaces apart and the adhesive forces between the lower
asperities trying to hold the surfaces together.
This
picture of adhesion developed by Tabor and Fuller would be fine {\it if}
real surfaces had roughness on a single length scale as assumed
in their study. However, with roughness occurring on many different length scales,
a qualitatively new picture emerges \cite{Persson},
where, e.g., the adhesion force may even vanish
(or at least be strongly reduced), if
the rough surface can be described as a self affine fractal with
fractal dimension $D_{\rm f}
>2.5$.
In fact even for surfaces with roughness on a single length scale,
the formalism used by Fuller and Tabor is
only valid at ``high'' surface roughness, where the area of real
contact (and the adhesion force) is very small.
The theory that will be presented below is
particularly accurate for ``small'' surface roughness,
where the area of real contact
equals the nominal contact area.

\begin{figure}[htb]
\begin{center}
\includegraphics[width=0.6\textwidth]{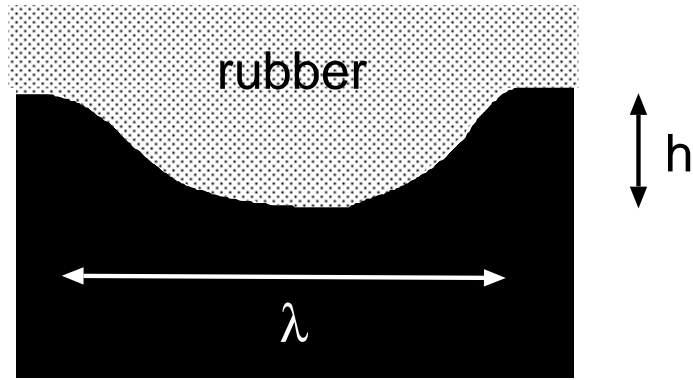} 
\end{center}
\caption{\label{InCavity}
A rubber surface is ``pulled'' into a cavity of the hard solid substrate
by the rubber-substrate adhesional interaction. The elastic energy
stored in the deformation field is of order $E\lambda h^2$.}
\end{figure}

\subsubsection{Qualitative discussion}

Let us estimate the energy necessary in order to deform a rubber block
so that the rubber fills out a substrate cavity of height $h$ and width $\lambda$.
The elastic energy stored in the deformation field in the rubber is given by
\[ U_{\rm el} \approx {1\over 2}\int \rmd^3x \ \sigma \epsilon \]
where the stress $\sigma \approx E \epsilon$, and $E$ is the elastic modulus. The
deformation field is mainly localized in a volume $\sim \lambda^3$
(see Fig.~\ref{InCavity}) where the strain $\epsilon \approx h/\lambda$. Thus we get
$U_{\rm el} \approx \lambda^3 E (h/\lambda)^2 = E \lambda h^2$.

Let us now consider the role of the rubber-substrate adhesion interaction.
If the elastic energy $U_{\rm el}\approx E \lambda h^2$ stored in the deformed
rubber is smaller than the gain in adhesion energy
$U_{\rm ad}\approx \Delta \gamma \lambda^2$,
where $\Delta \gamma =\gamma_1+\gamma_2-\gamma_{12}$
is the change of surface free energy (per unit area)
upon contact due to the rubber-substrate interaction (which usually
is mainly of the van der Waals type),
then (even in the absence of an external load $F_{\rm N}$) the rubber will
deform {\it spontaneously} to fill out the substrate cavities.
The condition $U_{\rm el} =U_{\rm ad}$ gives $h/\lambda \approx
(\Delta \gamma /E\lambda)^{1/2}$. For example, for very rough surfaces
with $h/\lambda \approx 1$, and with parameters typical for rubber
$E=1 \ {\rm MPa}$ and $\Delta \gamma = 3\,{\rm meV/\mbox{\AA}^2}$, the
adhesion interaction will be able to deform the rubber and completely fill
out the cavities if $\lambda < 0.1 \ {\rm \mu m}$. For very smooth surfaces
$h/\lambda \sim 0.01$ or smaller. In that case the rubber will be able
to follow the surface roughness profile up to the length scale
$\lambda \sim 1 \ {\rm mm}$ or longer.

The argument given above shows that for elastic solids with surface roughness
on a {\it single length scale} $\lambda$, the competition between adhesion and elastic
deformation is characterized by the parameter
$\theta =
Eh^2/\lambda \delta \approx U_{\rm el}/U_{\rm ad}$,
where $h$ is the amplitude of the
surface roughness and $\delta = 4(1-\nu^2)\Delta \gamma/E$ the
so called {\em adhesion length}, $\nu$ being the Poisson ratio
of the rubber.
The parameter $\theta$ is the ratio
between the elastic energy and the surface energy stored at the interface, assuming that
complete contact occurs. When $\theta \gg 1$ only partial contact occurs,
where the elastic solids make contact only close to the top of the highest asperities,
while complete contact occurs when $\theta \ll 1$.

\subsubsection{Pull-off force}

Consider a rubber ball (radius $R_0$) in adhesive contact with
a perfectly smooth and hard substrate. The elastic deformation of the rubber
can be determined by minimizing the total energy which
is the sum of the (positive) elastic energy stored in the deformation
field in the rubber ball, and the (negative)
binding energy between the ball and the substrate at
the contact interface.
The energy minimization
gives the pull-off force \cite{KLJ,[10]}
\begin{equation}
 F_{\rm c} = (3\pi/2) R_0 \Delta \gamma. \label{eq16}
\end{equation}

Consider now the same problems as above, but assume that
the substrate surface has a roughness described by the function
$z=h(\ve{x})$. Let us further assume a surface roughness power spectrum 
with a roll-off wavelength $\lambda_0 = 2\pi/q_0$
(see Fig.~\ref{Cq1}) smaller than the diameter of the nominal contact area
between the two solids.
In this case we can still use the result (\ref{eq16}),
but with $\Delta \gamma$ replaced by $\gamma_{\rm eff}$.
The effective interfacial energy $\gamma_{\rm eff}$ is the change in the interfacial
free energy when the elastic solid is brought in contact with the rough substrate.
$\gamma_{\rm eff}(\zeta )$ depends on the magnification
$\zeta$, and the interfacial energy which enter in the rubber ball pull-off experiment
is the macroscopic interfacial energy, i.e., $\gamma_{\rm eff}(\zeta)$ for $\zeta=1$.
If $A_0$ is the nominal contact area and $A_1$ the true atomic contact area, then
\begin{equation}
  A_0 \gamma_{\rm eff}(1) = A_1 \Delta \gamma -U_{\rm el}
  \label{eq17}
\end{equation}
where $U_{\rm el}$ is the elastic energy stored at the interface as a result of the
elastic deformations necessary in order to bring the solids
in atomic contact in the area $A_1$.

\subsubsection{Stress probability distribution}

The theory in Ref.~\cite{Persson} is based on the contact mechanics formalism described in Sec.~\ref{sec4.1}.
Thus, we focus on the stress probability distribution function $P(\sigma, \zeta)$
which satisfies Eq.~(\ref{eq5}):
\[ {\partial P \over \partial \zeta}  =
f(\zeta)
{\partial^2 P \over \partial \sigma^2} \]
We assume that detachment occurs when the local stress on the length scale $L/\zeta$ reaches
$-\sigma_{\rm a} (\zeta)$. Thus,
the following boundary condition applies to the present case
\[ P(-\sigma_{\rm a}(\zeta),\zeta)= 0 \]
This boundary condition replaces the condition $P(0,\zeta )=0$ valid in the absence of adhesion
(see Sec.~\ref{sec4.1}).

Let us consider the system on the characteristic length scale
$\lambda = L/\zeta$. The quantity $\sigma_{\rm a}(\zeta)$ is the stress necessary to
induce a detached area of width $\lambda$. This stress can be obtained from the theory of
cracks, where for
a penny-shaped crack of diameter $\lambda$
\begin{equation}
  \sigma_{\rm a} =
  \left[ {\pi \gamma_{\rm eff}(\zeta) E \over (1-\nu^2) \lambda} \right]^{1/2}
  = \left[ {\gamma_{\rm eff}(\zeta) E q \over 2 (1-\nu^2)} \right]^{1/2}
 \label{eq18}
\end{equation}
where $q=2\pi /\lambda = \zeta q_L$. In Ref.~\cite{Persson}
we derived two equations for $\gamma_{\rm eff}(\zeta)$
and $P(\zeta)$ which determine how these quantities depend on the magnification
$\zeta$; those equations are the basis for the numerical results
presented below.

\begin{figure}[htb]
\begin{center}
\includegraphics[width=0.80\textwidth]{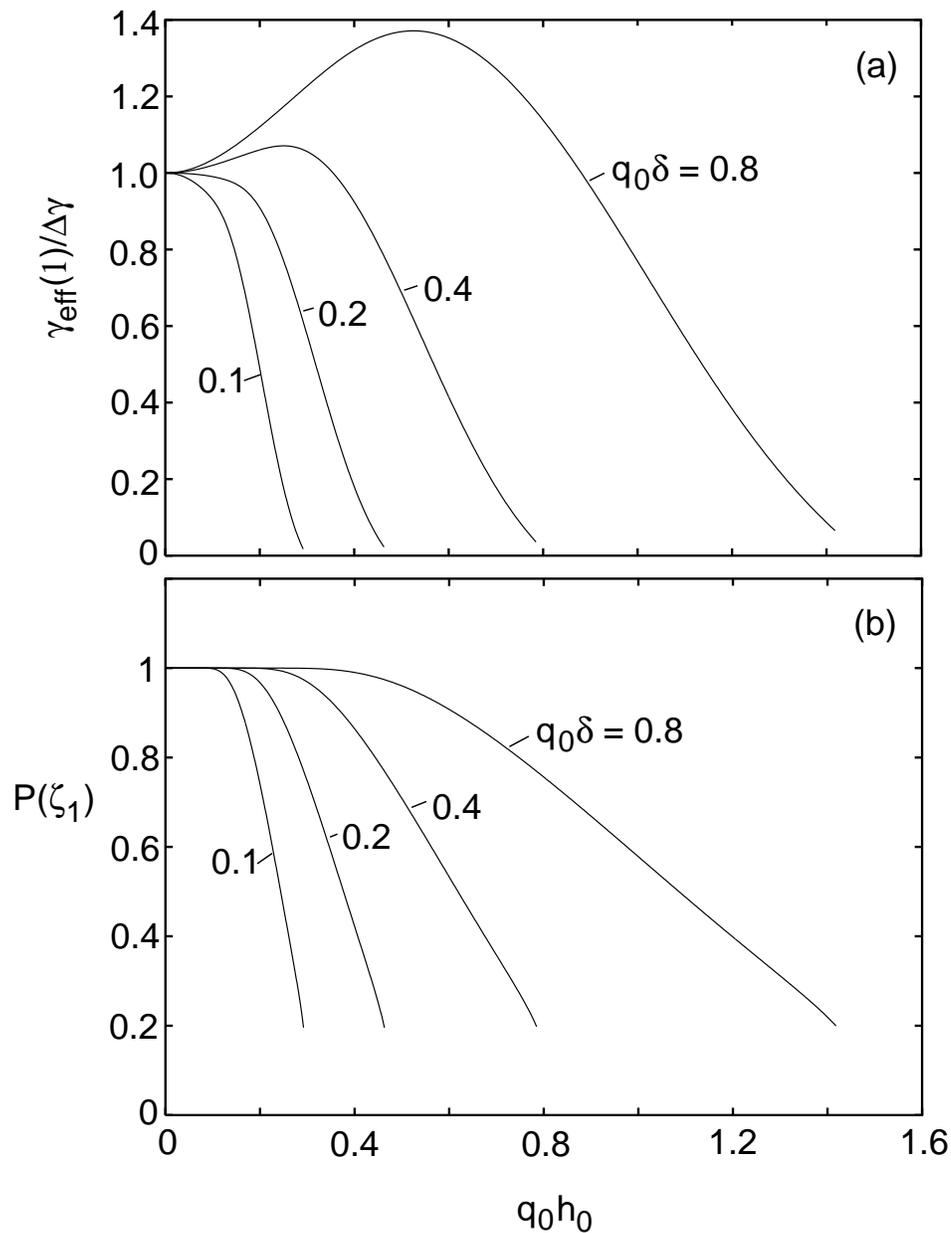} 
\end{center}
\caption{\label{Num1}
(a) Macroscopic interfacial energy for adhesion of rubber
to a fractal surface with $H=0.8$ and $q_1/q_0=\zeta_1=100$, as a function of $q_0h_0$. (b)
Normalized area of real contact, $P(\zeta_1)=A(\zeta_1)/A_0$,
as a function of $q_0h_0$. For $q_0 \delta = 0.1$, 0.2, 0.4
and 0.8 as indicated. }
\end{figure}

\subsubsection{Numerical results}

Fig.~\ref{Num1} shows
(a) the effective interfacial energy $\gamma_{\rm eff}(\zeta )$ ($\zeta=1$) and (b) the
normalized area of real contact, $P(\zeta_1)=A(\zeta_1)/A_0$,
as a function of $q_0h_0$. Results are shown for $q_0 \delta = 0.1$, 0.2, 0.4
and 0.8.
We will refer to $\gamma_{\rm eff}(1)$ at the magnification
$\zeta = 1$ as the {\it macroscopic} interfacial free energy
which can be deduced from, e.g., the pull off
force for a ball according to Eq.~(\ref{eq16}).
Note that for $q_0\delta =0.4$ and 0.8 the
macroscopic interfacial energy first increases with increasing amplitude $h_0$ of the surface
roughness, and then decreases. The increase in $\gamma_{\rm eff}$
arises from the increase in the surface area.
As shown in Fig.~\ref{Num1}(b), for small $h_0$ the two solids are
in complete contact, and, as expected, the complete contact remains to higher
$h_0$ as $\delta \sim \Delta \gamma/E$ increases.
Note also that the contact area is nonzero even when $\gamma_{\rm eff}(1)$ is virtually zero:
the fact that $\gamma_{\rm eff}(1)$ nearly vanishes does not imply that
the contact area vanishes (even in the absence of an external load), but rather that the
(positive) elastic energy stored at the interface exactly balances the (negative) adhesion
energy from the area of real contact.
{\em The stored elastic energy at the interface is returned back when removing the
block, and when $\gamma_{\rm eff} (1) \approx 0$ it
is just large enough to break the block-substrate bonding.}

\begin{figure}[htb]
\begin{center}
   \includegraphics[width=0.80\textwidth]{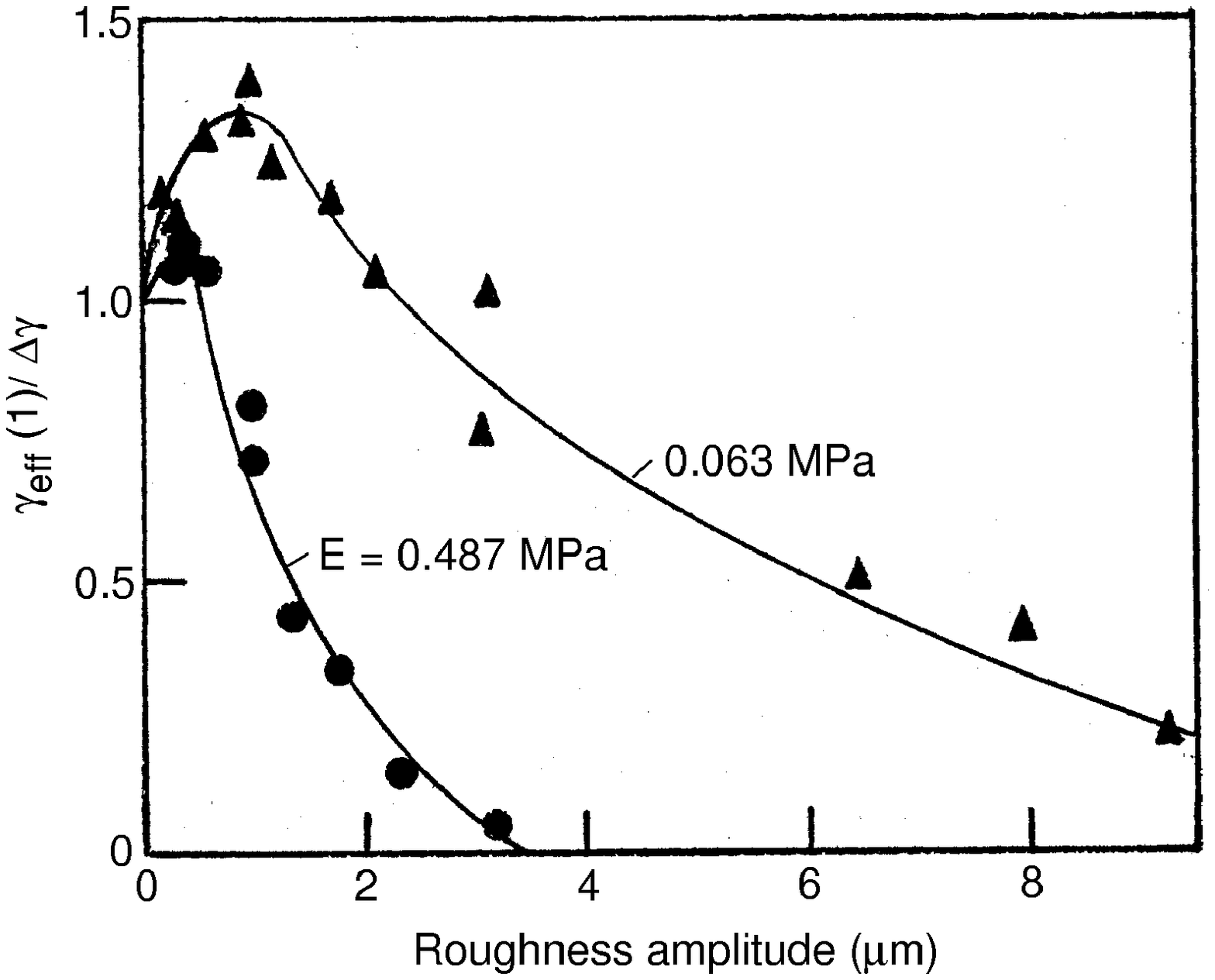} 
\end{center}
\caption{ \label{Briggs}
The macroscopic interfacial energy (obtained from the pull-off force) for a smooth rubber surface
(ball) in contact with a Perspex surface as a function of the roughness (center line average)
of the Perspex. Results are shown for a ``soft'' rubber ($E=0.063 \,{\rm MPa}$) and a ``hard''
rubber ($E=0.487 \,{\rm MPa}$). From \protect\cite{[8a]}.}
\end{figure}

\subsubsection{Experimental manifestations of adhesion}

Unfortunately, the surface
roughness power spectrum has not been measured for any of the real surfaces for which adhesion has
been studied in detail. Instead only the roughness amplitude (center line average)
and the radius of curvature of the largest surface asperities
was determined. Nevertheless, the experimental
data of Fuller, Tabor, Briggs, Briscoe and Roberts \cite{[3],[8a],[8b]}
are in good qualitative agreement
with our theoretical results. In Fig.~\ref{Briggs} we show the macroscopic interfacial energy
for ``hard'' and ``soft'' rubber in contact with Perspex, as a function the substrate
(Perspex) roughness amplitude as obtained by Briggs and Briscoe \cite{[8a]}.
It is not possible to compare
these results quantitatively with the theory developed above
since the power spectrum $C(q)$ was not measured for the Perspex substrate.
Even if the surfaces
would be self affine fractal as assumed above,
not only would the surface roughness amplitude change from one surface to another,
but so will the long distance cut off length $\lambda_0$ and hence also
the ratio $\zeta_1=q_1/q_0$.
In the experiments reported in Ref.~\cite{[8a]} the Perspex surfaces where
roughened by blasting with fine particles. The roughness could be varied
through the choice of the particles and the air pressure.

One practical problem in comparing the theory to 
experimental data is that most rubber materials have a wide distribution of relaxation times,
extending to extremely long times. This effect is well known in the context of
rubber friction (see Sec.~\ref{sec6.1}),
where measurements of the complex elastic modulus
show an extremely wide distribution of relaxation times,
resulting in large sliding friction even at very low sliding velocities,
$v <10^{-8}\,{\rm m/s}$.

\begin{figure}[htb]
\begin{center}
\includegraphics[width=0.90\textwidth]{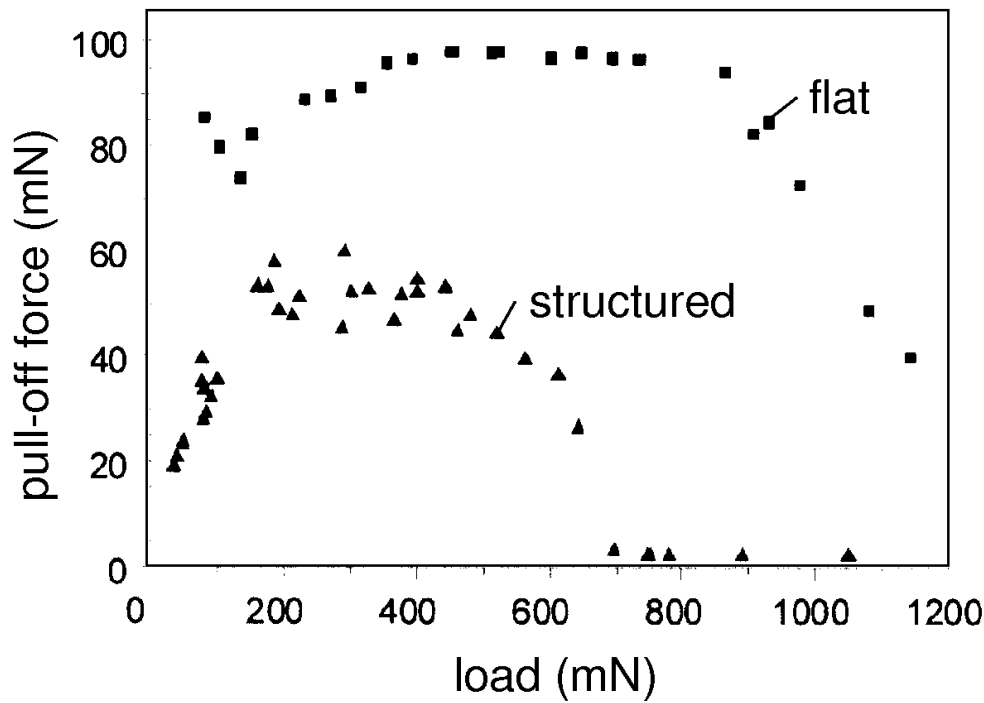} 
\end{center}
\caption{\label{Gorbis}
The pull-off force as a function of the squeeze force or load, for silicon
rubber in contact with a smooth glass surface. From
Ref.~\protect\cite{Gorbplus}.}
\end{figure}

\begin{figure}[htb]
\begin{center}
\includegraphics[width=0.70\textwidth]{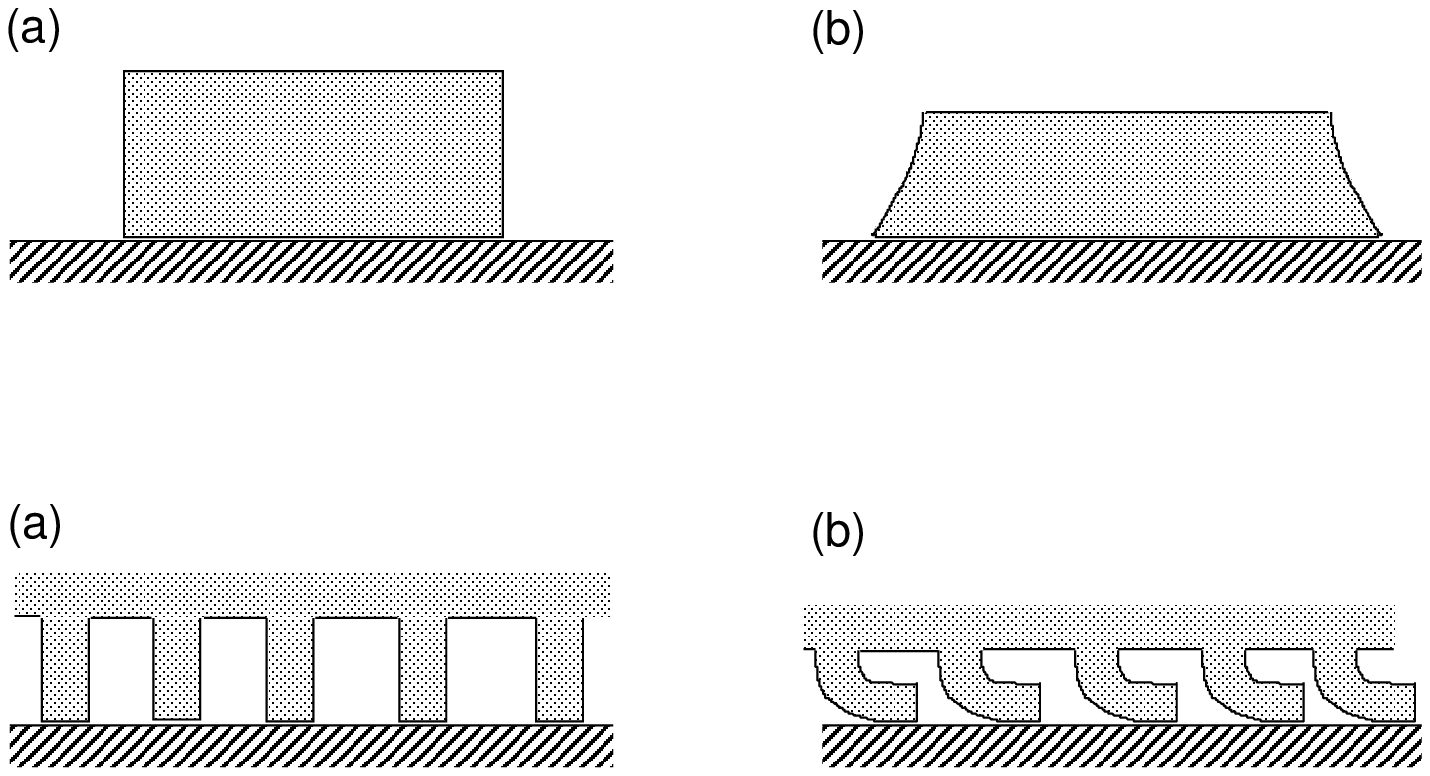} 
\end{center}
\caption{\label{Schem}
Elastic deformation of a rubber block with a smooth surface (top) and
a structured surface (bottom). (a) shows the initial state before applying a squeezing force,
and (b) the new state (without load)
after applying (and then removing)
a very large squeezing force. In state (b) a large elastic
energy is stored in the rubber which is ``given back'' during pull-off resulting in a nearly
vanishing pull-off force.}
\end{figure}

The effect of the stored elastic energy on adhesion has recently been studied
using a polyvinylsiloxane rubber block squeezed against a smooth glass surface
for a fixed time period before measuring the pull-off force \cite{Gorbplus}.
The square-symbols in Fig.~\ref{Gorbis}
show the pull-off force as a function of the squeezing force.
For squeezing forces $F_{\rm N}>850 \,{\rm mN}$ the pull off force decreases. This may be explained by
a drastic increase of the elastic energy stored in the rubber because of the
strong deformation of the rubber, see Fig.~\ref{Schem}(top), some of which remains even when the load is removed
as a result of the rubber-glass friction at the interface.
This energy, freed during the process of unloading, will help to break the adhesive bonds at
the interface. This effect is even stronger when the surface is structured. Thus, the
triangles in the figure shows the pull-off force when the rubber surface is covered by a
regular array of rubber cylindrical asperities. In this case the pull-off force drops to nearly
zero for $F_{\rm N} >700 \,{\rm mN}$. Visual inspection shows that in this 
case the cylindrical asperities at high
load bend and make contact with the glass on one side of the cylinder surface, see Fig.~\ref{Schem}(bottom).
This again stores a large amount of elastic
energy at the interface which is given back during pull-off, reducing
the pull-off force to nearly zero.

\subsubsection{The role of plastic yield on adhesion}

When the local stress in the asperity contact regions between
two solids becomes high enough, at least one of the
solids yields plastically. This will tend to increase the effective adhesion
(or pull-off force) for the following three reasons. First, the area of real
contact between the solids will increase as compared to the case
where the deformations are purely
elastic. Secondly, the amount of stored elastic
energy in the contact regions (to be given back at pull-off)
will be reduced because of the lowered elastic deformations.
Finally, for many materials plastic yield will strengthen
the junctions \cite{both2}. For example, most metals are protected by
thin oxide layers, and as long as these are intact the main interaction between the
surfaces in the contact areas may be of the van der Waals and electrostatic origin.
However, when plastic yield occurs it may break up the oxide films
resulting in direct metal-metal
contact and the formation of ``cold-welded'' junctions. When this occur, because of the high ductility
of many metals, during pull-off ``long'' metallic bridges may be formed between the solids
so that instead of having junctions popping one after another during pull-off, a large number of
adhesive junctions may simultaneously impede the surface separation
during pull-off, leading to a large pull-off force. However, experiment have shown \cite{Gellman} that
just squeezing before pull-off
will in general only result in very few cold welded junctions, while squeezing {\it and}
sliding will break up the oxide film,
resulting in the formation of many more cold welded contact regions, and will hence result
in a much larger pull-off force.

\subsection{The adhesion paradox}
\label{sec7.2}

\begin{figure}[htb]
\begin{center}
\includegraphics[width=0.80\textwidth]{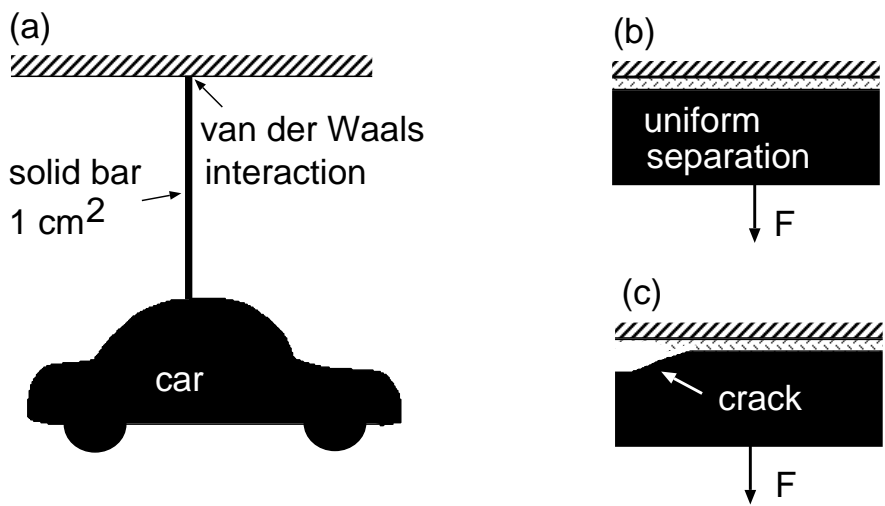} 
\end{center}
\caption{\label{car}
Even the weakest force in Nature which is of relevance in condensed matters physics,
namely the van der Waals force, is relative strong on a macroscopic scale. Thus,
for example, if the bond breaking were to occur uniformly over the contact area as in (b),
already a contact area of order $1 \,{\rm cm}^2$ could sustain the weight of
a car (i.e., a force of order
$10^4 \,{\rm N}$) [see (a)]. However, on a macroscopic scale the bond-breaking does not usually
occur uniformly over the contact area,
but by crack propagation, see (c), which drastically reduce the pull-off force.
Secondly, interfacial surface roughness drastically reduces the pull-off force.
Finally, the stored elastic energy will largely balance the adhesion energy (see text)}
\end{figure}

The biggest ``mystery'' related to adhesion is not why it is sometimes observed
but rather why it is usually not observed.
Even the weakest force in Nature of relevance in condensed matters physics,
namely the van der Waals force, is relatively strong on a macroscopic scale.
For example,
even a contact area of order $1 \,{\rm cm}^2$
could sustain the weight of a car (i.e., a force of order
$10^4 \,{\rm N}$) [see Fig.~\ref{car}(a)] even if only the van der Waals
interaction operated at the interface.
[Here we assumed that the bond breaking occur
uniformly over the contact area as illustrated in Fig.~\ref{car}(b).]
However, this is never observed in practice and this fact is referred to as the
{\it adhesion paradox}.

There are several reasons why adhesion is usually not observed between macroscopic bodies.
For example, on a macroscopic scale the bond-breaking usually does not occur
uniformly as in Fig.~\ref{car}(b),
but occurs by crack propagation, see Fig.~\ref{car}(c). The local stress at the crack tip is much
higher than the average stress acting in the contact area, and this
drastically reduces the pull-off force. Another reason, already addressed in Sec.~\ref{sec7.1}, is the influence
of surface roughness. Thus, for elastically hard surfaces the true atomic contact between the solids
at the interface is usually much smaller than the nominal contact area. In addition, the elastic energy
stored in the solids in the vicinity of the contact regions is given back during pull-off
and helps to break the interfacial bonds between the solids (see Sec.~\ref{sec7.1}).

It is interesting to note that for very small solid objects, typically of
order $100 \,{\rm \mu m}$ or smaller,
the bond breaking may occur uniformly over the contact area (no crack propagation)
so that adhesion between smooth surfaces
of small objects, e.g., in micromechanical applications (MEMS), may be much stronger
than for macroscopic bodies, and this fact
must be taken into account when designing MEMS \cite{WEAR2004,Gao}.

\subsection{Adhesion in rubber technology}
\label{sec7.3}

Most rubber compounds
of commercial use contains a large fraction (around $30 \%$) of filler
particles which usually are mixtures of carbon and silica particles. The
silica particles are very small (below micrometer-size) and very hard.
As a result, when rubber is sliding on a substrate, even if the latter is very
hard, e.g., stone or steel, the substrate will get polished by the rubber.
We have discussed this effect above for road surfaces (see Sec.~\ref{sec6.2}). Here we describe another
recent observation of the same effect with important practical implications.

One stage in the production of rubber for technological applications
involves the mixing in open mills, where unvulcanized
rubber (with filler particles) is fed between rotating steel cylinders.
The slip of the (silica-particle containing)
rubber relative to the steel walls during
the mixing process
result in a continuous ``polishing'' (or wear) of the steel surfaces.
This will slowly
increase the rubber-steel contact area, and finally the rubber may
adhere to
the steel surfaces, which is of course unwanted.

Using the
Atomic Force Microscopy (AFM) we measured the surface topography of the steel
cylinders, and found that during production the steel surfaces
are continuously being polished by the rubber containing silica particles.
This leads to a continuously increased contact
area between the steel and the rubber sheet, and to an
increased adhesive steel-rubber interaction. Using
a recently developed viscoelastic
contact mechanics theory (see Sec.~\ref{sec4.3})
we further found that the increase in the contact area can be very
large.

\begin{figure}[htb]
\begin{center}
\includegraphics[width=0.80\textwidth]{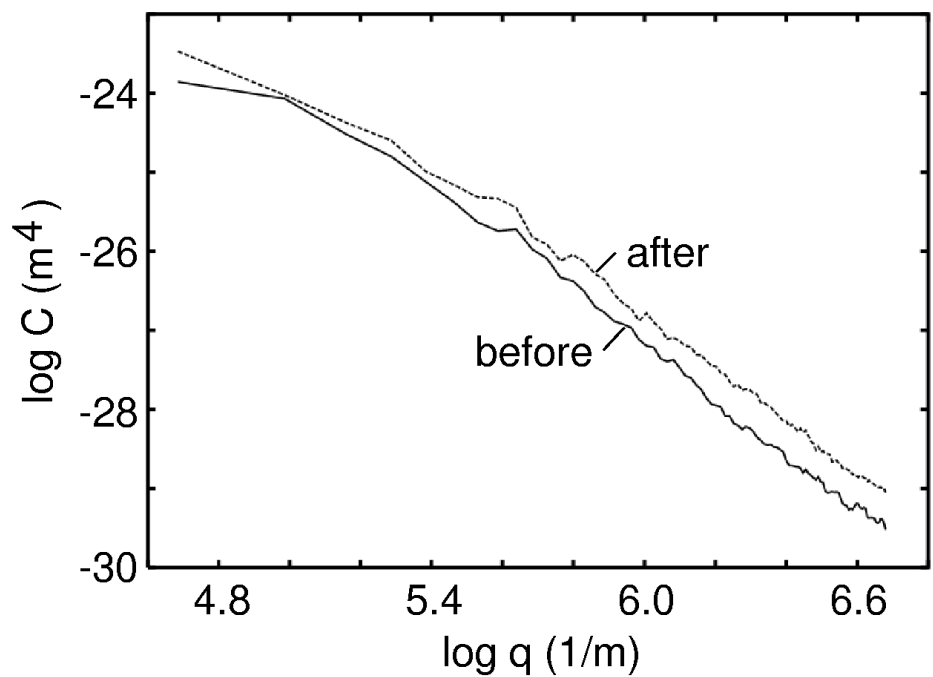} 
\end{center}
\caption{\label{mill}
The surface roughness power spectra of a
steel cylinder before and after lapping with
corundum paper. The lapping increases the surface roughness.}
\end{figure}

Also shown by AFM measurements was that
lapping of the steel surface by corundum paper {\em increases}
the steel surface roughness (see Fig.~\ref{mill}),
which is easy to understand when one recognizes that the corundum particles have on the average 
larger diameter than the silica particles in the rubber. Thus we conclude that lapping or ``polishing''
may increase the roughness of the steel cylinders, and hence reduce the adhesive
interaction to such a level that it becomes unimportant.

\begin{figure}[htb]
\begin{center}
  \includegraphics[width=0.40\textwidth]{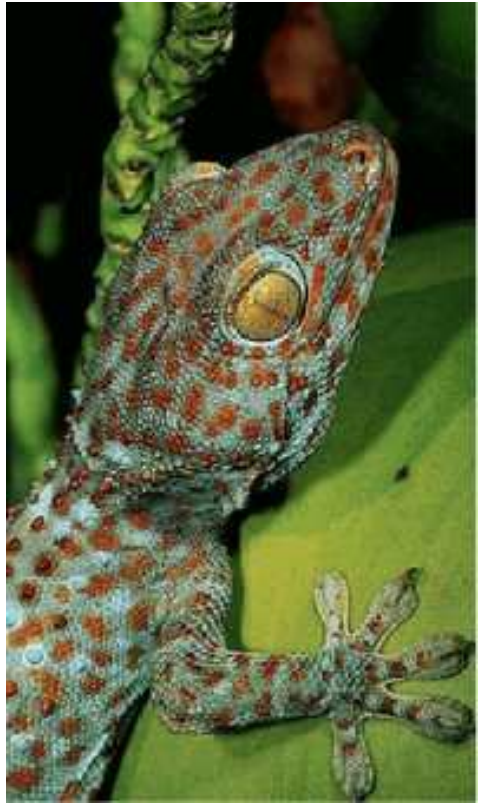} 
\end{center}
\caption{ \label{lizard}
A lizard adhering to a nearly vertical surface.}
\end{figure}

\subsection{Adhesion in biology}
\label{sec7.4}

How can a fly or a cricket walk on a glass window, or a
lizard (see Fig.~\ref{lizard}) move on a stone or concrete wall? These fundamental questions have
interested children and scientists for many years,
and recently very important experimental work has
been performed that gave a deeper
insight into these questions \cite{Scherge,Autumn}.
Here we focus mainly on dry adhesion, which seems to be relevant for
lizards, and we discuss the influence of surface roughness
on the adhesion between a lizard or a gecko toe and a rough hard substrate.

It has been demonstrated that a foot of a gecko can adhere to a
substrate with a force
$\sim 10 \,{\rm N}$ (corresponding to the weight of 1 kg!). The typical weight of a 
Tokay gecko lizard (Gecko gecko) is
approximately 40\,g meaning that only 1\% of the maximum adhering force
of its feet is required to
support the whole weight of the gecko. This raises the question of why geckos are apparently
so over-built. However, a gecko must be able to adhere to very rough
surfaces, and we will show below 
that the adhesion to rough surfaces can be
reduced significantly \cite{Per1,Per2}. 

The adhesion between two bodies in contact results almost entirely
from the regions where the surfaces are separated
by one nanometer or less. For hard solids this area of real contact is extremely small.
Furthermore, for elastically hard solids a large elastic energy is stored in the solids in the
vicinity of the contact regions and during separation of the solids this energy is released and
will hence reduce the separation force as explained earlier. Thus, strong adhesion is only possible if
at least one of the solids is elastically very soft, or if there is an elastically
very soft layer at the interface between the solids. Since the lizard skin comprises
a relative stiff material (keratin), with an elastic modulus of the order $1 \ {\rm GPa}$, i.e.,
about 1000 times higher than tire rubber, it is not immediately obvious why the lizard
can adhere to very rough stone walls.

\begin{figure}[htb]
\begin{center}
  \includegraphics[width=0.70\textwidth]{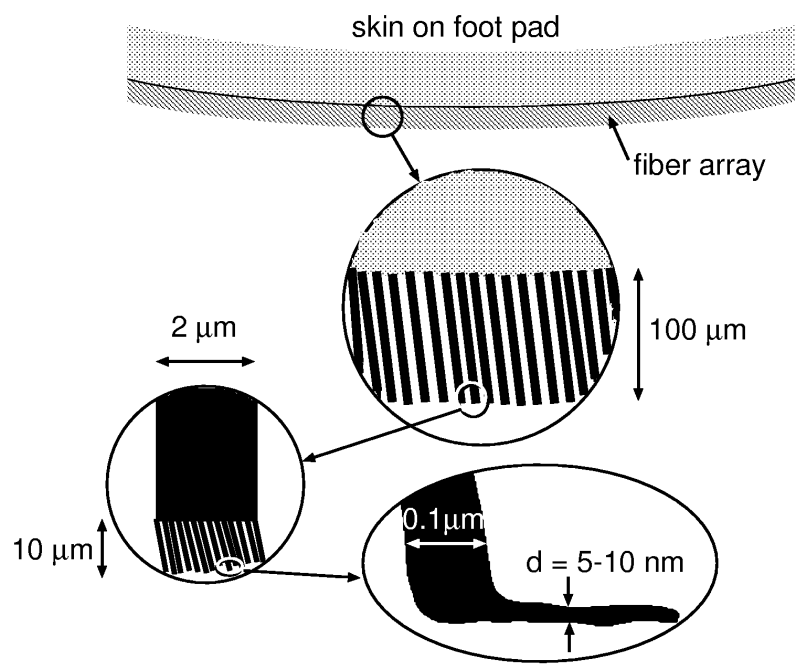} 
\end{center}
\caption{ \label{leafplate}
Schematic picture of the lizard adhesive system. The skin
of the lizard is covered by a dense layer of thin fibres or hair (setae)
(length $\approx 100 \,{\rm \mu m}$ and thickness of fibre of order
$\sim 4 \,{\rm \mu m}$). Each of these fibres branches out into about
1000 thinner fibres (length $\sim 10 \,{\rm \mu m}$ and width
of order $\sim 0.1 \,{\rm \mu m}$). Each of the thin fibres ends with
a thin ($5-10 \,{\rm nm}$) leaf-like plate (spatula).}
\end{figure}

\begin{figure}[htb]
\begin{center}
  \includegraphics[width=0.40\textwidth]{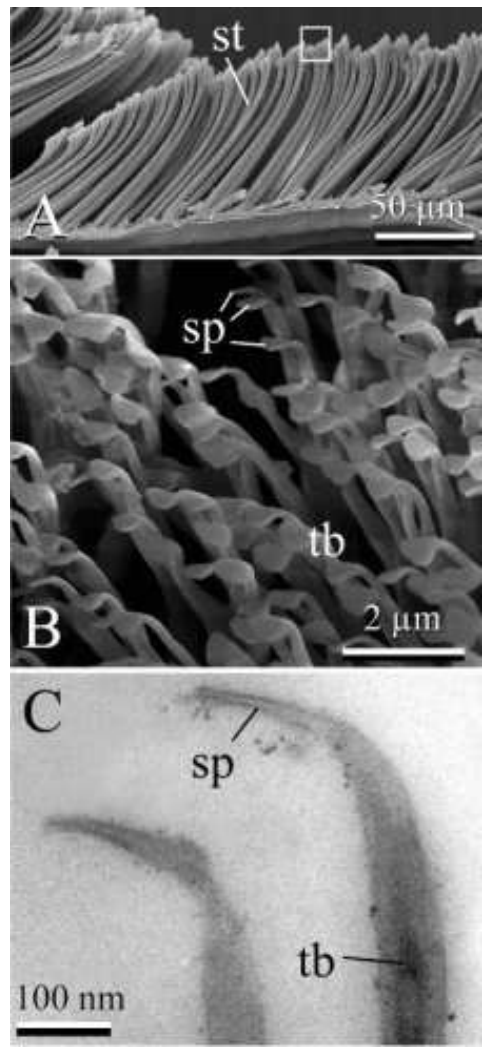} 
\end{center}
\caption{ \label{geckoplate}
Details of attachment system of the Tokay gecko (Gecko gecko).
{\bf A}. Scanning electron microscopy (SEM) micrograph of setae (st) located on thin keratin film.
{\bf B}.
Magnification (SEM micrograph) of the area surrounded by the
white rectangle in {\bf A},
showing terminal branches (tb) of setae with the
spatula (sp). {\bf C}. Transmission electron microscopy micrograph
of ultrathin section of two terminal branches (tb) with spatulae (sp).
 From \protect\cite{Per2}.}
\end{figure}

As it turns out, during millions of years of evolution and natural selection,
an extremely soft elastic layer has appeared on the lizard pad surface.
This layer is built in a hierarchical manner from fibres and plates
(Figs.~\ref{leafplate} and \ref{geckoplate}),
that reflect the
hierarchical nature
of most
natural surfaces
(to which the lizard must be able to adhere),
which have roughness on all length scales, from
the macroscopic scale (e.g., the size of the lizard toe pad)
down to the atomic scale.
The skin
of the lizard pad, which consist of a $\sim 100 \,{\rm \mu m}$ thick keratin layer,
is covered by a dense layer of fibres or hair (setae)
(length $\approx 100 \,{\rm \mu m}$ and thickness
$\sim 4 \,{\rm \mu m}$). Each of these fibres branches out into about
1000 thinner fibres (length $\sim 10 \,{\rm \mu m}$ and width
$\sim 0.1 \,{\rm \mu m}$), and each terminal fibre ends with
a thin ($5-10 \,{\rm nm}$) leaf-like plate (spatula). This
construction makes the lizard adhesive system
elastically very soft on all
relevant length scales (from mm to nm).

\begin{figure}[htb]
\begin{center}
  \includegraphics[width=0.40\textwidth]{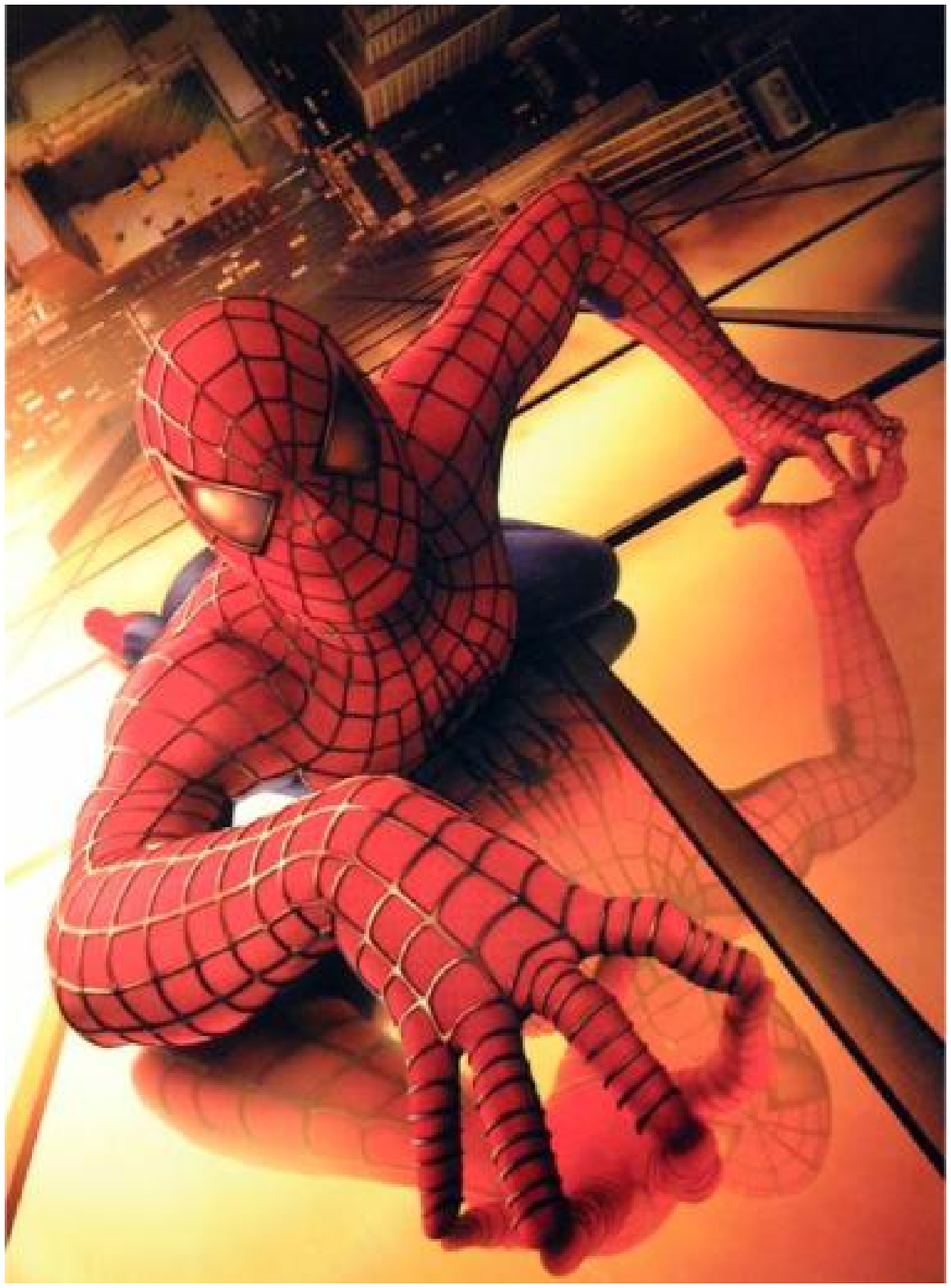} 
\end{center}
\caption{ \label{spider}
Spider man. Somewhere in our Universe there may be a planet
with `animals' as large as humans moving on vertical walls\ldots\
`SPIDER-MAN'. Motion Picture \copyright\ 2002 Columbia Pictures Industries,
Inc. Spider-Man Character \textregistered\ \&\ \copyright\ 2002 Marvel
Characters, Inc. All Rights Reserved. Courtesy of Columbia Pictures.
}
\end{figure}

\begin{figure}[htb]
\begin{center}
  \includegraphics[width=0.40\textwidth]{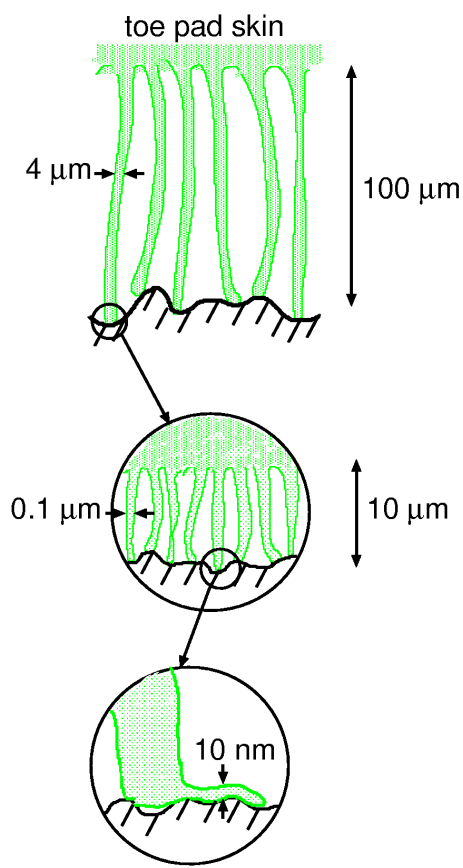} 
\end{center}
\caption{ \label{Hir}
The Lizard fibre-and-plate adhesive system can be easily elastically deformed to
bind to a rough substrate, even when the substrate has roughness on all length scales from,
say, $\sim 1 \,{\rm mm}$ to $\sim 1 \,{\rm nm}$.}
\end{figure}

The skin of the gecko toe-pad is able to deform and follow
the substrate roughness profile on length scales much longer
than the thickness $d \approx
100 \,{\rm \mu m}$ of the elastic keratin film, say
beyond $\sim 1000 \,{\rm \mu m}$.
At shorter length scales the keratin film,
because of its high elastic modulus (of order $1\,{\rm GPa}$),
can be considered rigid and flat.
Elastic deformation of the pad surface
on length scales shorter than $\sim 1000 \,{\rm \mu m}$,
involves the compliant setae fibre array system, with fibres of thickness
$\sim 4 \,{\rm \mu m}$.
In Ref.~\cite{Per1} one of us has shown that if
the surface roughness root-mean-square amplitude, measured over a patch $D\times D$ with
$D\approx 1000 \,{\rm \mu m}$,
is smaller than a characteristic length (the adhesion length)
(see Ref.~\cite{Per1}),
then the fibre array system is able to deform (without storing much elastic
energy) to follow the surface roughness in the wavelength range
$10 < \lambda < 1000 \,{\rm \mu m}$.
However,
if the setae fibre tips were instead blunt and compact,
they would not be able to
penetrate into the surface ``cavities'' with diameters less than a few
${\rm \mu m}$. Thus, negligible atomic contact would occur between the surfaces,
and adhesion would be negligible.
For this reason,
there is an array of $\sim 1000$
thinner fibres (diameter of order $\sim 0.1 \,{\rm \mu m}$)
at the tip of each long (thick) fibre.
These fibres are able to penetrate
into the surface roughness cavities down to length
scales of a few tenths of a micrometer, see Fig.~\ref{Hir}.
Again if the thin fibres had blunt and compact
tips made of the same ``hard'' keratin as the rest
of the fibre, one would still obtain very small adhesion, since much elastic
energy would be needed to deform the surfaces of the thin fibres to make
atomic contact with the substrate.
But in fact the top of the thin fibres end with thin leaf-like plates, which
can easily bend (without storing much elastic energy)
to follow the surface roughness profile \cite{Per2,Carbone}.
The calculations presented in Ref.~\cite{Per2}
show that, for rough surfaces with the
fractal dimension $D_{\rm f} > 2.3$, very small
spatula-substrate adhesion may occur in most cases. However, natural surfaces 
tend to have a fractal
dimension of order 2.2, and
adhesion may be appreciable even for very
rough surfaces
in these cases.
Experiments to test the theoretical results are in progress.

There is an interesting aspect of fibre adhesion which differs from that
involving compact solids.  Fibre adhesion depends only on the height
probability distribution $P_h$ and not on the power spectra $C(q)$.  The reason
is that the elastic energy stored in a bent fibre depends only on the
separation between the surfaces at the interface which is determined by the
surface roughness height distribution $P_h$. Thus, the elastic energy stored in
a bent fibre does not depend directly on $C(q)$.  On the other hand for compact
solids the elastic energy is determined only by $C(q)$.

Lizards are the heaviest living objects on this
planet that are able to adhere to, for example, a rough vertical stone wall. Since
the surface area of a body increases less than the volume, or mass
as the size of the body increases, the adhesive system
in large living bodies, such as lizards, must be much more effective (per unit
attachment area) than in smaller living objects such as flies or beetles.
This implies that lizards have the most effective adhesive systems
found in the biological evolution for the purpose of locomotion. This is
confirmed by electron microscopy studies. Thus,
the spatula are thinner in lizards than in beetles. Also the diameter of the
terminal branches is
smaller. This implies that less elastic energy per unit surface area will be
stored in the lizard adhesive system,  resulting in a stronger adhesion
for lizards than for beetles.

\begin{figure}[htb]
\begin{center}
  \includegraphics[width=0.40\textwidth]{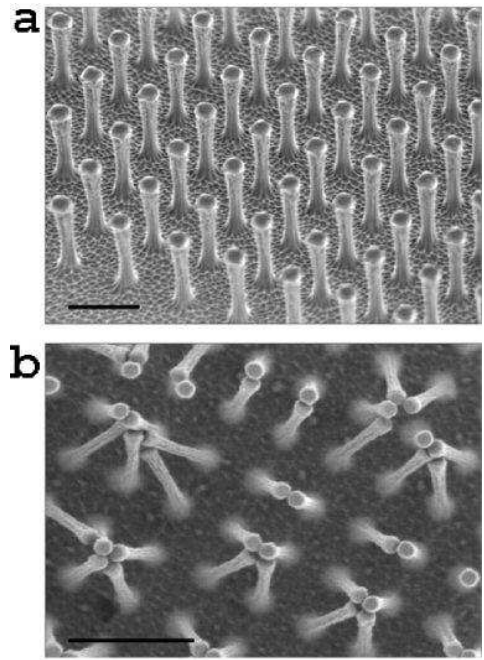} 
\end{center}
\caption{ \label{boundle}
When the fibres are too thin and long, or made of materials elastically 
too soft (as in the present case of rubber shown here) the standing-up
fibre array system in (a) is unstable, and after contact with a substrate
the fibres end up in the bundled-up state (b).  Adapted from
\protect\cite{Geim}. Reproduced with permission.}
\end{figure}

The construction of man-made adhesives based
on fibre and plate arrays might constitute
an attractive alternative to the usual
adhesives based on thin polymer films (see Fig.~\ref{spider} for one ``application'').
Some pioneering experiments have indeed shown enhanced adhesion for fibre
array systems, but no man-made systems with
the hierarchic nature found in biological systems
have so far been produced.
In addition, if the fibres are too long and thin, or the fibre material is (elastically)
too soft, the attractive fibre-fibre interaction will result in
elastic instabilities, leading to fibre bundles
or fibre ``condensation''
into compact layers. This effect was in fact
observed in the first man-made fibre adhesion system (which used fibres made from
rubber which is 1000 elastically softer than keratin used
in most biological applications), see Fig.~\ref{boundle},
and was at the same time predicted theoretically \cite{Per1}.
It turns out
that the fibre arrays in Lizards are close to this instability,
but, unsurprisingly, on the correct side of it,
and no fibre bundling or condensation occur in these biological systems.

\subsection{The role of liquids in adhesion between rough solid surfaces}
\label{sec7.5}

As explained in Sec.~\ref{sec7.1}, surface roughness reduces
the adhesion between clean surfaces.
First, it lowers the area of real contact.
Since the adhesion interaction comes almost entirely from
the area where the solids make atomic contact, it is clear that
the surface roughness can drastically reduce the adhesion.
Secondly, elastic deformation energy is stored
in the vicinity of the asperity contact regions. During pull-off
the elastic energy
is ``returned back'' to the
system, usually resulting in a drastic reduction in the effective adhesion and the
pull-off force.

Most surfaces have at least nano-scale roughness, and hard solids in the normal
atmosphere have
at least a monolayer of liquid-like ``contamination'' molecules, e.g., water and
hydrocarbons.
Small amounts of a wetting
lubricant or contamination liquids between rough solid walls may
drastically
enhance the adhesion. Thus, for surfaces with nanoscale roughness, a monolayer
of wetting
liquid may result in the formation of a large number of
nano-bridges between the solids, which increases the pull-off force. This effect is
well known
experimentally. For example, the adhesion force which can be detected between gauge
blocks
(steel blocks with very smooth surfaces) is due to the formation of many very small
capillary bridges
made of water or organic contamination. For thicker lubrication or contamination
films the
effective adhesion will be more long-ranged but the pull-off force may be smaller.
The thickness of the lubricant or contamination layer for which the pull-off force is
maximal will in general depend on the nature of the surface roughness, but is likely to be of
order the root-mean-square roughness amplitude. In fact, it is an interesting and important
problem to find out at exactly what
liquid thickness the pull-off force is maximal.

Some insects such as
flies or crickets inject a thin layer of a wetting liquid in the contact region between the
insect attachment surfaces and the (rough) substrate. The optimum amount of injected
liquid will depend on the nature of the substrate roughness, and it is likely that the
insect can regulate the amount of injected liquid by a feedback system involving the
insect nerve system.

Here we consider the adhesion between two solid elastic walls with nanoscale roughness,
lubricated by octane \cite{Israel,PM2004,Sam}.
We consider
two types of substrates (bottom surface) -- flat and nano-corrugated
(corrugation amplitude $1 \,{\rm nm}$ and wavelength of the corrugation in $x$ and $y$ direction,
$4 \,{\rm nm}$) -- and varied the
lubricant coverage from $\sim 1/8$ to
$\sim 4$ monolayers of octane. The upper surface (the block) is assumed to be
atomically smooth but
with a uniform cylindrical curvature with a
radius of curvature $R \approx 100 \,{\rm nm}$ (see Fig.~\ref{Bos3} below).
The simulation results presented here were obtained using standard molecular
dynamics calculations \cite{Sam}.

\begin{figure}
\begin{center}
  \includegraphics[width=0.80\textwidth]{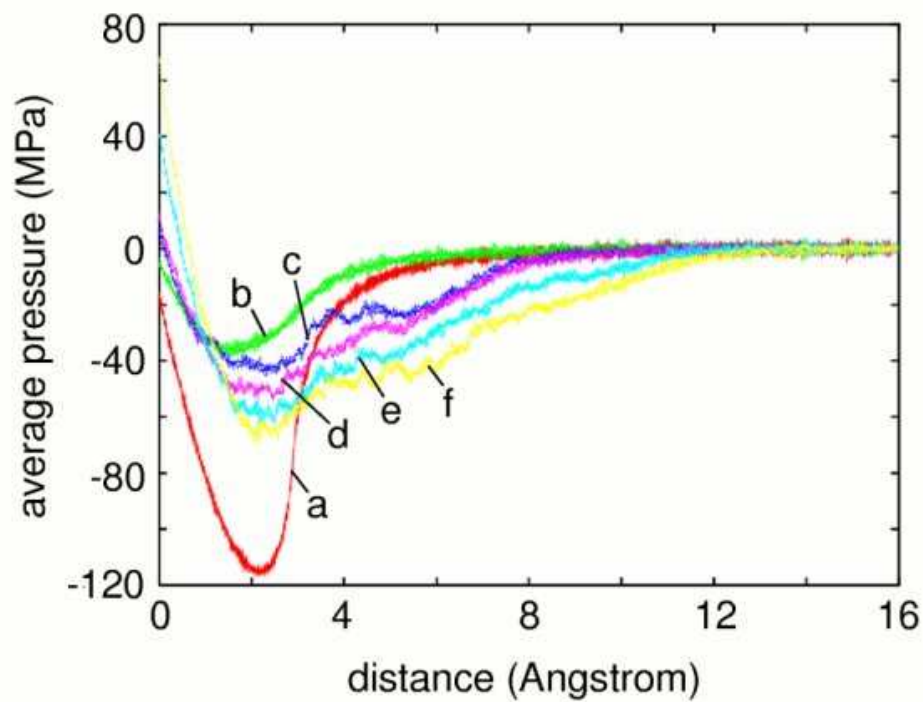} 
\end{center}
\caption{\label{fig:Fig_1}
Simulated adhesion between two solid elastic walls, 
one of which with nanoscale roughness, and with a variable amount of wetting lubricant in between. 
The variation of the average pressure
during retraction develops as a block moves a distance of 16\,\AA\
away from the
substrate. Octane C$_8$H$_{18}$ was used as lubricant. Pull-off (retraction)
velocity was $v_{\rm z} = 1 \,{\rm m/s}$.
(a) For the flat substrate without lubricant.
(b) For the corrugated substrate without lubricant.
Curves (c)--(f) show results for the corrugated substrate
with about 1/8, 1/4, 1/2 and 1 monolayer of octane in the
contact region, respectively.
For clarity, the curve for the flat substrate (a) is displaced
to the right, by 2\,\AA.}
\end{figure}

Fig.~\ref{fig:Fig_1} shows
the variation of the average pressure during retraction as the block moves
a distance of 16\,\AA\
away from the substrate. The pull-off
(retraction) velocity was $v_z = 1 \,{\rm m/s}$. We varied the lubricant
coverage from 0 to 1 monolayer in the contact region.
The pull-off force is maximal when the adsorbate coverage is of the order of one
monolayer [curve (f)]. However, the pull-off force is still smaller than
for a {\em flat} substrate without lubricant [curve (a)].
As a function of the
octane lubricant coverage and for the corrugated substrate,
the pull-off force first increases as the coverage increases from zero to
$\sim 1$ monolayer, and then decreases as the coverage is increased beyond
monolayer coverage (not shown).

\begin{figure}
\begin{center}
\includegraphics[width=0.450\textwidth]{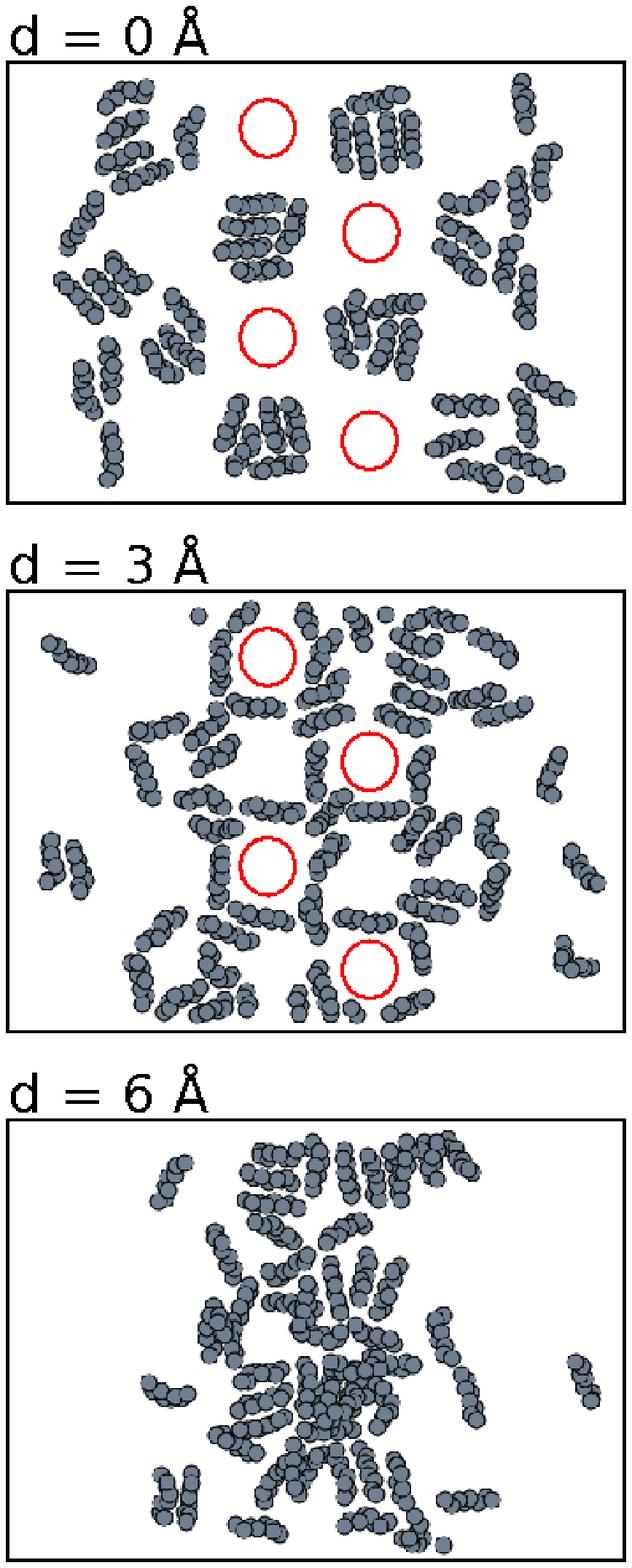} 
\end{center}
\caption{\label{Bos2}
Snapshot pictures (for three different
block positions $d = 0$, $3$ and $6$\,\AA) of the
lubricant layer during retraction.
We only show the lubricant molecules in the central part of the contact area
between the block and the substrate surfaces (top view, surfaces parallel to
the plane of the image). Corrugated substrate with about 1/4 monolayer
of octane in the contact region. The circles indicate
the position of several asperity tops of the corrugated
substrate surface.}
\end{figure}

At low octane coverage, the octane molecules
located in the substrate corrugation wells during squeezing are pulled out of
the wells during pull-off, forming a network of nano capillary bridges around
the substrate nanoasperities, thus
increasing adhesion between two surfaces, see Figs.~\ref{Bos2} and \ref{Bos3}.
For greater lubricant coverages
a single capillary bridge is formed.

Let us discuss the nature of the adhesion
for the corrugated substrate, with
about $1/4$ monolayer
of octane in the contact region.
Fig.~\ref{Bos2} shows snapshot pictures of the
lubricant layer during retraction, as the block moves
away from the
substrate for three different block positions $d = 0$, $3$ and
$6$\,\AA.
Only the central part of the contact between the block and
the substrate is shown, top view, after removing
the block and substrate atoms. In the beginning ($d = 0$\,\AA) octane molecules
are located in the substrate corrugation wells,
or cavities with direct metal--metal contact between the
block and the top of the substrate nano asperities (see Fig.~\ref{Bos3}).
During retraction ($d = 3$\,\AA) the
octane molecules are pulled out of the wells forming an almost symmetric network
of nano-bridges around the asperity tops,
increasing the adhesion between the two surfaces.
This
configuration corresponds to the maximal adhesion force, see curve (d)
in Fig.~\ref{fig:Fig_1}. Thus maximal adhesion is achieved via the
formation of many small capillary nano-bridges,
involving just a few molecules for each bridge (see Fig.~\ref{Bos3}).
Further retraction ($d = 6$\,\AA) results
in the collapse of the nano-bridges and the
formation of a single ``large''
capillary bridge in the centre
of the contact region.

\begin{figure}
\begin{center}
\includegraphics[width=0.450\textwidth]{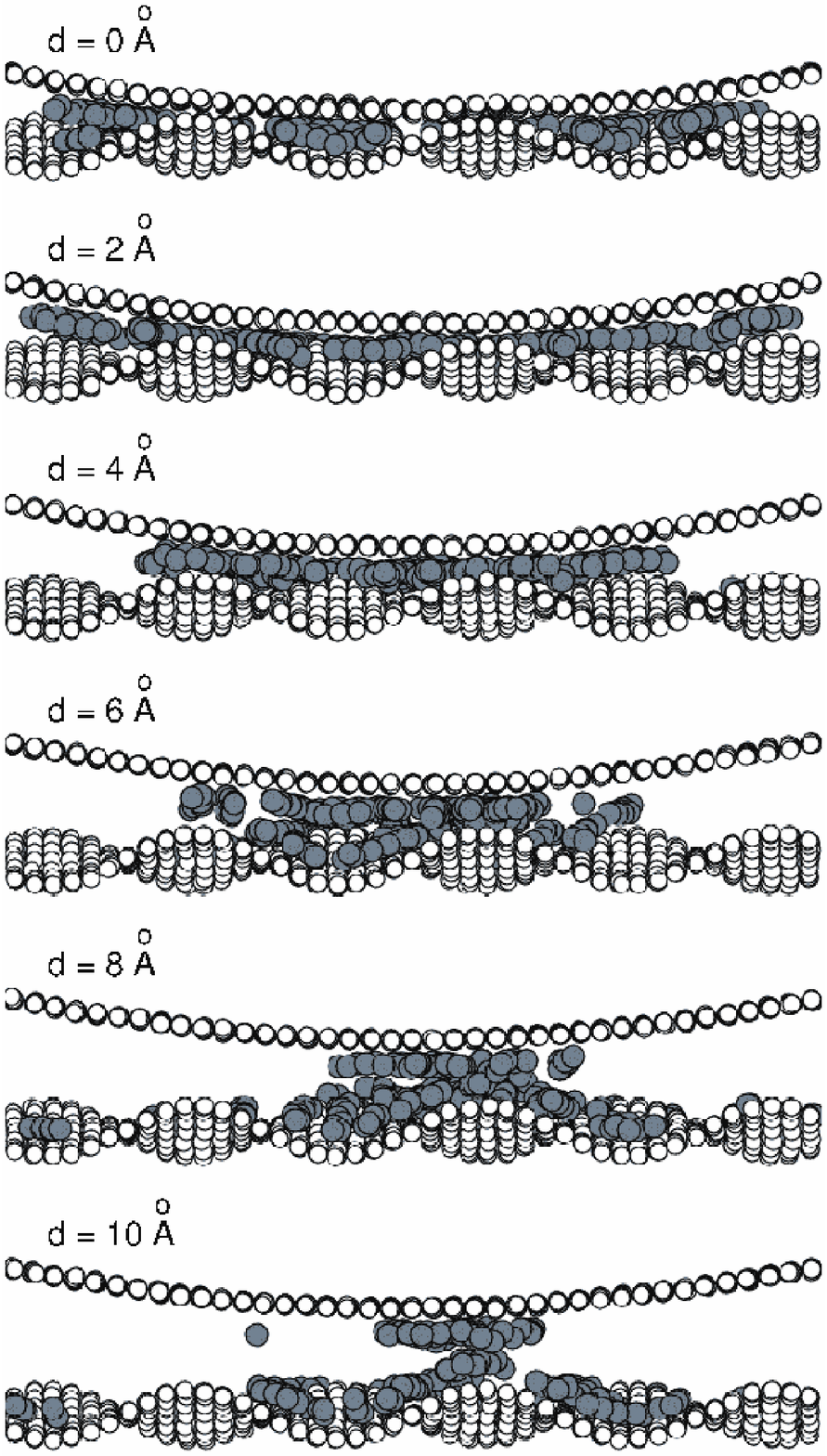} 
\end{center}
\caption{\label{Bos3}
Snapshot pictures (for six different
block positions) during retraction.
The snapshot pictures show the side view of the central
$108\,\mbox{\AA}\times 50\,\mbox{\AA}$ section (in the $xy$-plane) of the contact
area. Octane C$_8$H$_{18}$ was used as lubricant. Pull-off (retraction) velocity was
$v_{\rm z} = 1 \,{\rm m/s}$. For the corrugated substrate with about
$1/4$ monolayer of octane in the contact region.}
\end{figure}

\section{Summary and outlook}
\label{sec8}

Surface roughness has a huge influence on many common phenomena. It is the main reason
for why macroscopic bodies usually do not adhere to each other with any measurable strength.
For example, if the floor and the sole of the shoes were atomically smooth and clean,
it would be impossible to walk. The (near) absence of adhesion in most
situations is crucial for the function of many man-made and nature-made constructions.

The surface to volume ratio of solid objects increases as the lateral size
of the object decreases. The role of surface roughness becomes therefore more important 
as the size of objects decreases.
The present drive toward the miniaturization of mechanical devices,
e.g., Micro-Electro-Mechanical Systems (MEMS), 
requires a better
understanding of the role of surface roughness on, e.g., contact mechanics and adhesion.

Surface roughness is also of great importance for the function of many biological systems.
Flies, bugs, crickets and lizards have developed very soft layers on their feet
organs which allow them to attach and move on both very smooth and rough vertical solid walls,
e.g. stone walls or leafs. Another example is non-wetting coatings on plant surfaces based
on surface roughness on many different length scales (the so called
Lotus effect \cite{Inter1}).

The roughness of surfaces can nowadays be studied rather straightforwardly using standard 
equipments based on optical methods and on cantilever methods, e.g., the atomic force 
microscope (AFM). These methods
cover the whole length scale from atomic dimension to macroscopic distances. Thus, the AFM
can probe the surface profile from $\sim 1 \,{\rm nm}$ to $100 \,{\rm \mu m}$ and optical methods
from $\sim 1 \,{\rm \mu m}$ to kilometers. For randomly rough surfaces, the most important quantity
which can be deduced from the measured height profile is the surface roughness power spectrum. 
We have shown throughout this paper how the roughness power spectrum determines the contact 
mechanics and adhesion for solid objects in direct contact. It also governs rubber friction 
on rough substrates, e.g., friction of tires on a road surfaces, and influence
other phenomena of technological importance, e.g., the roughness
induced leaking of sealings.

Thus study and characterization of surface roughness is important not only
for understanding many natural and biological phenomena, but also for many 
technological processes. The present drive toward miniaturization and
the design of optimal systems by transfer of ideas from biology to
materials science (bionics) \cite{Inter1,Inter2},
is likely to accelerate the interest and efforts to study and
predict the influence of surface roughness on many phenomena.

\ack

We thank  Benz, Rosenberg and Israelachvili for unpublished experimental
information about surface topography of polymer films.
Work in SISSA was  sponsored through INFM PRA NANORUB, through MIUR COFIN 2003
and COFIN 2004, as well as FIRB RBAU01LX5H and FIRB RBAU017S8R, and by Regione
Friuli Venezia Giulia.
B.P.\ thanks EC for a ``Smart QuasiCrystals'' grant under the EC Program
``Promoting Competitive and Sustainable GROWTH''.

\appendix

\section{More about surface roughness}
\label{AppendixA}

Assume that a randomly rough surface is
described by the height profile $z=h(\ve{x})$, where $\ve{x} =(x,y)$ is
a 2D vector in the surface plane $z=0$, conveniently chosen so that $\langle h \rangle = 0$
and so that $\langle (h-\langle h \rangle)^2 \rangle$ is minimal. The statistical
properties of randomly rough surfaces is completely specified when all the correlation functions
\[
  \langle h (\ve{x}_1) h(\ve{x}_2) \rangle, \qquad
  \langle h(\ve{x}_1) h(\ve{x}_2) h(\ve{x}_3) \rangle, \  \ldots,
\]
are known. Here $\langle \ldots \rangle$ stands for ensemble averaging.
In most cases one assume that the correlation
functions involving an odd number of $h$-functions vanishes, while the correlation functions
involving an even number of $h$-functions can be decomposed into a product of
pair correlation functions, e.g., if we denote
$h(\ve{x}_1)=h_1$ and so on,
\[
 \langle h_1 h_2 h_3 h_4 \rangle = 
 \langle h_1 h_2 \rangle \langle h_3 h_4 \rangle +
 \langle h_1 h_3 \rangle \langle h_2 h_4 \rangle +
 \langle h_1 h_4 \rangle \langle h_2 h_3 \rangle
\]
In this case the surface is completely specified
by the surface roughness power spectrum $C(q)$, which is the Fourier transform of the
pair correlation function $\langle h(\ve{x}) h(\ve{0})\rangle$,
and the surface height
distribution $P_h$ is Gaussian (see below).

It is possible to generate surface roughness profiles, which are very
similar to experimentally observed surface profiles, as follows:
The surface height over a $L\times L$ square area can be expressed through
its Fourier series:
\begin{equation}
 h(\ve{x}) =\sum_{\ve{q}} B(\ve{q}) \rme^{\rmi[\ve{q}\cdot \ve{x}+\phi(\ve{q})]}
 \label{eqA1}
\end{equation}
where $\ve{q}$ spans all the vectors whose components are whole multiples
of $2\pi/L$. Since $h(\ve{x})$ is real,
$B(-\ve{q})=B(\ve{q})$ and $\phi(-\ve{q})= -\phi(\ve{q})$.
If $\phi(\ve{q})$ are independent random variables, uniformly distributed
in the interval $[0,2\pi[$, then one can easily show that higher order
correlation functions can be decomposed into a product of pair correlations
in the way described above.
Here we will demonstrate that in this case the height
probability distribution $P_h$ is always Gaussian, while any power spectrum
can be arbitrarily imposed by choosing properly the amplitudes $B(\ve{q})$

Let us consider a randomly rough surface described by equation (\ref{eqA1}).
Provided that the phases $\phi(\ve{q})$ are uniformly distributed and
independent, the statistical properties of the surface are translationally
invariant, thereafter
\[ \langle h(\ve{x}_1) h(\ve{x}_2) \rangle = C(\ve{x}_1-\ve{x}_2). \]
The surface roughness power spectrum is defined by
\begin{equation}
  C(\ve{q})={1\over (2\pi )^2} \int \rmd^2x
   \ C(\ve{x}) \rme^{-\rmi\ve{q}\cdot \ve{x}}
 \label{eqA2}
\end{equation}
By substituting (\ref{eqA1}) in (\ref{eqA2}) and using
\[
 \langle \rme^{\rmi\phi(\ve{q'})} \rme^{\rmi\phi(\ve{q''})}\rangle =
 \delta_{\ve{q'},-\ve{q''}}
\]
it follows
\[
  C(\ve{q}) =
  {1\over (2\pi )^2} \int \rmd^2x \ \sum_{\ve{q'}}
    |B(\ve{q'})|^2 \rme^{\rmi(\ve{q}-\ve{q'})\cdot \ve{x}}
  = \sum_{\ve{q'}} |B(\ve{q'})|^2 \delta (\ve{q}-\ve{q'}) \]
When the sampling of the $\ve{q}$-space is dense enough we can approach
the continuous limit by replacing
\[ \sum_{\ve{q}} \rightarrow {A\over (2\pi)^2}\int \rmd^2q, \]
where $A$ is the nominal surface area. This gives
\[ C(\ve{q})={A\over (2\pi)^2} |B(\ve{q})|^2 \]
Thus, if we choose
\[ B(\ve{q})= (2\pi/L) [C(\ve{q})]^{1/2}, \]
where $L=A^{1/2}$, then the surface roughness profile (\ref{eqA1}) has the
surface roughness power density $C(\ve{q})$. We can also guarantee that the
statistical properties of the rough surface are isotropic by imposing
$B(\ve{q})=B(q)$, then $C(\ve{q})=C(q)$ is a function of the
magnitude $q=|\ve{q}|$, but not of the direction of $\ve{q}$.
(The condition above shows the proportionality relation between the Fourier
transform of the height correlation and the amplitude of the corresponding
Fourier component of the surface profile. Such result, known as
Wiener-Khintchine theorem, is further discussed in \ref{AppendixC})

Let us prove that surfaces whose
correlation functions $\langle h_1 h_2 \ldots h_n \rangle$ vanish
for odd $n$ while for even $n$ 
can be decomposed into a product of
pair correlation functions, have Gaussian height probability distributions.
First note that the height probability distribution
\begin{eqnarray}
  P_h = \langle \delta [h-h(\ve{x})] \rangle & = &
  {1\over 2 \pi}\int \rmd\alpha 
   \ \langle \rme^{\rmi\alpha [h-h(\ve{x})]}\rangle \nonumber \\
  & = &
  {1\over 2 \pi}\int \rmd\alpha \ \rme^{\rmi\alpha h}
    \langle \rme^{-\rmi\alpha h(\ve{x})}\rangle
  \label{eqA3}
\end{eqnarray}
But
\[
 \langle \rme^{-\rmi\alpha h(\ve{x})}\rangle =
 \sum_n {(-\rmi\alpha)^n\over n!} \langle [h(\ve{x})]^n\rangle
\]
However, $\langle [h(\ve{x})]^n\rangle $ vanishes for odd $n$. Thus, writing $n=2m$ ($m=0,1,...$)
gives
\begin{equation}
  \langle \rme^{-\rmi\alpha h(\ve{x})}\rangle =
  \sum_m {(-\rmi\alpha)^{2m}\over (2m)!} \langle [h(\ve{x})]^{2m}\rangle
  \label{eqA4}
\end{equation}
We will now make use of the following equation
\begin{equation}
  \langle [h(\ve{x})]^{2m}\rangle =
   {(2m)! \over m! 2^m} \langle h^2\rangle^m
  \label{eqA5}
\end{equation}
The prefactor $(2m)!/m! 2^m$ is easy to understand: when decomposing
$\langle h_1 h_2 \ldots h_{2m}\rangle$ there are
$(2m)!$ possible ordering of $h_1$, $h_2$, \ldots, $h_{2m}$.
However, many of these products of pair correlation functions are identical. Thus, since
in the
pair correlation function $\langle h_i h_j\rangle$
the order of $h_i$ and $h_j$
is irrelevant,
we must divide $(2m)!$ with
$2^m$. Furthermore, permutation of
the $m$ different pair correlation functions
in a product term results in
$m!$ identical terms,
which is the origin of the $1/m!$ factor in (\ref{eqA5}).
Substituting (\ref{eqA5}) into (\ref{eqA4}) gives
\[
  \langle \rme^{-\rmi\alpha h(\ve{x})}\rangle =
  \sum_m {(-\rmi\alpha)^{2m}\over m!  2^m} (\langle h^2\rangle)^m
  = \rme^{-{1\over 2} \alpha^2 \langle h^2\rangle}
\]
Substituting this result into (\ref{eqA3}) and performing the
integration over $\alpha$ gives
\[ P_h
= {1\over (2\pi )^{1/2} \sigma } \rme^{-h^2/2\sigma^2}, \]
where $\sigma$ is the rms roughness amplitude,
$\sigma^2 = \langle [h(\ve{x})]^2\rangle$.

\begin{figure}
\begin{center}
\includegraphics[width=0.80\textwidth]{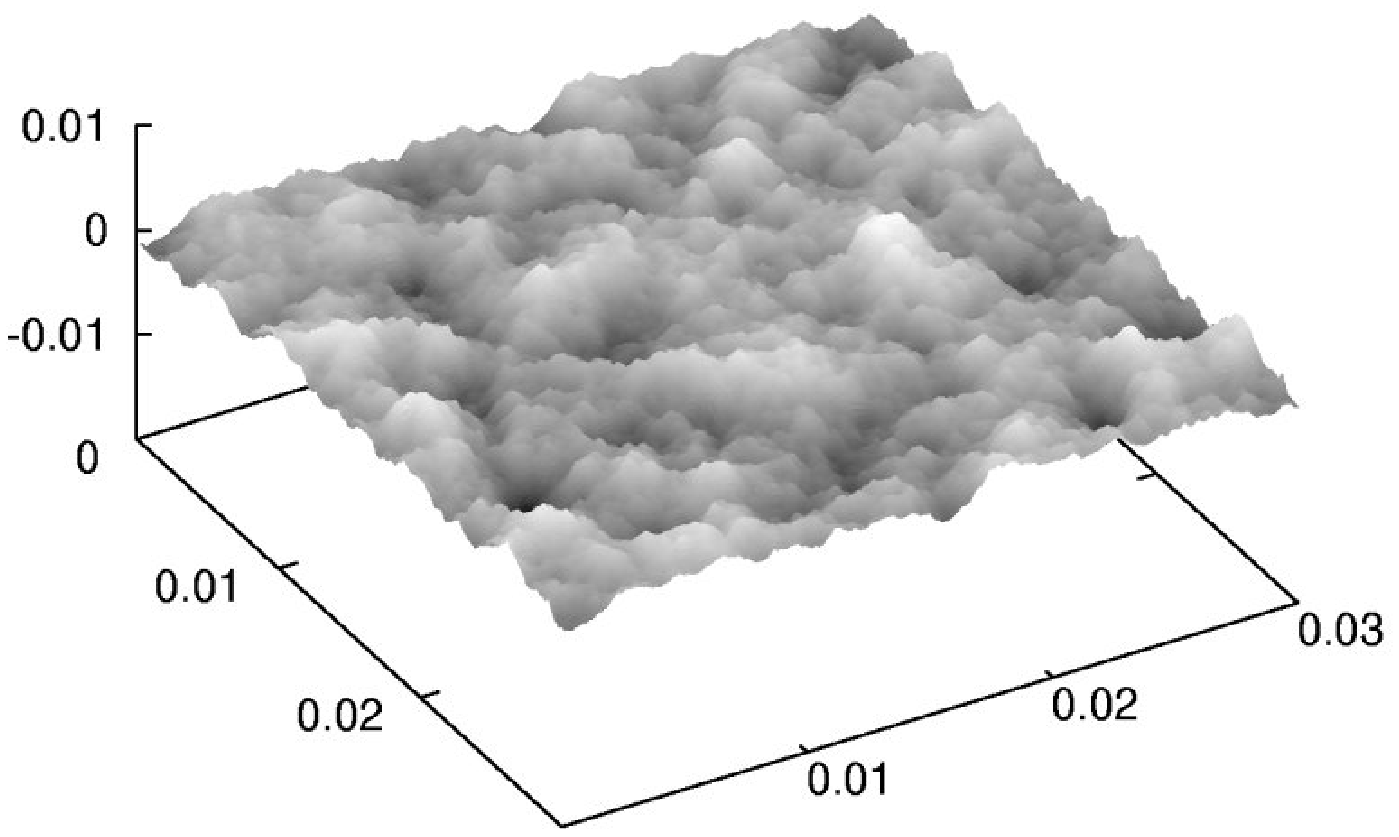} 
\end{center}
\caption{\label{generatedsurface}
Computer generated rough substrate. The surface has root mean square
roughness 0.001\,m, roll-off wave vector $q_0 = 1000$\,m$^{-1}$, and Hurst
exponent $H=0.8$. Length in the picture are in meters.}
\end{figure}

Fig.~\ref{generatedsurface} shows a randomly Gaussian random surface
generated with the method discussed above, through formula (\ref{eqA1}).
The power spectrum is imposed to be proportional to $q^{-3.6}$
for large wave vectors, so that the Hurst exponent is 0.8 and the
fractal dimension is 2.2.

\section{Hurst exponent and fractal dimension}
\label{AppendixB}

A surface is said to be self-similar if it has the same statistical
properties as a magnified version of
itself. Thus, if $z=h(x,y)$ is the equation of a self-similar
surface, then its magnified version $z=\lambda h(x/\lambda, y/\lambda)$
cannot be distinguished from the original. A self-affine surface is
analogous, except that the magnification factor along the $z$ direction
differs from the in-plane magnification factor $\lambda$. In particular
the Hurst exponent $H$ defines the scaling factor $\lambda^{H}$ along the
$z$ direction in order to recover the same statistical properties.
The transformed surface $z=\lambda^{H}h(x/\lambda,y/\lambda)$ ``looks''
exactly as the original one $z=h(x,y)$.
The Hurst exponent is in between 0 and 1, the latter being the case of
a self-similar surface.

The general form of the power spectrum for an isotropic self-affine fractal surface is
easy to derive. We have
\[
  C(q)={1\over (2\pi )^2}\int \rmd^2x \ \langle h(\ve{x})h(\ve{0})\rangle
  \rme^{-\rmi\ve{q}\cdot \ve{x}}
\]
Now let us write $\ve{x}=\ve{x'}/\lambda$ so that
\[
 C(q)={1\over (2\pi )^2}\int \rmd^2x' \ \lambda^{-2} \langle h(\ve{x'}/\lambda)h(\ve{0})\rangle
\rme^{-\rmi\ve{q}\cdot \ve{x'}/\lambda}
\]
But since for a self affine fractal surface
\[ \langle \lambda^H h(\ve{x'}/\lambda)\lambda^Hh(\ve{0})\rangle =
\langle h(\ve{x'})h(\ve{0})\rangle
\]
we get
\[ C(q)={1\over (2\pi )^2}\int \rmd^2x' \ \lambda^{-2-2H} \langle h(\ve{x'})h(\ve{0})\rangle
\rme^{-\rmi\ve{q}\cdot \ve{x'}/\lambda}
\]
Thus, if we choose $\lambda = q$ and denote $\hat \ve{q} = \ve{q}/q$ we get
\[
  C(q)= q^{-2(1+H)} {1\over (2\pi )^2}\int \rmd^2x' \ \langle h(\ve{x'})h(\ve{0})\rangle
\rme^{-\rmi\hat \ve{q}\cdot \ve{x'}}
\]
Thus the power spectrum of a self-affine surface
decreases as $q^{-2(H+1)}$ with increasing wave vector $q=|\ve{q}|$.

The Hurst exponent is directly related to the fractal dimension $D_{\rm f}$ through the
formula $D_{\rm f}=3-H$. The proof is straightforward: consider a patch of the
surface with extension $L\times L$ in the two in-plane directions $x$ and $y$.
The fractal dimension can be defined through the number of
cubes of size $\Delta$ required to cover completely the surface: $N(\Delta)
\propto \Delta^{-D_{\rm f}}$ for $\Delta\to 0$. To cover the $L\times L$ area a number
of cubes $(L/\Delta)^2$ is required, that is the fractal dimension cannot be
smaller than 2. In any $\Delta\times\Delta$ sub-domain the $z$-coordinate of the
surface spans a range of values $\Delta z$ proportional to $\Delta^{H}$,
because of the self-affine property. Actually $\Delta z = K (\Delta/L)^{H}$,
where $K$ is the range of $z$ values spanned by $h(x,y)$ over the whole
$L\times L$ domain.  Provided that $\Delta$ is small enough, the corresponding
$\Delta z$ gets larger than $\Delta$ itself, since $H<1$. Thereafter in any
sub-domain of size $\Delta$ we have to employ $K(\Delta/L)^{H}/\Delta \propto
\Delta^{H-1}$ cubes, and the total number of cubes necessary to cover the
surface is thus proportional to $\Delta^{-(3-H)}$, i.e.
\[
 D_{\rm f}=3-H.
\]

\section{Moments of power spectra}
\label{AppendixC}

\begin{figure}
\begin{center}
\includegraphics[width=0.70\textwidth]{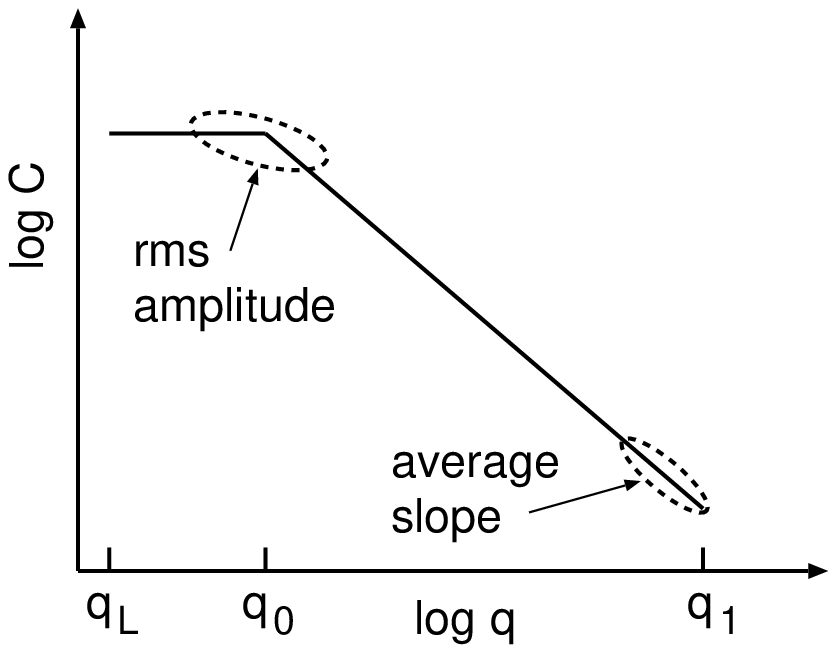} 
\end{center}
\caption{\label{Cq2}
Surface roughness power spectra of a surface which is self affine fractal for
$q_1>q>q_0$.
The {\em rms} roughness amplitude, and the
average slope (and the average curvature), are determined mainly by
the encircled regions of the power spectrum.}
\end{figure}

We consider rough surfaces for which the statistical properties are
isotropic and translational invariant. Thus the surface roughness power
spectrum
\begin{equation}
  C(q)= {1\over (2\pi )^2 } \int \rmd^2x
   \ \langle h(\ve{x}+\ve{x}')h(\ve{x}')\rangle
   \rme^{-\rmi\ve{q}\cdot \ve{x}}
 \label{eqC1}
\end{equation}
is independent of $\ve{x'}$ and of the orientation of the wavevector
$\ve{q}$. If we express $h(\ve{x})$ through its Fourier transform
\begin{equation}
 h(\ve{x}) = \int \rmd^2q \ h(\ve{q}) e^{i\ve{q}\cdot \ve{x}}
 \label{eqC2}
\end{equation}
and we substitute it into (\ref{eqC1}) we get
\[
  C(q) = {1\over (2\pi )^2 } \int \rmd^2q' \int \rmd^2q'' \int \rmd^2x
    \ \langle h(\ve{q'})h(\ve{q''})\rangle 
    \rme^{\rmi (\ve{q'}-\ve{q}) \cdot \ve{x}}
    \rme^{\rmi (\ve{q'}+\ve{q''}) \cdot \ve{x'}} \]
Generally $h(\ve{q})$ is not an ordinary function but is a distribution,
thus $\langle h(\ve{q'}) h(\ve{q''})\rangle$ can diverge. However
we can replace $h(\ve{q})$ with $h_A(\ve{q})$, the Fourier transform of
$h(\ve{x})$ restricted to a square area $A$, i.e.,
\[
 h_A(\ve{q}) = {1\over (2\pi )^2 }
                 \int_{-\sqrt{A}/2}^{+\sqrt{A}/2} \rmd x_1
                 \int_{-\sqrt{A}/2}^{+\sqrt{A}/2} \rmd x_2
                   \ h(\ve{x}) \rme^{-\rmi\ve{q}\cdot\ve{x}},
\]
and the equations above still hold in the limit $A\to\infty$.
\[
  C(q)= {1\over (2\pi )^2 } \int \rmd^2q' \int \rmd^2q'' \int \rmd^2x
    \ \langle h_{A}(\ve{q'})h_{A}(\ve{q''})\rangle 
    \rme^{\rmi (\ve{q'}-\ve{q}) \cdot \ve{x}}
    \rme^{\rmi (\ve{q'}+\ve{q''}) \cdot \ve{x'}}
\]
To simplify this equation we use the standard relation
\begin{equation}
 \int \rmd^2x \ \rme^{\rmi\ve{q}\cdot \ve{x}} =
 (2\pi)^2 \delta (\ve{q})
 \label{eqC3}
\end{equation}
to get
\begin{equation}
  C(q) =\int \rmd^2q''
   \ \langle h_{A}(\ve{q})h_{A}(\ve{q''})\rangle
   \rme^{\rmi (\ve{q}+\ve{q''}) \cdot \ve{x'}}
  \label{eqC4}
\end{equation}
This expression is independent of $\ve{x'}$, at least in the limit $A\to\infty$,
hence
$\langle h(\ve{q})h(\ve{q''})\rangle =
 \lim_{A\to\infty} \langle h_{A}(\ve{q})h_{A}(\ve{q''})\rangle$ must vanish when
$\ve{q}+\ve{q''}\ne\ve{0}$.
Moreover the isotropy implies that $\langle h(\ve{q})h(-\ve{q})\rangle $
depends on $\ve{q}$ only through its modulus $q$; thereafter:
\begin{equation}
  \langle h(\ve{q})h(\ve{q'})\rangle = C(q) \delta (\ve{q}+\ve{q'})
  \label{eqC5}
\end{equation}
Since (\ref{eqC4}) is independent of $\ve{x'}$ we may integrate over
$\ve{x'}$ and divide by the surface area $A$ to get
\begin{eqnarray}
 C(q) & = &
   {1\over A} \int \rmd^2 x' \int \rmd^2q''
   \ \langle h_{A}(\ve{q})h_{A}(\ve{q''})\rangle
   \rme^{\rmi (\ve{q}+\ve{q''}) \cdot \ve{x'}} \nonumber \\
 & = & {(2\pi)^2 \over A}
   \ \langle h(\ve{q})h(-\ve{q})\rangle, \nonumber
\end{eqnarray}
where we have again used (\ref{eqC3}).
Since $h(\ve{x})$ is real, $h_{A}(-\ve{q})= h_{A}^*(\ve{q})$, and we can write
\begin{equation}
 C(q) =
  {(2\pi)^2 \over A} \ \langle |h_{A}(\ve{q})|^2 \rangle
 \label{eqC6}
\end{equation}
Equations (\ref{eqC5}) and (\ref{eqC6}) are very important and useful
equations of general validity.
As an application, let us calculate the average surface slope. Using
(\ref{eqC2}) we get
\[
  \langle (\nabla h )^2 \rangle =
  \int \rmd^2q \rmd^2q' \ (-\ve{q}\cdot \ve{q'})
  \langle h(\ve{q})h(\ve{q'})\rangle \rme^{\rmi(\ve{q}+\ve{q'})\cdot x}
\]
Using (\ref{eqC5}) this gives
\[
  \langle (\nabla h )^2 \rangle =
  \int \rmd^2 q \ q^2 C(q)= 2\pi \int \rmd q \ q^3 C(q)
\]
For a self affine fractal surface $C(q)\sim q^{-2(1+H)}$ and
\[
  \langle (\nabla h )^2 \rangle \sim \int_{q_0}^{q_1} \rmd q \ q^{1-2H}
  \sim q_1^{2(1-H)}-q_0^{2(1-H)}\approx q_1^{2(1-H)}
\]
if $q_1\gg q_0$. Thus the average slope (and the average curvature) is
determined by the shortest wavelength roughness components, as indicated
in Fig.~\ref{Cq2}. On the other hand, as shown in Sec.~\ref{sec2}, the
{\em rms}-roughness amplitude is determined by mainly by the longest
surface roughness wavelength components, i.e., the region around
$q\sim q_0$ (see Fig.~\ref{Cq2}).

\section{Numerical recipes for calculating power spectra}
\label{AppendixD}

Here we describe how the
roughness power spectrum $C(q)$ can be numerically determined from the surface
height $h(\ve{x})$
measured over a square area $0<x<L$ and $0<y<L$.

The surface roughness power spectrum is given by
\[
  C(\ve{q}) = {1\over (2\pi)^2} \int \rmd^2x \ \langle h(\ve{x})
  h(\ve{0})\rangle \rme^{-\rmi \ve{q}\cdot \ve{x}},
\]
which also can be written as (see \ref{AppendixC})
\[
  C(\ve{q})={(2\pi )^2\over A}\langle |h_{A}(\ve{q})|^2\rangle,
\]
where $A=L^2$ is the surface area under study and
\[
  h_{A}(\ve{q}) = {1\over (2\pi )^2}\int_{A} \rmd^2x \ h(\ve{x})
   \rme^{-\rmi\ve{q}\cdot \ve{x}}.
\]
In order to calculate $C(q)$ numerically from the measured height profile
$h(\ve{x})$, we assume that the measurement points form a square mesh
with the lattice constant $a$, and with $N$ data points along the $x$
and $y$-coordinates.
In other words, $h(\ve{x})$ is sampled and its values are known
only in the points
\[ \ve{x}= (n_x , n_y ) a = \ve{n} a = \ve{x}_{\ve{n}}, \]
where $n_x = 1, 2, \ldots, N$ and $n_y=1, 2, \ldots, N$ are integers.
The integral for the Fourier transform $h_{A}$ can be approximated by a
discrete sum:
\[
  h_{A}(\ve{q}) \approx
  {a^2\over (2\pi )^2} \sum_{\ve{n}}
      h_{\ve{n}} \rme^{-\rmi(q_x n_x a+q_y n_y a)}
\]
where $h_{\ve{n}} = h(\ve{x}_{\ve{n}})$. The Fourier transform
can be sampled too, by using a grid with lattice spacing
$q_L = 2\pi/L = 2\pi / (Na)$,
since the finite size of the surface is limiting the resolution in
the $\ve{q}$-space. Thus $\ve{q}=(q_x,q_y)= (2\pi m_x/L, 2\pi m_y/L)$
where $m_x$ and $m_y$ are integer numbers between $0$ and $N-1$ (any
larger value of $m_x$ or $m_y$ corresponds to wavelengths shorter than
$a$, and does not carry extra information because of the finite
resolution in the real space).
Now we can write:
\begin{equation}
 h_{A}(\ve{q}) \approx 
  {a^2\over (2\pi )^2} \sum_{\ve{n}}
      h_{\ve{n}} \rme^{-\rmi \frac{2\pi}{N} (m_x n_x + m_y n_y)} =
  {a^2\over (2\pi )^2}  H_{\ve{m}}, 
 \label{eqD1}
\end{equation}
where $H_{\ve{m}}$ is the 2-dimensional {\em Discrete Fourier Transform}
of $h_{\ve{n}}$. The big advantage of evaluating the power spectrum
through the Fourier transform of $h(\ve{x})$, instead of carrying on
the calculation of the correlation, is that there exists an efficient
algorithm for the Discrete Fourier Transform: the well known
Fast Fourier Transform (FFT) method. Indeed the computational time scales
as $N^2 \log(N)$ (in two dimensions), while the convolution integral
to estimate the 2-dimensional correlation would imply a computational
time that scales as $N^4$.

A crude result can be obtained by substituting (\ref{eqD1}) into
(\ref{eqC6}): the power spectrum obtained so far would be identical
to the one resulting from the Fourier transform of the correlation,
provided that the correlation is calculated applying periodic boundary
conditions over the surface of area $A$. But real surfaces are
portions of larger samples, where there is no match at the boundary:
the periodic boundary conditions introduce discontinuities,
corresponding to spurious Fourier components at the larger wave vectors
of the spectrum. Thereafter it is necessary to filter them out.
A typical solution consists in multiplying the height profile $h_{\ve{n}}$
by a {\em windowing function} that goes gradually to $0$ at the edge of
the surface area and is almost constant around the centre; and then
carrying on the Fast Fourier Transform with the adjusted data.
This procedure is described in detail in Ref.~\cite{Numerical}.

Note also that from
\[ \langle h(\ve{x})h(\ve{0})\rangle = \int d^2q \ C(\ve{q}) e^{i\ve{q}\cdot \ve{x}} \]
it follows that
\begin{equation} \langle h^2\rangle = \int d^2q \ C(\ve{q}).\label{eqD2} \end{equation}
But the mean of the square of the surface roughness amplitude can also be obtained directly
from the experimental data via
\begin{equation} \langle h^2 \rangle = {1\over N^2 } \sum_{\ve{n}} (h_{\ve{n}} -\bar h)^2. \label{eqD3} \end{equation}
It is a good check of the numerical accuracy
in the evaluation of the surface roughness power spectra to show that
the $\langle h^2\rangle$ calculated from (\ref{eqD2}) and (\ref{eqD3})
gives identical results.

Let us assume that the surface roughness is isotropic. In this case
$C(\ve{q})$ is independent of the direction of $\ve{q}$, i.e.,
$C(\ve{q})=C(q)$. We can make use of this fact to reduce the statistical
noise in $C(q)$ by performing an angular average in (\ref{eqD2}).
Let us for simplicity write
$C(\ve{q}) = C(\ve{m} q_L)$ as $C(m_x,m_y)$, where
$q_L = 2\pi/L$ is the lattice spacing in the $\ve{q}$-space,
$m_x=0$, $1$, \ldots, $N-1$ and similarly for $m_y$.
In an analogous way we write $C(q)=C(m q_L)$ as $C(m)$, $m=1,2,..., N/2$.
We have
\begin{equation}
 \int \rmd^2 q \ C(\ve{q}) = 2\pi \int_0^{q_1} \rmd q \ qC(q).
 \label{eqD4}
\end{equation}
The discretized version of (\ref{eqD4}) takes the form
\begin{equation} q_L^2 \sum_{m_x=0}^{N/2}\sum_{m_y=0}^{N/2}
\bar C(m_x,m_y) = 2\pi q_L^2 \sum_m m C(m)\label{eqD5} \end{equation}
where
\begin{eqnarray}
 \bar C(m_x,m_y) & = & C(m_x,m_y)+C(N-m_x, m_y) \nonumber \\
                 & + & C(m_x,N-m_y)+C(N-m_x,N-m_y) \nonumber
\end{eqnarray}
corresponds to $C(q_x,q_y)+C(-q_x,q_y)+C(q_x,-q_y)+C(-q_x,-q_y)$
(remember that the Discrete Fourier Transform is periodic with period
$N$).
Eq.~(\ref{eqD5}) is obeyed if we define
\[ C(m) =
{1\over  2\pi m} \sum_{-1/2 < |\ve{m'}| -m < 1/2} \bar  C(m'_x,m'_y) \]
With this definition the ``sum-rule''
\[ \langle h^2\rangle = 2\pi \int dq \ qC(q) = 2 \pi q_L^2 \sum_{m=1}^{N/2} m C(m) \]
will be exactly obeyed.

\begin{figure}[thb]
\begin{center}
\includegraphics[width=0.80\textwidth]{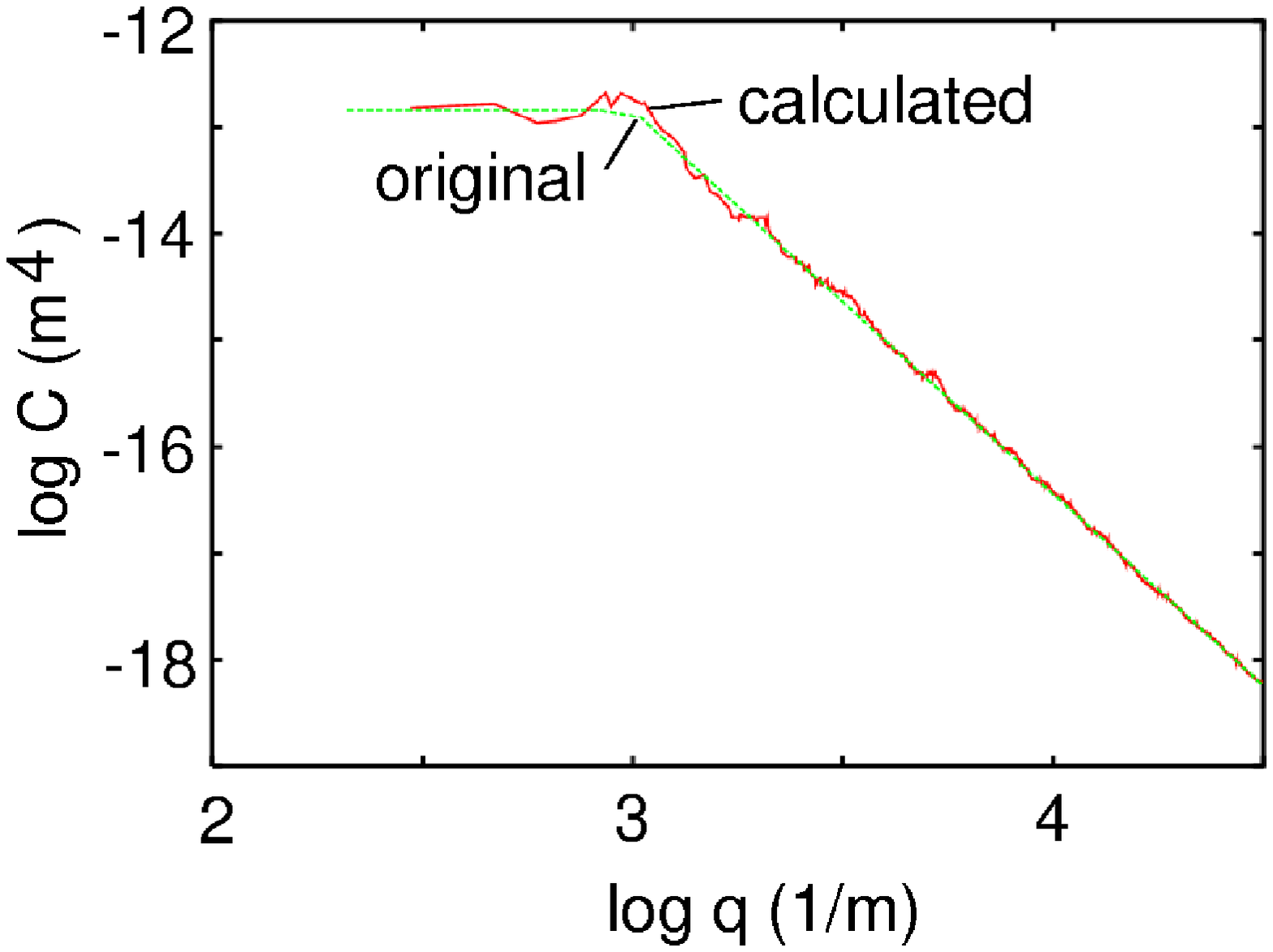} 
\end{center}
\caption{\label{spectrumgenerated}
Red line: calculated power spectrum for the surface shown
in Fig.~\protect\ref{generatedsurface}. Green line: the
original power spectrum, employed to generate the surface.
The surface extends over an area of $0.03\,$m $\times 0.03$\,m
and the sampling length is $a=0.0001$\,m.}
\end{figure}

As an application in Fig.~\ref{spectrumgenerated} we show the spectrum calculated
with this method for the same surface described in Fig.~\ref{generatedsurface}.
The agreement between the original spectrum and the calculated one provides
evidence of the accuracy that can be achieved.

\end{document}